\documentclass[fleqn,10pt]{wlscirep}
\usepackage[utf8]{inputenc}
\usepackage[T1]{fontenc}
\usepackage{bm}
\usepackage{amsmath}
\usepackage{multirow}
\usepackage{caption}
\usepackage{subcaption}
\usepackage{lineno}
\usepackage{color}
 \usepackage{url}
\usepackage{tabularray}
\bibstyle{naturemag}
\newcommand\at[2]{\left.#1\right|_{#2}}
\usepackage{algorithm,algpseudocode}
\usepackage{algpseudocodex}
\newcounter{algsubstate}
\renewcommand{\thealgsubstate}{\alph{algsubstate}}
\newenvironment{algsubstates}
  {\setcounter{algsubstate}{0}%
   \renewcommand{\State}{%
     \stepcounter{algsubstate}%
     \Statex {\footnotesize\thealgsubstate:}\space}}
  {}

\usepackage{fancyhdr}

\title{Improving deep neural network generalization and robustness to background bias via layer-wise relevance propagation optimization}

\author[1,2,*]{Pedro R.A.S. Bassi}
\author[3]{Sergio S.J. Dertkigil}
\author[1,4,*]{Andrea Cavalli}
\affil[1]{Alma Mater Studiorum - University of Bologna, Bologna, Italy}
\affil[2]{Center for Biomolecular Nanotechnologies, Istituto Italiano di Tecnologia, 73010, Arnesano (LE), Italy}
\affil[3]{School of Medical Sciences, University of Campinas (UNICAMP), Campinas (SP), Brazil}
\affil[4]{Istituto Italiano di Tecnologia, 16163, Genova (GE), Italy}
\affil[*]{e-mail: pedro.salvadorbassi2@unibo.it, andrea.cavalli@unibo.it}

\begin{abstract}

Features in images' backgrounds can spuriously correlate with the images' classes, representing background bias. They can influence the classifier's decisions, causing shortcut learning (Clever Hans effect). The phenomenon generates deep neural networks (DNNs) that perform well on standard evaluation datasets but generalize poorly to real-world data. Layer-wise Relevance Propagation (LRP) explains DNNs' decisions. Here, we show that the optimization of LRP heatmaps can minimize the background bias influence on deep classifiers, hindering shortcut learning. By not increasing run-time computational cost, the approach is light and fast. Furthermore, it applies to virtually any classification architecture. After injecting synthetic bias in images' backgrounds, we compared our approach (dubbed ISNet) to eight state-of-the-art DNNs, quantitatively demonstrating its superior robustness to background bias. Mixed datasets are common for COVID-19 and tuberculosis classification with chest X-rays, fostering background bias. By focusing on the lungs, the ISNet reduced shortcut learning. Thus, its generalization performance on external (out-of-distribution) test databases significantly surpassed all implemented benchmark models.

\end{abstract}

\begin{document}

\rhead{Preprint. The corresponding article was published by Nature Communications and is available at: \href{https://doi.org/10.1038/s41467-023-44371-z}{https://doi.org/10.1038/s41467-023-44371-z} or \href{https://www.nature.com/articles/s41467-023-44371-z}{https://www.nature.com/articles/s41467-023-44371-z}}

\flushbottom
\maketitle

\thispagestyle{empty}

\section{Introduction}

Deep neural networks (DNNs) revolutionized image classification. Counting on millions of trainable parameters, the models proved capable of analyzing entire images, becoming a new standard in many different fields. However, the features contained in images' backgrounds may have a strong and detrimental effect on the classification process. Background features can unintentionally correlate with the images' classes, thus representing background bias, also called spurious correlations. Trained in such environments, a DNN can learn to base its decisions not only on relevant image regions, but also on background features. The influence of background bias over the classifier deteriorates its capacity to analyze the images' relevant features and reduces the trustworthiness of its decisions. The biased model will perform artificially well on the training dataset, and on evaluation databases that contain the same background biases. This is a common condition for test datasets that are independent and identically distributed (i.i.d.) in relation to the training data (e.g., evaluation databases created by randomly slipping a dataset into a training and a testing partition). Nevertheless, the DNN will not be adequate for a practical application, which may not portray the same background biases. This scenario favors shortcut learning (or Clever Hans effect)\cite{ShortcutLearning}, a condition where deep neural networks learn decision rules that perform well on standard benchmarks, but poorly on real-world applications. Unlike an overfitted DNN, with shortcut learning the model will perform well on an i.i.d. test dataset, but it will fail to generalize and be accurate on out-of-distribution (o.o.d.) databases. Furthermore, the image features that cause shortcut learning can be difficult for a person to identify\cite{ShortcutLearning}. This study presents a DNN architecture and training strategy to more efficiently deal with one of the main causes of shortcut learning: background bias.

Layer-wise relevance propagation (LRP)\cite{LRP} is an explanation technique designed to create heatmaps for deep classifiers. LRP heatmaps are graphics that explain the model's behavior by making explicit how each part of an input image influenced the DNN output. We can create a heatmap for an input image, explaining the classifier score for a given class. Positive and negative heatmap values (dubbed relevance) indicate image areas that increased or decreased the classifier's confidence in the class, respectively. Areas with relevance closer to zero were less important for the classifier's decision. Thus, high relevance magnitude in the image background indicates strong background bias influence over the classifier's output.

This study proposes producing differentiable LRP heatmaps during training and optimizing them with a loss function. The function, named heatmap loss, employs ground-truth semantic segmentation masks to identify heatmap regions that correspond to the input image’s background, and minimizes the LRP relevance magnitude in these regions. We called the minimization of the heatmap loss, alongside a classification loss, Background Relevance Minimization (BRM). BRM hinders classifier's decision rules based on background features. Since the training technique uses segmentation masks and LRP heatmaps to teach the classifier which image regions it must focus on, it can be regarded as an explanation- and segmentation-based spatial attention mechanism. A DNN trained with BRM can implicitly distinguish the input image’s foreground and background features (i.e., implicitly segmenting a figure), and make decisions based only on the foreground. Accordingly, we named a classifier trained with BRM an Implicit Segmentation Neural Network (ISNet). 

By minimizing the influence of background features over the DNN's outputs, the ISNet hinders the shortcut learning caused by background bias, improving out-of-distribution (o.o.d.) generalization. Despite its name, the ISNet is a classifier, not a segmenter. It does not produce segmentation outputs. After the training procedure, the ground-truth foreground masks and the creation of LRP heatmaps are no longer needed. Thus, the ISNet introduces no computational cost at run-time. We use BRM to train a classifier, defined by a backbone architecture, and the resulting ISNet has the same structure as its backbone. The ISNet's run-time efficiency is important for deploying DNNs in mobile or less powerful devices. Furthermore, the ISNet is versatile, because virtually any classification architecture can be used as its backbone.

A segmentation-classification pipeline is a standard strategy to avoid the influence of background bias over a classifier's decisions. It is a sequence of two networks. The first DNN segments the image's foreground. Afterwards, its output is used to erase the background, creating a segmented image that the second DNN classifies. Running two large DNNs sequentially leads to heavy memory and time requirements for the pipeline. We consider it as a benchmark in this study. Moreover, we also compare the ISNet to DNNs optimizing alternative explanation heatmaps. Hierarchical Attention Mining\cite{HAM} (HAM, in Appendix \ref{supApplications}) and Guided Attention Inference Network\cite{GAIN} (GAIN) optimize Grad-CAM\cite{GradCAM}, Right for the Right Reasons\cite{RRR} (RRR) optimizes input gradients\cite{saliency}, and the ISNet Grad*Input is an ablation experiment where we substitute the ISNet's LRP heatmaps by Gradient*Input\cite{GradInput} explanations. We also contrast the ISNet to a standard deep classifier, and to other state-of-the-art DNN architectures designed to minimize shortcut learning or control classifier attention. They are: Attention Gated Networks\cite{AGNet} (AG-Sononet) and the Vision Transformer\cite{VisionTransformer}, exemplifying attention mechanisms that do not learn from semantic segmentation masks; and a multi-task U-Net, which simultaneously produces classification scores and semantic segmentation outputs. 

We present multiple classification experiments, designed to assess the ISNet's capacity of hindering the shortcut learning caused by background bias. Currently, the most popular open databases of COVID-19 chest X-rays contain either no or few COVID-19 negative/control cases\cite{GitCovidSet}\textsuperscript{,}\cite{BrixiaSet}. Due to this limitation, to this date most studies resorted to mixed source datasets to train DNNs to differentiate between COVID-19 patients, healthy people, and patients with other diseases (e.g., non-COVID-19 pneumonia). In these datasets, different classes come from different databases assembled in distinct hospitals and cities. The dissimilar sources may contain different biases, which may help DNNs to classify the images according to their source dataset, rather than the disease symptoms. One study\cite{critic} accurately determined the source datasets after removing all lung regions from the images, proving the presence of background bias. A review\cite{ShortcutCovid} concluded that, if models are allowed to analyze the whole X-ray or a rectangular bounding box around the lungs, they tend to strongly focus on areas outside of the lungs. Thus, they fail to generalize or achieve satisfactory performance on external, o.o.d. datasets, with dissimilar sources in relation to the training images. Moreover, the review identified that the problem is a cause of shortcut learning\cite{ShortcutLearning}. A paper\cite{NatureCovidBias} utilized external datasets to evaluate traditional COVID-19 detection DNNs, whose reported results had been highly positive, and it demonstrated a strong drop in their performances. For these reasons, researchers have resorted to testing on o.o.d. datasets to understand the bias and generalization capability of DNNs trained on mixed databases. The approach shows reduced and more realistic performances in relation to the standard (i.i.d.) evaluation\cite{ShortcutCovid}\textsuperscript{,}\cite{bassi2021covid19}\textsuperscript{,}\cite{covidSegmentation}. A common conclusion of these works is that using a segmentation-classification pipeline (segmenting the lungs before classification) improves generalization capability, reducing the bias induced by mixed training datasets\cite{bassi2021covid19}\textsuperscript{,}\cite{covidSegmentation}. The ISNet shall be able to focus only on the lungs, and we consider that the task of COVID-19 detection using the usual mixed datasets of chest X-rays will be useful to demonstrate its benefits.

We trained the ISNet on a mixed dataset based on one of the largest open COVID-19 chest X-ray databases\cite{BrixiaSet}, with the objective of distinguishing COVID-19 positive cases, normal images, and non-COVID-19 pneumonia. The two diseases have similar symptoms, making their differentiation non-trivial, and both produce signs in chest X-rays. Examples of COVID-19 signs are bilateral radiographic abnormalities, ground-glass opacity, and interstitial abnormalities\cite{clinicalCOVID}. Examples of pneumonia signals are poorly defined small centrilobular nodules, airspace consolidation, and bilateral asymmetric ground-glass opacity\cite{clinicalPneumonia}. We evaluate the optimized DNN in a cross-dataset approach, using images collected from external locations in relation to the training samples. Evaluation with an o.o.d. dataset\cite{ShortcutCovid} shall allow us to assess whether the ISNet can reliably ignore the image background, reducing shortcut learning and increasing generalization performance.

The dataset mixing issue is not exclusive to COVID-19 detection. Instead, the technique is necessary whenever researchers need to use a classification database that does not contain all the required classes. Sometimes, such databases are the largest ones, a desirable quality for deep learning. This is the case in COVID-19 detection, and we find another example in tuberculosis (TB) detection. To the best of our knowledge, the tuberculosis X-ray dataset from TB Portals\cite{TBPortals} is the open-source database with the largest number of tuberculosis-positive X-rays to date. However, the dataset is composed of TB-positive cases only. Thus, dataset mixing is required to use the TB Portals data for training tuberculosis detection DNNs. Moreover, a recent review\cite{TBReview} showed that many studies that classify tuberculosis with neural networks use dataset mixing. Furthermore, very few works evaluate DNN generalization to o.o.d. test datasets\cite{TBReview}, and a study suggested that TB classification performance may strongly drop when DNNs are tested with datasets from external sources (o.o.d.)\cite{TBBadGeneralization}. Another paper exposed strong DNN attention outside of the lungs in TB-detection with mixed databases and advised the utilization of lung segmentation before classification\cite{TBSegmentation}. Although the World Health Organization states that chest radiography is essential for the early detection of TB, they provide no recommendation on the use of computer aided detection as of 2016\cite{who}. The reasons for this decision were the small number of studies, methodological limitations, and limited generalizability of the findings\cite{who}. Consequently, we hypothesize that the tuberculosis detection task is prone to producing background bias and shortcut learning, especially when mixed datasets are employed. For this reason, we included the application in this study. We classify X-rays as normal or tuberculosis-positive. Our training dataset mixes the TB Portals database\cite{TBPortals} with healthy X-rays from the CheXPert dataset\cite{irvin2019chexpert}. To analyze the extent of shortcut learning, we evaluate the DNNs on an i.i.d. test dataset, and on an o.o.d. database\cite{ChineseDataset1}. We utilize the TB detection task to demonstrate that background bias and shortcut learning are not exclusive to COVID-19 detection, and to assess the ISNet's capacity of addressing the problem in diverse unfavorable scenarios.

Before showing the aforementioned applications, our Results Section begins with a set of experiments using artificial bias. They consist in training the neural networks in natural and medical image datasets containing synthetic background bias, defined as a geometrical shape whose format (square, triangle or circle) is correlated with the image's classification label. The strong artificial background bias attracts the attention of standard classifiers, making them lose focus on the image's true region of interest and causing evident shortcut learning. The artificial bias is controllable, allowing the creation of diverse evaluation scenarios. We employ them to quantitatively compare the different neural networks' capacity of hindering the shortcut learning induced by background bias. Moreover, LRP heatmaps should be able to clearly show attention to the geometrical shapes in the models affected by shortcut learning. We experimented with adding synthetic background bias to 3 diverse databases. The first is the aforementioned COVID-19 dataset, also used in experiments without synthetic bias. The second is a facial attribute estimation dataset, extracted from CelebA\cite{celebA}. Since a study\cite{celebA} indicated that the classification of more attributes causes the classifiers to naturally focus more on the faces, we opted to classify only 3 facial attributes (rosy cheeks, high cheekbones and smiling), increasing the difficulty of learning an adequate attention profile. The third database is a subset of the Stanford Dogs\cite{StanfordDogs} dog breed classification dataset, comprising the following breeds: Pekingese, Tibetan Mastiff, and Pug. Stanford Dogs contains bounding-boxes, but no ground-truth dog segmentation target\cite{StanfordDogs} (necessary for ISNet training). Thus, we converted its bounding-boxes to foreground masks, employing a pretrained general purpose semantic segmenter, DeepMAC\cite{deepMAC}. Upon visual inspection, the quality of the masks was high. The three datasets allow us to assess the ISNet's background bias resistance in very diverse scenarios. 

Appendix \ref{dataset} explains all datasets and their limitations. Appendix \ref{ImpDetails} displays implementation details for the diverse tasks. Moreover, Appendix \ref{supApplications} introduces two additional applications, CheXPert\cite{irvin2019chexpert} (a large single-source X-ray database displaying multiple conditions) classification, and the facial attribute estimation task without synthetic bias. Both datasets do not present strong background bias. Thus, the experiments better delimit the ISNet use-case scenario.

The ISNet PyTorch code is available at \url{https://github.com/PedroRASB/ISNet}\cite{CodeGit}. Summarizing, in this study, we:

\begin{enumerate}
  \item Proposed a classifier architecture (ISNet) that, without a segmenter at run-time, reliably ignored images' backgrounds. In relation to the eight implemented state-of-the-art DNNs, the ISNet displayed superior capacity of hindering the shortcut learning caused by background bias, improving generalization. We justified this empirical result (Section \ref{results}) with an in-depth theoretical comparison between the DNNs (Appendix \ref{SOTACompare}). Moreover, the ISNet is flexible (accepting virtually any classifier backbone) and introduces no extra computational cost at run-time.
  \item Proposed the optimization of LRP heatmaps to improve and control DNNs, introducing the concept of background relevance minimization. The technique produced a theoretically founded (Section \ref{mathBackground}) explanation-based spatial attention mechanism that learns from foreground segmentation masks.
  \item Presented an efficient and automatically differentiable PyTorch implementation of LRP.
\end{enumerate}

\section{Results}
\label{results}

\subsection{Synthetic Background Bias}
\label{resultsSynth}

In these experiments, training images contained synthetic background bias designed to cause shortcut learning. We tested the neural networks on 3 datasets: a biased set, which contains the same background geometrical shapes found in training; the standard set, with no synthetic bias; and a deceiving bias dataset, which has geometrical shapes, but the correlation between them and the classification labels deceivingly change (e.g., a circle that was correlated with class 1 in training will be associated to class 2 during testing). Appendix \ref{dataset} provides more details. A comparison of a DNN's performance on the three testing scenarios provides a quantitative assessment of shortcut learning. In relation to the biased test performance, the higher the influence of the geometrical shapes on the classifier's outputs, the larger the F1-Score reduction when they are removed from the evaluation dataset (standard test) or substituted by deceiving bias. Table \ref{synth} presents the test average F1-Scores for all DNNs in the three testing environments.

\begin{table}[!h]
\centering
\small
\caption{Test macro-average F1-Scores for neural networks trained in datasets with synthetic background bias\textsuperscript{1}}
  \small\textsuperscript{1:} In the multi-class single-label experiments (Stanford Dogs and COVID-19 detection), scores are reported as mean and standard deviation. In facial attribute estimation (multi-label problem), they are displayed as mean and 95\% confidence intervals. Appendix \ref{statisticalMethods} provides more details about the statistical analysis in this study.
\begin{tblr}{
  width = \linewidth,
  colspec = {Q[412]Q[175]Q[175]Q[177]},
  row{2} = {c},
  row{13} = {c},
  row{24} = {c},
  cell{2}{1} = {c=4}{0.939\linewidth},
  cell{13}{1} = {c=4}{0.939\linewidth},
  cell{24}{1} = {c=4}{0.939\linewidth},
  hlines,
  vlines,
}
Model                                                               & {Biased \\Test maF1} & {Standard \\Test maF1} & {Deceiving Bias\\~Test maF1} \\
Stanford Dogs with Synthetic Background Bias               &                      &                        &                              \\
ISNet                                                      & 0.548 +/-0.035       & 0.553 +/-0.035         & 0.548 +/-0.035               \\
ISNet Grad*Input                                                    & 0.55 +/-0.034        & 0.545 +/-0.034         & 0.545 +/-0.034               \\
Standard Classifier                                                 & 0.926 +/-0.019       & 0.419 +/-0.034         & 0.071 +/-0.017               \\
Segmentation-classification Pipeline                                & 0.519 +/-0.035       & 0.519 +/-0.035         & 0.518 +/-0.035               \\
Multi-task U-Net                                                    & 0.522 +/-0.036       & 0.455 +/-0.036         & 0.38 +/-0.035                \\
AG-Sononet                                                          & 0.956 +/-0.015       & 0.214 +/-0.027         & 0.019 +/-0.009               \\
Extended GAIN                                                       & 0.935 +/-0.017       & 0.445 +/-0.034         & 0.1 +/-0.019                 \\
RRR                                                                 & 0.851 +/-0.025       & 0.548 +/-0.034         & 0.299 +/-0.025               \\
Vision Transformer (ViT-B/16)                                                  & 0.637 +/-0.034       & 0.419 +/-0.032         & 0.399 +/-0.032               \\
{Standard Classifier Reference\\(trained without synthetic bias)}   & -                    & 0.556 +/-0.035         & -                            \\
COVID-19 Detection with Synthetic Background Bias          &                      &                        &                              \\
ISNet                                                      & 0.775 +/-0.008       & 0.775 +/-0.008         & 0.775 +/-0.008               \\
ISNet Grad*Input                                                    & 0.542 +/-0.01        & 0.544 +/-0.01          & 0.417 +/-0.01                \\
Standard Classifier                                                 & 0.775 +/-0.008       & 0.434 +/-0.01          & 0.195 +/-0.004               \\
Segmentation-classification Pipeline                                & 0.618 +/-0.009       & 0.619 +/-0.009         & 0.618 +/-0.009               \\
Multi-task U-Net                                                    & 0.667 +/-0.01        & 0.341 +/-0.007         & 0.156 +/-0.004               \\
AG-Sononet                                                          & 0.943 +/-0.005       & 0.386 +/-0.008         & 0.047 +/-0.003               \\
Extended GAIN                                                       & 0.41 +/-0.009        & 0.306 +/-0.006         & 0.219 +/-0.003               \\
RRR                                                                 & 0.464 +/-0.009       & 0.458 +/-0.008         & 0.426 +/-0.008               \\
Vision Transformer (ViT-B/16)                                                  & 0.685 +/-0.009       & 0.496 +/-0.009         & 0.327 +/-0.009               \\
{Standard Classifier Reference\\(trained without synthetic bias)}   & -                    & 0.546 +/-0.01          & -                            \\
Facial Attribute Estimation with Synthetic Background Bias &                      &                        &                              \\
ISNet                                                      & 0.807 +/-0.027       & 0.807 +/-0.027         & 0.807 +/-0.027               \\
ISNet Grad*Input                                                    & 0.496 +/-0.02        & 0.499 +/-0.02          & 0.503 +/-0.021               \\
Standard Classifier                                                 & 0.974 +/-0.012       & 0.641 +/-0.054         & 0.398 +/-0.019               \\
Segmentation-classification Pipeline                                & 0.794 +/-0.031       & 0.794 +/-0.031         & 0.794 +/-0.031               \\
Multi-task U-Net                                                    & 0.985 +/-0.008       & 0.665 +/-0.129         & 0.351 +/-0.015               \\
AG-Sononet                                                          & 0.985 +/-0.009       & 0.616 +/-0.094         & 0.326 +/-0.016               \\
Extended GAIN                                                       & 0.886 +/-0.023       & 0.773 +/-0.034         & 0.633 +/-0.03                \\
RRR                                                                 & 0.794 +/-0.024       & 0.77 +/-0.032          & 0.557 +/-0.025               \\
Vision Transformer (ViT-B/16)                                                  & 0.675 +/-0.023       & 0.645 +/-0.03          & 0.531 +/-0.023               \\
{Standard Classifier Reference\\(trained without synthetic bias)}   & -                    & 0.802 +/-0.028         & -                            
\end{tblr}
\label{synth}
\end{table}

Data augmentation in COVID-19 detection and facial attribute estimation consisted on random rotations, translations and flipping (Appendix \ref{ProcessingAgumentation}). Thus, the applications exemplify standard data augmentation practices. Meanwhile, we utilized no data augmentation in Stanford Dogs. As the augmentation procedures may totally or completely remove the synthetic background bias from the image, this choice makes the Stanford Dogs experiment a scenario of more extreme background bias. The ISNet is resistant to image flipping, rotations, and translations. These operations did not negatively affect the training procedure, nor did they worsen the validation error during preliminary tests with an augmented validation dataset. 

The three synthetic bias applications represent very distinct scenarios, all of them considering high-resolution images (224x224) and deep classifiers. COVID-19 detection is a challenging and contemporary biomedical classification task, with images' foregrounds defined as the lungs. Stanford Dogs and CelebA are natural image RGB datasets, where the foregrounds are the dogs or faces. CelebA presents challenging in-the-wild pictures, where faces can appear in multiple poses\cite{celebA}. Stanford Dogs represents a difficult fine-grained classification task, with large intra-class variance, low inter-class variance, high background variety, and possible occlusion\cite{StanfordDogs}. Like the CelebA faces, dogs can appear in multiple poses and distances. Thus, in the two datasets, foregrounds strongly vary throughout the figures. Dataset sizes also substantially change, with 501 images in the dogs dataset, 13932 in COVID-19 detection, and 30000 for facial attribute estimation. As the three tasks represent very diverse scenarios, the DNNs' performances vary across the experiments. However, the distinct applications allow us to draw more reliable conclusions by analyzing repeating patterns in the experiments' results.

First, all datasets have a strong tendency to cause shortcut learning, as intended. This is proved by the large performance drops seen for the standard classifiers in Table \ref{synth}. The baseline model represents a common DNN, without any mechanism to avoid background attention. In the most extreme case, Stanford Dogs, the model's macro-average F1-Score (maF1) falls from 0.926 +/-0.019 (with geometrical shapes in the test dataset) to 0.546 +/-0.034 (with no synthetic bias), and finally to 0.071 +/-0.017 (with deceiving background bias). Putting these results into perspective, from the 201 evaluation samples, in the biased test the model correctly classifies 192, in the unbiased test, 89, and, in the deceiving test, 10.

Second, the segmentation-classification pipeline and the ISNet were the only models not affected by background bias in any of the three experiments in Table \ref{synth}. They show no reduction in maF1 when the synthetic background bias was removed or substituted by deceiving bias. On the other hand, all other models display maF1 drop across the columns in Table \ref{synth}. Even considering the interval estimates, none of the other benchmark DNNs have overlapping maF1 intervals in all three testing scenarios. Thus, the bias influence over these classifiers is evident. Conversely, the ISNet successfully minimized background attention and the consequent shortcut learning. Moreover, the ISNet's resistance to background bias is not accompanied by an accuracy drop: it has the highest average F1-Scores in the three diverse tasks (standard and deceiving bias tests in Table \ref{synth}). It could even surpass the segmentation-classification pipeline, a much larger and slower model (Appendix \ref{speed} presents a speed and size comparison).

As further proof of the ISNet capacity of avoiding background bias attention while retaining high accuracy, its test maF1 scores, when trained with the synthetically biased data, match or surpass a standard classifier trained in datasets without any synthetic background bias (Table \ref{synth}). We did not find relevant natural background bias in Stanford Dogs or CelebA. Accordingly, in these two cases, the ISNet matched a standard classifier trained in an unbiased environment, indicating that the insertion of synthetic bias could not reduce its accuracy. Meanwhile, the ISNet also hindered the shortcut learning caused by the naturally occurring background bias in the COVID-19 dataset. This result explains why it strongly surpassed the standard classifier trained in the COVID-19 database without any synthetic bias. I.e., in this scenario the standard model suffers shortcut learning caused by non-synthetic background bias (resulting from dataset mixing), which the ISNet hinders (Section \ref{covidResults}).

In Stanford Dogs, the standard classifier is a VGG-19\cite{vggOriginal}. This architecture is also used as the classification backbone for the ISNet, ISNet Grad*Input, segmentation-classification pipeline (defined as  a U-Net\cite{unet} segmenter followed by VGG-19 classifier), extended GAIN, and RRR. In the two other datasets (and in the remaining experiments in this study), a DenseNet121\cite{DenseNet} substitutes the VGG-19 as the standard classifier and the backbone for the aforementioned DNNs. The remaining benchmark networks (multi-task U-Net, AG-Sononet and Vision Transformer) utilize their original architectures in all tasks (Appendix \ref{SOTACompare}). The DenseNet and VGG show how the diverse DNN architectures behave with different types of backbone. A Densely Connected Convolutional Network\cite{DenseNet} (DenseNet121) exemplifies very deep (121 layers) architectures with skip connections, and a VGG-19\cite{vggOriginal} represents DNNs with fewer layers (19) and no skip connections. The VGG was chosen as the shallower model because it is an influential and popular architecture, with a simple yet effective design. DenseNets are very deep but efficient, carrying a small number of parameters in relation to their depth\cite{DenseNet}. This characteristic resonates with the efficiency focus of the ISNet design. Another reason for the DenseNet121 choice is that the architecture is among the most popular for the classification of lung diseases in chest X-rays\cite{XRayArchitectures}, a task that is considered in multiple experiments throughout this work. Evidencing the ISNet's versatility, it surpassed all benchmark DNNs and was resistant to shortcut learning with both backbones (Table \ref{synth}).

\subsection{COVID-19 Detection}
\label{covidResults}

For COVID-19 detection we employ an external (o.o.d.) test dataset, whose images come from distinct hospitals and cities in relation to the training database (Appendix \ref{COVIDdataset}). Tables \ref{performance} and \ref{performanceSup} show the DNNs' test performances. 

\begin{table}[!h]
\centering
\small
\caption{Test F1-Scores and ROC-AUC for the deep neural networks in COVID-19 detection (o.o.d. evaluation)\textsuperscript{2}}
  \small\textsuperscript{2:} Class ROC-AUC scores are calculated with a one-versus-rest approach and accompanied by 95\% confidence intervals. Mean AUC is provided as point estimates and we calculate it with a pairwise technique\cite{MulticlassAUC}, instead of averaging the class scores. Other metrics are reported as: mean +/-std, [95\% HDI]. Both mean and standard deviation (std) are extracted from the metric's probability distribution, according to Bayesian estimation. 95\% HDI indicates the 95\% highest density interval, an interval containing 95\% of the metric's probability mass. Furthermore, any point inside the interval has a probability density that is higher than that of any point outside. Appendix \ref{statisticalMethods} explains the statistical methods in detail.

\label{performance}
\begin{tblr}{
  width = \linewidth,
  colspec = {Q[315]Q[152]Q[152]Q[152]Q[163]},
  hlines,
  vlines,
}
Model and Metric                          & Normal                             & Pneumonia                          & COVID-19                           & {Mean\\(macro-average)}            \\
{ISNet\\F1-Score}                         & {0.555 +/-0.022,\\{[}0.512,0.597]} & {0.858 +/-0.007,\\{[}0.844,0.871]} & {0.907 +/-0.006,\\{[}0.896,0.918]} & {0.773 +/-0.009,\\{[}0.755,0.791]} \\
{U-Net+DenseNet121\\ F1-Score}            & {0.571 +/-0.018,\\{[}0.535,0.607]} & {0.586 +/-0.013,\\{[}0.561,0.611]} & {0.776 +/-0.008,\\{[}0.76,0.792]}  & {0.645 +/-0.009,\\{[}0.626,0.663]} \\
{DenseNet121\\ F1-Score}                  & {0.444 +/-0.02,\\{[}0.403,0.482]}  & {0.434 +/-0.015,\\{[}0.405,0.463]} & {0.76 +/-0.008,\\{[}0.744,0.775]}  & {0.546 +/-0.01,\\{[}0.527,0.565]}  \\
{Multi-task U-Net\\ F1-Score}             & {0.419 +/-0.025,\\{[}0.369,0.469]} & {0.119 +/-0.011,\\{[}0.098,0.14]}  & {0.585 +/-0.009,\\{[}0.566,0.602]} & {0.374 +/-0.01,\\{[}0.355,0.394]}  \\
{AG-Sononet\\ F1-Score}                   & {0.124 +/-0.015,\\{[}0.096,0.153]} & {0.284 +/-0.015,\\{[}0.255,0.312]} & {0.659 +/-0.009,\\{[}0.641,0.676]} & {0.356 +/-0.008,\\{[}0.34,0.372]}  \\
{Extended GAIN\\F1-Score}                 & {0.203~+/-0.019,\\{[}0.166,0.24]}  & {0.485~+/-0.013,\\{[}0.46,0.511]}  & {0.711~+/-0.009,\\{[}0.693,0.728]} & {0.466~+/-0.009,\\{[}0.449,0.485]} \\
{RRR\\F1-Score}                           & {0.36~+/-0.018,\\{[}0.325,0.394]}  & {0.552~+/-0.013,\\{[}0.526,0.577]} & {0.737~+/-0.009,\\{[}0.72,0.755]}  & {0.55 +/-0.009,\\{[}0.532,0.568]}  \\
{Vision Transformer (ViT-B/16)\\F1-Score} & {0.382~+/-0.017,\\{[}0.348,0.415]} & {0.474~+/-0.013,\\{[}0.448,0.499]} & {0.525~+/-0.011,\\{[}0.503,0.548]} & {0.46~+/-0.009,\\{[}0.443,0.478]}  \\
ISNet AUC                                 & 0.931 +/-0.01                      & 0.962 +/-0.006                     & 0.976 +/-0.005                     & 0.952                              \\
U-Net+DenseNet121~AUC                     & 0.888 +/-0.019                     & 0.78 +/-0.016                      & 0.846 +/-0.013                     & 0.833                              \\
DenseNet121~AUC                           & 0.804 +/-0.023                     & 0.805 +/-0.015                     & 0.86 +/-0.013                      & 0.808                              \\
Multi-task U-Net~AUC                      & 0.721 +/-0.034                     & 0.412 +/-0.019                     & 0.487 +/-0.02                      & 0.553                              \\
AG-Sononet~AUC                            & 0.451 +/-0.028                     & 0.681 +/-0.019                     & 0.658 +/-0.018                     & 0.591                              \\
Extended GAIN~AUC                         & 0.7 +/-0.025                       & 0.756 +/-0.016                     & 0.806 +/-0.016                     & 0.724                              \\
RRR AUC                                   & 0.782~+/-0.02                      & 0.736~+/-0.017                     & 0.835~+/-0.014                     & 0.775                              \\
Vision Transformer (ViT-B/16) AUC         & 0.755~+/-0.032                     & 0.645~+/-0.019                     & 0.619~+/-0.019                     & 0.683                              
\end{tblr}
\end{table}

\begin{table}
\centering
\small
\caption{Test precision, recall and specificity for the deep neural networks in COVID-19 detection (o.o.d. evaluation)\textsuperscript{3}}
  \small\textsuperscript{3:}Metrics are reported as: mean +/-std, [95\% HDI], according to Bayesian estimation. Appendix \ref{statisticalMethods} provides more details about the statistical analysis.
\label{performanceSup}
\begin{tblr}{
  width = \linewidth,
  colspec = {Q[285]Q[160]Q[160]Q[160]Q[173]},
  hlines,
  vlines,
}
Model and Metric                             & Normal                             & Pneumonia                          & COVID-19                           & {Mean\\(macro-average)}             \\
{ISNet \\Precision}                          & {0.544 +/-0.026,\\{[}0.494,0.594]} & {0.794 +/-0.01, \\{[}0.774,0.814]} & {0.993 +/-0.002,\\{[}0.988,0.997]} & {0.777 +/-0.009,\\{[}0.759,0.795]}  \\
{U-Net+DenseNet121\\ Precision}              & {0.446 +/-0.019,\\{[}0.408,0.483]} & {0.791 +/-0.015,\\{[}0.763,0.82]}  & {0.723 +/-0.011,\\{[}0.702,0.744]} & {0.653 +/-0.009,\\{[}0.636,0.67]}   \\
{DenseNet121\\ Precision}                    & {0.364 +/-0.02,\\{[}0.324,0.402]}  & {0.827 +/-0.018,\\{[}0.792,0.861]} & {0.649 +/-0.01,\\{[}0.629,0.67]}   & {0.614 +/-0.009,\\{[}0.594,0.631]}  \\
{Multi-task U-Net\\ Precision}               & {0.552 +/-0.033,\\{[}0.488,0.617]} & {0.232 +/-0.02,\\{[}0.194,0.272]}  & {0.469 +/-0.01,\\{[}0.449,0.489]}  & {0.418 +/-0.013,\\{[}0.392,0.444]}  \\
{AG-Sononet\\ Precision}                     & {0.104 +/-0.013,\\{[}0.079,0.129]} & {0.665 +/-0.025,\\{[}0.616,0.715]} & {0.549 +/-0.01,\\{[}0.528,0.569]}  & {0.439 +/-0.01,\\{[}0.419,0.459]}   \\
{Extended GAIN\\Precision}                   & {0.189~+/-0.019,\\{[}0.152,0.225]} & {0.603~+/-0.016,\\{[}0.571,0.636]} & {0.642~+/-0.011,\\{[}0.62,0.664]}  & {~0.478 +/-0.009,\\{[}0.461,0.496]} \\
{RRR \\Precision}                            & {0.262~+/-0.015,\\{[}0.232,0.293]} & {0.728~+/-0.016,\\{[}0.697,0.758]} & {0.723~+/-0.011,\\{[}0.701,0.745]} & {0.571 +/-0.008\\{[}0.555,0.587]}   \\
{Vision Transformer (ViT-B/16)\\Precision}   & {0.268~+/-0.015,\\{[}0.239,0.297]} & {0.552~+/-0.016,\\{[}0.521,0.584]} & {0.572~+/-0.014,\\{[}0.544,0.598]} & {0.464~+/-0.009,\\{[}0.447,0.481]}  \\
{ISNet\\Recall}                              & {0.566 +/-0.026,\\{[}0.515,0.616]} & {0.933 +/-0.007,\\{[}0.919,0.946]} & {0.835 +/-0.01,\\{[}0.816,0.853]}  & {0.778 +/-0.009,\\{[}0.76,0.796]}   \\
{U-Net+DenseNet121\\ Recall}                 & {0.796 +/-0.021,\\{[}0.756,0.837]} & {0.466 +/-0.014,\\{[}0.439,0.494]} & {0.838 +/-0.009,\\{[}0.819,0.856]} & {0.7 +/-0.009,\\{[}0.683,0.717]}    \\
{DenseNet121\\ Recall}                       & {0.57 +/-0.026,\\{[}0.518,0.618]}  & {0.294 +/-0.013,\\{[}0.27,0.32]}   & {0.916 +/-0.007,\\{[}0.902,0.93]}  & {0.594 +/-0.01,\\{[}0.574,0.612]}   \\
{Multi-task U-Net\\ Recall}                  & {0.338 +/-0.024,\\{[}0.29,0.386]}  & {0.08 +/-0.008,\\{[}0.066,0.095]}  & {0.776 +/-0.011,\\{[}0.755,0.797]} & {0.398 +/-0.009,\\{[}0.38,0.416]}   \\
{AG-Sononet\\ Recall}                        & {0.156 +/-0.019,\\{[}0.12,0.192]}  & {0.18 +/-0.011,\\{[}0.16,0.201]}   & {0.824 +/-0.01,\\{[}0.805,0.843]}  & {0.387 +/-0.008,\\{[}0.371,0.402]}  \\
{Extended GAIN\\Recall}                      & {0.22~+/-0.021,\\{[}0.178,0.261]}  & {0.406~+/-0.014,\\{[}0.379,0.432]} & {0.796~+/-0.01,\\{[}0.775,0.816]}  & {0.474~+/-0.009,\\{[}0.456,0.492]}  \\
{RRR~\\Recall}                               & {0.574~+/-0.025,\\{[}0.524,0.624]} & {0.445~+/-0.014,\\{[}0.417,0.471]} & {0.753~+/-0.011,\\{[}0.731,0.775]} & {0.59~+/-0.01,\\{[}0.57,0.611]}     \\
{Vision Transformer (ViT-B/16)\\Recall}      & {0.665~+/-0.024,\\{[}0.616,0.712]} & {0.415~+/-0.014,\\{[}0.388,0.442]} & {0.486~+/-0.013,\\{[}0.461,0.511]} & {0.522~+/-0.01,\\{[}0.501,0.542]}   \\
{ISNet\\Specificity}                         & {0.937 +/-0.005,\\{[}0.928,0.946]} & {0.834 +/-0.009,\\{[}0.817,0.851]} & {0.995 +/-0.002,\\{[}0.991,0.998]} & {0.922 +/-0.003,\\{[}0.916,0.928]}  \\
{U-Net+DenseNet121\\ Specificity}            & {0.869 +/-0.006,\\{[}0.857,0.882]} & {0.915 +/-0.006,\\{[}0.903,0.928]} & {0.708 +/-0.011,\\{[}0.686,0.73]}  & {0.831 +/-0.004,\\{[}0.823,0.839]}  \\
{DenseNet121\\ Specificity}                  & {0.869 +/-0.006,\\{[}0.856,0.881]} & {0.958 +/-0.005,\\{[}0.949,0.967]} & {0.549 +/-0.012,\\{[}0.525,0.573]} & {0.792 +/-0.004,\\{[}0.784,0.8]}    \\
{Multi-task U-Net\\ Specificity}             & {0.964 +/-0.004,\\{[}0.957,0.971]} & {0.818 +/-0.009,\\{[}0.801,0.835]} & {0.201 +/-0.01,\\{[}0.182,0.22]}   & {0.661 +/-0.004,\\{[}0.653,0.669]}  \\
{AG-Sononet\\ Specificity}                   & {0.823 +/-0.007,\\{[}0.808,0.837]} & {0.938 +/-0.006,\\{[}0.927,0.948]} & {0.384 +/-0.012,\\{[}0.361,0.408]} & {0.715 +/-0.004,\\{[}0.707,0.722]}  \\
{Extended GAIN\\Specificity}                 & {0.875~+/-0.006,\\{[}0.863,0.888]} & {0.817~+/-0.009,\\{[}0.799,0.834]} & {0.597~+/-0.012,\\{[}0.573,0.62]}  & {0.763~+/-0.004,\\{[}0.754,0.772]}  \\
{RRR\\Specificity}                           & {0.787~+/-0.008,\\{[}0.772,0.802]} & {0.886~+/-0.007,\\{[}0.872,0.9]}   & {0.737~+/-0.011,\\{[}0.716,0.758]} & {0.803~+/-0.004,\\{[}0.795,0.812]}  \\
{Vision Transformer (ViT-B/16)\\Specificity} & {0.761~+/-0.008,\\{[}0.745,0.776]} & {0.769~+/-0.01,\\{[}0.75,0.788]}   & {0.669~+/-0.012,\\{[}0.646,0.692]} & {0.733~+/-0.005,\\{[}0.723,0.742]}  
\end{tblr}
\end{table}

The ISNet obtained the best o.o.d. test performance in Table \ref{performance} and \ref{performanceSup}, surpassing all benchmark DNNs' average performance metrics. Moreover, the ISNet results' 95\% highest density intervals do not overlap with any other network for any average performance measurement. The ground-truth foreground masks used for training were automatically generated by  a U-Net, trained for lung segmentation in another study\cite{bassi2021covid19}. Therefore, the performances achieved by the ISNet and by the alternative segmentation-classification pipeline could possibly increase even more, provided a dataset containing a large amount of chest X-rays accompanied by manually segmented lungs.

Tables \ref{performance} and \ref{performanceSup} results may seem worse than other COVID-19 detection studies, which report remarkably high performances, strongly surpassing expert radiologists (e.g., F1-Scores close to 100\%). However, evidence suggests that, currently, such results are obtained when the training and test datasets come from the same distribution/sources (i.i.d. datasets)\cite{ShortcutCovid}. Moreover, studies showed that these strong performances may be boosted by bias and shortcut learning, preventing the neural networks from generalizing, or achieving comparable results in the real-world\cite{ShortcutCovid}\textsuperscript{,}\cite{bassi2021covid19}\textsuperscript{,}\cite{NatureCovidBias}. Instead, the performances in Table \ref{performance} are comparable to other works that evaluate their DNNs in external (o.o.d.) databases\cite{ShortcutCovid}\textsuperscript{,}\cite{bassi2021covid19}\textsuperscript{,}\cite{NatureCovidBias}. For example, an article\cite{NatureCovidBias} reported AUC of 0.786 +/-0.025 on an external COVID-19 X-ray dataset, considering a DenseNet121 and no lung segmentation. Here, the DenseNet121 obtained 0.808 AUC, which falls into their reported confidence interval. Another paper\cite{bassi2021covid19} evaluates COVID-19 detection and utilizes lung segmentation before classification with a DenseNet201\cite{DenseNet}. They achieved maF1 of 0.754, with 95\% HDI of [0.687,0.82], evaluating on an external dataset. The ISNet maF1 95\% HDI, [0.755,0.791], fits inside their reported 95\% HDI. We must note that the aforementioned studies use different databases. Thus, caution is required when directly comparing the numerical results.

COVID-19 detection with mixed datasets is a task known for background bias and common shortcut learning, which results in subpar o.o.d. generalization\cite{NatureCovidBias}\textsuperscript{,}\cite{ShortcutCovid}. Accordingly, the results in Tables \ref{performance} and \ref{performanceSup} are consistent with our findings from the synthetic bias experiments. Firstly, the standard classifier (DenseNet121) displayed unimpressive generalization, achieving only 0.546 +/-0.01 o.o.d. maF1. Moreover, in Table \ref{synth}, the ISNet and the alternative segmentation-classification pipeline consistently were the two models with the highest resistance to background bias attention and the best generalization. Accordingly, in Tables \ref{performance} and \ref{performanceSup}, the two models had superior o.o.d. maF1 in COVID-19 detection. Therefore, the results in the COVID-19 detection confirm that the task is prone to shortcut learning, which the ISNet and pipeline could better mitigate. However, the ISNet could surpass even the large pipeline, showing no HDI superposition with its results. We analyze this finding in Appendix \ref{baselineComparisons}, where we also theoretically justify why the ISNet surpassed all other benchmark models.

To illustrate that it is not possible to simply train a model with segmented images, and then use it without segmentation at run-time, we tested the segmentation-classification pipeline after removing its segmenter (U-Net). Thus, we simulated a DenseNet121 trained on segmented images and used to classify unsegmented ones. As expected, this resulted in a dramatic performance drop: maF1 fell from 0.645 +/-0.009 to 0.217 +/-0.003 (changing its 95\% HDI from [0.626,0.663] to [0.211,0.224]). Therefore, unlike the ISNet, the pipeline needs a segmenter at run-time. Table \ref{confusion} displays the confusion matrices for all DNNs. 

\begin{table}[!h]
\centering
\small
\caption{Test confusion matrices for the deep neural networks in COVID-19 detection (o.o.d. evaluation)}
\label{confusion}
\begin{tblr}{
  width = \linewidth,
  colspec = {Q[125]Q[85]Q[125]Q[115]Q[25]Q[125]Q[85]Q[125]Q[115]},
  cell{1}{1} = {r=2}{},
  cell{1}{2} = {c=3}{0.325\linewidth},
  cell{1}{6} = {r=2}{},
  cell{1}{7} = {c=3}{0.325\linewidth},
  cell{3}{1} = {c=4}{0.45\linewidth,c},
  cell{3}{6} = {c=4}{0.45\linewidth,c},
  cell{7}{1} = {c=4}{0.45\linewidth,c},
  cell{7}{6} = {c=4}{0.45\linewidth,c},
  cell{11}{1} = {c=4}{0.45\linewidth,c},
  cell{11}{6} = {c=4}{0.45\linewidth,c},
  cell{15}{1} = {c=4}{0.45\linewidth,c},
  cell{15}{6} = {c=4}{0.45\linewidth,c},
  cell{16}{1} = {},
  cell{16}{2} = {},
  cell{16}{3} = {},
  cell{16}{4} = {},
  cell{16}{6} = {},
  cell{16}{7} = {},
  cell{16}{8} = {},
  cell{16}{9} = {},
  cell{17}{1} = {},
  cell{17}{2} = {},
  cell{17}{3} = {},
  cell{17}{4} = {},
  cell{17}{6} = {},
  cell{17}{7} = {},
  cell{17}{8} = {},
  cell{17}{9} = {},
  cell{18}{1} = {},
  cell{18}{2} = {},
  cell{18}{3} = {},
  cell{18}{4} = {},
  cell{18}{6} = {},
  cell{18}{7} = {},
  cell{18}{8} = {},
  cell{18}{9} = {},
  vlines,
  hline{1,3-19} = {1-4,6-9}{},
  hline{2} = {2-4,7-9}{},
}
True Class                   & Predicted Class &           &          &  & True Class                       & Predicted Class &           &          \\
                             & Normal          & Pneumonia & COVID-19 &  &                                  & Normal          & Pneumonia & COVID-19 \\
ISNet               &                 &           &          &  & Multi-task U-Net        &                 &           &          \\
Normal                       & 210             & 157       & 3        &  & Normal                           & 125             & 42        & 203      \\
Pneumonia                    & 81              & 1210      & 4        &  & Pneumonia                        & 62              & 103       & 1130     \\
COVID-19                     & 93              & 156       & 1266     &  & COVID-19                         & 38              & 300       & 1177     \\
U-Net + DenseNet121 &                 &           &          &  & Attention Gated Sononet &                 &           &          \\
Normal                       & 296             & 9         & 65       &  & Normal                           & 57              & 2         & 311      \\
Pneumonia                    & 271             & 604       & 420      &  & Pneumonia                        & 346             & 233       & 716      \\
COVID-19                     & 95              & 149       & 1271     &  & COVID-19                         & 151             & 114       & 1250     \\
DenseNet121         &                 &           &          &  & Extended GAIN           &                 &           &          \\
Normal                       & 211             & 16        & 143      &  & Normal                           & 81              & 145       & 144      \\
Pneumonia                    & 306             & 381       & 608      &  & Pneumonia                        & 241             & 526       & 528      \\
COVID-19                     & 63              & 62        & 1390     &  & COVID-19                         & 108             & 200       & 1207     \\
RRR        &                 &           &          &  & Vision Transformer      &                 &           &          \\
Normal                       & 213             & 84        & 73       &  & Normal                           & 247             & 50        & 73       \\
Pneumonia                    & 355             & 576       & 364      &  & Pneumonia                        & 279             & 538       & 478      \\
COVID-19                     & 243             & 130       & 1142     &  & COVID-19                         & 393             & 385       & 737      
\end{tblr}
\end{table}

\subsection{Tuberculosis Detection}
\label{TBResults}

Table \ref{tb} reports performances for tuberculosis detection, using an external (o.o.d.) test dataset (Appendix \ref{TBDatasetSec}). On the i.i.d. test dataset all models had mean AUC over 0.9. Moreover, considering i.i.d. evaluation, the segmentation-classification pipeline achieved maF1 (with 95\% confidence interval) of 0.955 +/-0.016, the ISNet 0.974 +/-0.012, the extended GAIN 0.982 +/-0.009, the Vision Transformer 0.926 +/-0.02, RRR 0.839 +/-0.028, and all other DNNs' mean maF1 scores surpassed 0.985. Our i.i.d. test results are in line with other studies that detected tuberculosis with DNNs, most of which report very high AUC and F1-Score\cite{TBReview}. We could not find studies employing a training dataset like ours and o.o.d. testing, as the evaluation methodology is rare in tuberculosis detection.

\begin{table}[!h]
\centering
\small
\caption{Performance metrics for the deep neural networks in tuberculosis detection (o.o.d. evaluation)\textsuperscript{4}}
  \small\textsuperscript{4:} The cells display the metrics' mean and 95\% confidence intervals
\label{tb}
\begin{tabular}{|l|l|l|l|} 
\hline
Model and Metric                        & Normal          & Tuberculosis    & \begin{tabular}[c]{@{}l@{}}Mean\\ (macro-average)\end{tabular}  \\ 
\hline
ISNet Precision                         & 0.744 +/-0.045  & 0.734 +/-0.043  & 0.739 +/-0.044                                                  \\ 
\hline
U-Net+DenseNet121 Precision             & 0.63 +/-0.061   & 0.573 +/-0.043  & 0.601 +/-0.052                                                  \\ 
\hline
DenseNet121 Precision                   & 0.578 +/-0.055  & 0.564 +/-0.046  & 0.571 +/-0.05                                                   \\ 
\hline
Multi-task U-Net Precision              & 0.515 +/-0.048  & 0.539 +/-0.053  & 0.527 +/-0.05                                                   \\ 
\hline
AG-Sononet Precision                    & 0.731 +/-0.06   & 0.599 +/-0.041  & 0.665 +/-0.05                                                   \\ 
\hline
Extended GAIN Precision                 & 0.576 +/-0.04   & 0.766 +/-0.06   & 0.671 +/-0.05                                                   \\ 
\hline
RRR Precision                           & 0.663 +/-0.049  & 0.664 +/-0.046  & 0.663 +/-0.048                                                  \\ 
\hline
Vision Transformer (ViT-B/16) Precision & 0.52 +/-0.044   & 0.56 +/-0.059   & 0.54 +/-0.052                                                   \\ 
\hline
ISNet Recall                            & 0.714 +/-0.046  & 0.762 +/-0.042  & 0.738 +/-0.044                                                  \\ 
\hline
U-Net+DenseNet121 Recall                & 0.409 +/-0.05   & 0.768 +/-0.042  & 0.589 +/-0.046                                                  \\ 
\hline
DenseNet121 Recall                      & 0.479 +/-0.05   & 0.659 +/-0.047  & 0.569 +/-0.048                                                  \\ 
\hline
Multi-task U-Net Recall                 & 0.586 +/-0.05   & 0.468 +/-0.05   & 0.527 +/-0.05                                                   \\ 
\hline
AG-Sononet Recall                       & 0.406 +/-0.05   & 0.855 +/-0.035  & 0.631 +/-0.042                                                  \\ 
\hline
Extended GAIN Recall                    & 0.883 +/-0.032  & 0.372 +/-0.048  & 0.627 +/-0.04                                                   \\ 
\hline
RRR Recall                              & 0.642 +/-0.049  & 0.685 +/-0.046  & 0.663 +/-0.048                                                  \\ 
\hline
Vision Transformer (ViT-B/16) Recall    & 0.679 +/-0.047  & 0.395 +/-0.049  & 0.537 +/-0.048                                                  \\ 
\hline
ISNet F1-Score                          & 0.729 +/-0.046  & 0.748 +/-0.043  & 0.738 +/-0.044                                                  \\ 
\hline
U-Net+DenseNet121 F1-Score              & 0.496 +/-0.056  & 0.656 +/-0.044  & 0.576 +/-0.05                                                   \\ 
\hline
DenseNet121 F1-Score                    & 0.524 +/-0.052  & 0.608 +/-0.047  & 0.566 +/-0.05                                                   \\ 
\hline
Multi-task U-Net F1-Score               & 0.548 +/-0.049  & 0.501 +/-0.052  & 0.524 +/-0.05                                                   \\ 
\hline
AG-Sononet F1-Score                     & 0.522 +/-0.057  & 0.704 +/-0.04   & 0.613 +/-0.048                                                  \\ 
\hline
Extended GAIN F1-Score                  & 0.697 +/-0.039  & 0.501 +/-0.056  & 0.599 +/-0.048                                                  \\ 
\hline
RRR F1-Score                            & 0.652 +/-0.049  & 0.674 +/-0.046  & 0.663 +/-0.048                                                  \\ 
\hline
Vision Transformer (ViT-B/16) F1-Score  & 0.589 +/-0.046  & 0.463 +/-0.054  & 0.526 +/-0.05                                                   \\ 
\hline
ISNet AUC                               & 0.809 +/-0.031  & 0.809 +/-0.031  & 0.809 +/-0.031                                                  \\ 
\hline
U-Net+DenseNet121 AUC                   & 0.667 +/-0.039  & 0.667 +/-0.039  & 0.667 +/-0.039                                                  \\ 
\hline
DenseNet121~AUC                         & 0.576 +/-0.04   & 0.576 +/-0.04   & 0.576 +/-0.04                                                   \\ 
\hline
Multi-task U-Net~AUC                    & 0.549 +/-0.041  & 0.549 +/-0.041  & 0.549 +/-0.041                                                  \\ 
\hline
AG-Sononet~AUC                          & 0.717 +/-0.037  & 0.717 +/-0.037  & 0.717 +/-0.037                                                  \\ 
\hline
Extended GAIN~AUC                       & 0.676 +/-0.038  & 0.676 +/-0.038  & 0.676 +/-0.038                                                  \\ 
\hline
RRR AUC                                 & 0.728 +/- 0.036 & 0.728 +/- 0.036 & 0.728 +/- 0.036                                                 \\ 
\hline
Vision Transformer (ViT-B/16) AUC       & 0.558 +/-0.041  & 0.558 +/-0.041  & 0.558 +/-0.041                                                  \\
\hline
\end{tabular}
\end{table}

Like in COVID-19 detection and all experiments with synthetic bias, the standard classifier (DenseNet121) o.o.d. generalization was underwhelming in TB detection (0.566 +/-0.05 maF1), and the ISNet was the best performing model on the tuberculosis o.o.d. dataset. Furthermore, the proposed model showed no confidence interval overlap with the other DNNs' AUCs. Here, the segmentation-classification pipeline results were not as promising as in COVID-19 detection. Appendix \ref{heatmapAnalysis} analyses this finding. The ISNet performance on the i.i.d. evaluation database is among the lowest, while it was the best performing model on the o.o.d. dataset. Shortcut learning is characterized by high accuracy on standard benchmarks (i.i.d. datasets), but impaired o.o.d. generalization, and poor real-world performance\cite{ShortcutLearning}. Thus, the ISNet quantitative results are coherent with a reduction in shortcut learning, without overall accuracy degradation. Therefore, TB classification is an additional example of a real-world application that can be heavily affected by background bias and shortcut learning, representing another notable use-case for the ISNet.

\subsection{Heatmaps}

\begin{figure}[!h]
\includegraphics[width=0.63\textwidth]{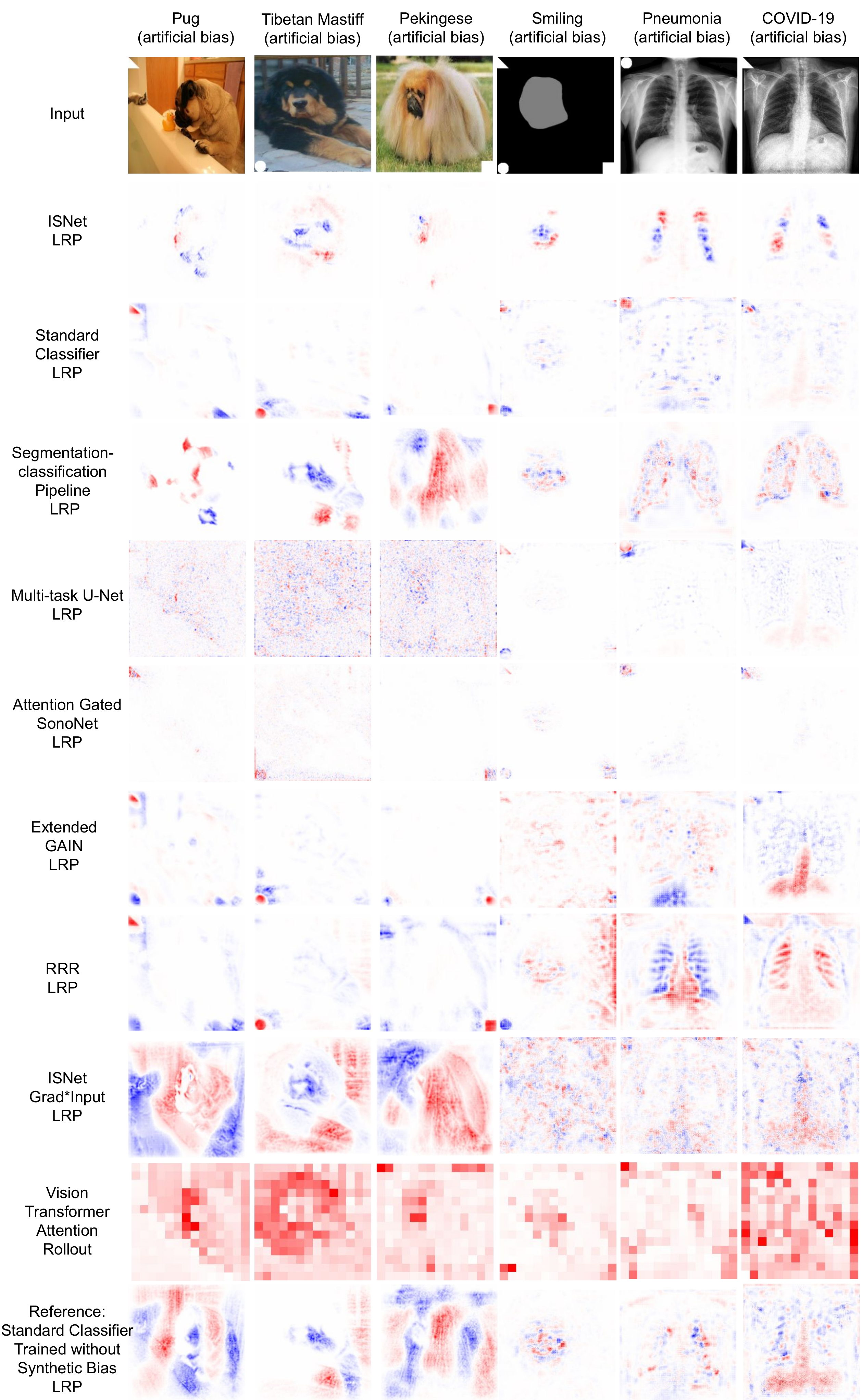}
\centering
\caption{\textbf{Heatmaps (Layer-wise Relevance Propagation/LRP for convolutional networks and attention rollout for Vision Transformer) for positive COVID-19 and Pneumonia X-rays and photographs, extracted from the synthetically biased test datasets (biased test).} Last row displays classifier trained without the synthetic bias (and analyzing images without the bias), for reference. The Image's true class is stated above the figures, the DNN that produced the heatmap is identified on the left. The triangle (background bias) indicates the classes COVID-19, smiling or Pug. The circle pneumonia, high cheekbones, and Tibetan Mastiff. The square rosy cheeks and Pekingese. Red colors in the LRP maps indicate areas the DNN associated to the image's true class, while blue colors are areas that reduced the network confidence for the class. For attention rollout, red shows the DNN attention. White represents areas with little influence over the classifiers. In the heatmaps, focus on the images' foregrounds (dogs, faces or lungs) is desirable. For privacy, the face picture was substituted by a representation of the face (gray) and bias (white) locations, but classifiers received the real picture}
\label{triangle}
\end{figure}

Figure \ref{triangle} presents heatmaps for the experiments with synthetic background bias. They show that only the ISNet and the segmentation-classification pipeline consistently and effectively minimized the influence of the background bias over the classifier. The results in Table \ref{synth} support the heatmaps, by quantitatively proving that the two models were the only ones never influenced by the artificial bias.

\begin{figure}[!h]
\includegraphics[width=0.55\textwidth]{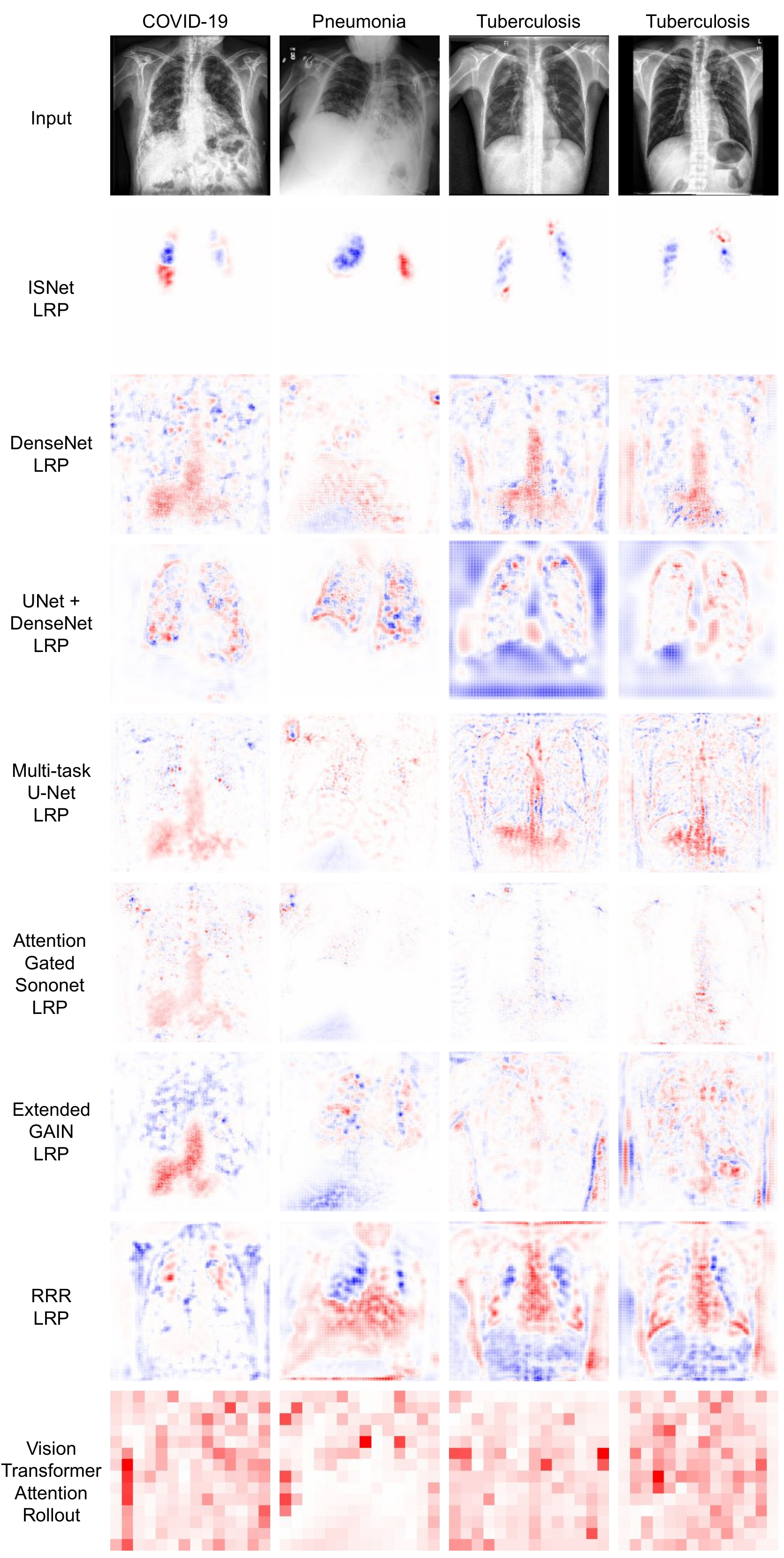}
\centering
\caption{\textbf{Heatmaps (Layer-wise Relevance Propagation/LRP for convolutional networks and attention rollout for Vision Transformer) for positive COVID-19, Pneumonia and tuberculosis.} The Image's true class is stated above the figures, the DNN that produced the heatmap is identified on the left. For LRP, red colors indicate areas that the DNN associated to the true class, blue colors are areas that decreased the network confidence for the class. For attention rollout, red indicates the DNN attention. White represents areas with little influence over the classifiers. Attention outside of the lungs (foreground) is undesirable}
\label{maps}
\end{figure}

Figure \ref{maps} shows heatmaps for the COVID-19 and TB detection applications (without synthetic bias). The tasks use mixed training datasets, which are known to cause background bias and shortcut learning\cite{bassiCovid}\textsuperscript{,}\cite{critic}\textsuperscript{,}\cite{NatureCovidBias}\textsuperscript{,}\cite{ShortcutCovid}. Accordingly, the LRP heatmaps for a standard classifier (DenseNet121) demonstrate a significant influence of background features over the classifier's decisions, indicating shortcut learning. Supporting this finding, Tables \ref{performance} and \ref{tb} show that the model's generalization performance was impaired. It achieved only 0.546 +/-0.01 and 0.566 +/-0.05 average F1-Scores in the COVID-19 and TB o.o.d. test datasets, respectively. Conversely, the heatmaps in Figure \ref{maps} indicate that the ISNet is the DNN with the least amount of background attention in the two tasks. Quantitatively supporting the information in the heatmaps, the ISNet's o.o.d. generalization performance surpassed all other models in TB and COVID-19 detection (Tables \ref{performance} and \ref{tb}), indicating that it could better minimize the influence of background bias over the classification decisions.

Appendix \ref{heatmapAnalysis} thoroughly analyses the heatmaps in view of the quantitative results in Tables \ref{synth} to \ref{tb}, and uses this investigation to compare the ISNet and the benchmark DNNs in more detail. Moreover, it also presents Grad-CAM explanations for the ISNet. Finally, Appendix \ref{radiologistAnalysis} compares the ISNet's LRP heatmaps to X-rays where a radiologist, who had no access to the DNN, marked the lung diseases' lesions. The comparison demonstrates a correlation between the marked regions and the areas that influenced the ISNet's decisions. In summary, the ISNet diverted DNN focus from background bias to the lesions. The high-resolution marked X-rays and ISNet heatmaps are available and individually analyzed in Supplementary Data 1\cite{SupData}, and exemplified in Figure \ref{rad}.

\section{Discussion}

In three synthetic bias applications, considering diverse tasks, dataset sizes, and classifier backbones, we quantitatively demonstrated that the artificial background bias could not influence the ISNet's decisions. Therefore, the model hindered shortcut learning and improved generalization. The COVID-19 and Tuberculosis classification tasks exemplify realistic and contemporary scenarios where dataset mixing is commonly employed, frequently causing background bias, shortcut learning and impaired o.o.d. generalization\cite{bassiCovid}\textsuperscript{,}\cite{critic}\textsuperscript{,}\cite{NatureCovidBias}\textsuperscript{,}\cite{ShortcutCovid}\textsuperscript{,}\cite{TBSegmentation}. The two applications displayed shortcut learning in this study, as indicated by the standard classifiers' (DenseNet121) unimpressive o.o.d. generalization and strong background attention. Consequently, being able to hinder background attention and shortcut learning, the ISNet was consistently the model with the best o.o.d. generalization in the two tasks, like in the problems with synthetic background bias. Accordingly, the ISNet's LRP heatmaps indicate a minimization of the influence of background features on classifier's outputs. 

In the applications with artificial bias and on COVID-19 detection, the segmentation-classification pipeline was surpassed only by the ISNet. However, its generalization capacity was less promising in tuberculosis detection. The synthetic bias experiments quantitatively demonstrated that, besides the ISNet and the segmentation-classification pipeline, the remaining benchmark models could not effectively hinder the shortcut learning caused by background bias. Accordingly, their generalization capacity did not match the ISNet's, and their heatmaps showed significant background attention. 

We justify these empirical findings with a theoretical analysis in Appendix \ref{SOTACompare} and Section \ref{mathBackground}. It explains the benchmark models' drawbacks, which the ISNet does not share. In summary, the segmentation-classification pipeline is robust to background bias, but it is computationally expensive even at run-time, and it may form decision rules based on images' foreground shape, possibly hindering generalization. In the multi-task DNN, foreground features can guide the creation of the segmentation output, while background features heavily influence the DNN's classification output. Therefore, the model can precisely segment the foreground, while background bias influences its classification scores. Attention mechanisms that do not learn from foreground segmentation masks (e.g., AG-Sononet\cite{AGNet} and Vision Transformer\cite{VisionTransformer}) cannot reliably differentiate background and foreground features. Thus, they may not hinder background bias attention and shortcut learning. Input gradients and Gradient*Input explanations are much noisier than LRP for deep neural networks. Accordingly, considering deep backbones and high-resolution images, the ISNet can more effectively and stably minimize background attention in relation to DNNs optimizing input gradients (RRR\cite{RRR}) and Gradient*Input (ISNet Grad*Input). Finally, Grad-CAM optimization (GAIN\cite{GAIN}) can produce spurious Grad-CAM heatmaps, which deceivingly display no background attention, while background bias influences the classifier's outputs. Thus, minimization of Grad-CAM backgrounds may not suppress shortcut learning. Summarizing, in relation to the state-of-the-art, we empirically and theoretically demonstrate the ISNet's superior capacity of avoiding the influence of background bias, thus hindering shortcut learning and improving generalization. 

Besides its superior resistance to background bias, the ISNet introduces no increment in run-time computational cost to its backbone classifier. E.g., by replacing  a U-Net followed by a DenseNet121 (segmentation-classification pipeline) with an ISNet containing a DenseNet121 backbone, we obtain a model that is about 70\% to 108\% faster at run-time and has almost 80\% less parameters. Indeed, the ISNet matches a standard DenseNet121 run-time speed and size (Appendix \ref{speed}). Accordingly, the proposed architecture is an efficient technique to suppress background attention and shortcut learning, increasing confidence in a DNN's decisions. Moreover, it accepts virtually any backbone classifier and represents an interpretable attention mechanism: we know that it works by hindering attention outside of a region of interest, which is clearly defined by the ground-truth segmentation targets used during training. In summary, this study empirically and theoretically demonstrated that the ISNet's optimization of LRP heatmaps is a flexible approach, which produces deep classifiers that are resistant to background bias while retaining high accuracy and efficiency.

\section{Methods} 
\subsection{Layer-wise Relevance Propagation}
\label{LRP}
Since DNNs are complex and nonlinear structures with millions of parameters, it is difficult to explain their decisions. Layer-wise relevance propagation\cite{LRP} (LRP) is an explanation technique tailored for deep models, providing heatmaps to interpret DNNs. A past work qualitatively and quantitatively demonstrated that LRP explanations provide higher resolution heatmaps and more interpretable information than attention mechanisms and the corresponding attention heatmaps\cite{LRPVsAttention}. Furthermore, in relation to standard attention, LRP heatmaps can reveal additional evidence used by the classifier to make a decision\cite{LRPVsAttention}. Other studies compared explanation techniques and found LRP to be among the most robust, surpassing Grad-CAM and Gradient*Input\cite{LRPRobustness}\textsuperscript{,}\cite{LRPvsGrad}. From a theoretical perspective, LRP is rooted on the Deep Taylor Decomposition framework\cite{LRPBook}\textsuperscript{,}\cite{LRPZb}: it explains a classifier's decision by approximating a series of local Taylor expansions, performed at each DNN neuron (Section \ref{mathBackground}).

For each class, an LRP heatmap explains the influence of the input image regions on the classifier's confidence for that class. LRP is based on a semi-conservative propagation\cite{LRP} of a quantity called relevance through the DNN layers, starting from one of the outputs of DNN's last layer neurons (logits), and ending at the input layer, where the heatmap is produced. The meaning of the relevance in the heatmap is determined by the choice of the logit where the propagation starts. Positive values indicate that an input image pixel positively influenced the logit, increasing the classifier confidence in the class it represents. Meanwhile negative values indicate areas that reduced this confidence (e.g., regions that the classifier related to other classes in a multi-class, single-label problem). The relevance's magnitude indicates how important the image regions were for the classifier's decision. 

After choosing the explained output neuron, we define its relevance as its output (the logit, prior to nonlinear activation), and set the relevance of all other last-layer neurons to zero. Afterwards, diverse rules propagate the relevance through each DNN layer, one at a time, until the model's input. The choice of propagation rules influences the heatmap's interpretability, noisiness, and the stability of the propagation\cite{LRPBook}. The most basic rule is called LRP-0. We define the k-th output of a fully-connected layer, $z_{k}$, before the layer's non-linear activation, as:

\begin{equation}
z_{k}=\sum_{j} w_{jk}a_{j}
\end{equation}

Where $w_{jk}$ represents the layer's weight connecting its input j ($a_{j}$) to output k ($z_{k}$). The output k bias parameter is represented as $w_{0k}$, and the equation assumes $a_{0}=1$. LRP-0 propagates the relevance from the layer output, $R_{k}$, to its input, $R_{j}$, according to the following equation\cite{LRPBook}:

\begin{equation}
\label{LRP0}
R_{j}=\sum_{k}\frac{w_{jk}a_{j}}{z_{k}}R_{k}
\end{equation}

LRP-0 redistributes the relevance from the layer's k-th output ($R_{k}$) to its inputs ($a_{j}$) according to how much they contributed to the k-th output ($z_{k}$). A second rule, LRP-$\varepsilon$, changes LRP-0 to improve the relevance propagation stability, noisiness, and the explanation's contextualization and coherence (Section \ref{mathBackground}). It adds a small constant, $\varepsilon$, to $z_{k}$. Being $\mathrm{sign}(\cdot)$, a function evaluating to 1 for positive or zero arguments, and to -1 otherwise, LRP-$\varepsilon$ is defined as:

\begin{equation}
\label{LRPeEquation}
R_{j}=\sum_{k}\frac{w_{jk}a_{j}}{z_{k}+\mathrm{sign}(z_{k})\varepsilon}R_{k} \mbox{, where } \varepsilon>0
\end{equation}

It is possible to adopt a different rule for the DNN input layer, taking into account the input space domain. For images, a rule called LRP-z$^{\mathrm{B}}$\cite{LRPBook} considers the maximum and minimum pixel values allowed in the figures (Appendix \ref{lrpZb}). LRP has defined rules for the most common DNN layers. The technique is scalable and can be efficiently implemented and applied to virtually any neural network architecture. Since convolutions have equivalent fully-connected layers, the rules explained here directly apply to them. However, efficient implementations of LRP for convolutions and other layers are presented in Appendix \ref{layers}.

\subsection{Background Relevance Minimization and ISNet}
\label{IsNetSec}

Layer-wise Relevance Propagation was created to show how input features influenced classifier's decisions\cite{LRP}. With the ISNet, we suggest directly optimizing LRP to improve classifiers' behavior. During training, background relevance minimization (BRM) penalizes undesired relevance in the training images' LRP heatmaps, constraining the classifier to learn decision rules that do not rely on background features. Algorithm \ref{BRMAlgo} explains BRM, the ISNet training procedure. The heatmap loss ($L_{LRP}$, Algorithm \ref{BRMAlgo}, step 2d) employs gold standard foreground segmentation masks to identify and penalize the background relevance in LRP heatmaps. The masks are only necessary for training. They are figures valued one in the image’s foreground, and zero in the background. When not available, a pretrained segmenter can create them. E.g., a U-Net\cite{unet} pretrained to segment a specific type of foreground, or a general model pretrained for novel class segmentation (like DeepMAC\cite{deepMAC}). Appendix \ref{ImpDetails} details training settings, hyper-parameters, data processing, and augmentation used for the applications in this study, considering both the ISNet and the benchmark models. 

\begin{algorithm}
  \caption{Background Relevance Minimization: the ISNet Training Procedure\label{BRMAlgo}}
  \begin{algorithmic}[1]
    \State Initialization: randomly initialize the backbone classifier's trainable parameters, $\mathbf{\theta}$.
    \State Training epoch: for every randomly drawn mini-batch (images $\mathbf{X}$, segmentation ground-truth foreground masks $\mathbf{M}$, and classification labels $\mathbf{Y}$) in the training dataset:
    \begin{algsubstates}
        \State Preprocess and (optional) augment the mini-batch.
        \State Classifier forward pass: classify the B (mini-batch size) images with the backbone classifier, assigning probabilities ($\mathbf{\hat{Y}}$) for the K possible classes.
        \State Layer-wise Relevance Propagation: create $B\times K$ differentiable LRP heatmaps in parallel, $\mathbf{H}$, explaining the K classifier's logits for the B input images. Employ LRP-$\varepsilon$ (we set $\varepsilon=0.01$) throughout the entire DNN, except for its first layer, where we use LRP-z$^{\mathrm{B}}$.
        \State Loss calculation: calculate classification loss (e.g., cross-entropy) according to the classifier output and the classification labels, $L_{C}(\mathbf{\hat{Y}},\mathbf{Y})$. Calculate the heatmap loss according to the LRP heatmaps and the foreground masks, $L_{LRP}(\mathbf{H},\mathbf{M})$. Linearly combine both to produce the ISNet loss, using a balancing hyper-parameter, P: $L_{IS}=(1-P).L_{C}+P.L_{LRP}, \mbox{ where } 0 \le P \le 1$. Increasing P increases the ISNet resistance to background bias, but it may reduce training speed.
        \State Gradient backward pass: use automatic backpropagation to calculate the gradient of the ISNet loss with respect to the backbone classifier's trainable parameters, $\Delta_{\mathbf{\theta}}L_{IS}$.
        \State Optimizer step: update the backbone classifier's trainable parameters according to the gradient $\Delta_{\mathbf{\theta}}L_{IS}$. We use stochastic gradient descent with momentum as the optimizer.
    \end{algsubstates}
    \State Validation epoch. For every mini-batch (images $\mathbf{X}$, segmentation ground-truth foreground masks $\mathbf{M}$, and classification labels $\mathbf{Y}$) in the hold-out validation dataset:
    \begin{algsubstates}
        \State Perform the image processing, classifier forward pass, LRP, and loss calculation as described in steps 2a (except for data augmentation), 2b, 2c, and 2d. Monitor the average validation ISNet loss at the end of each epoch.
    \end{algsubstates}
    \State Repeat steps 2 and 3 until the maximum number of epochs (N) is reached. At the end of the training procedure, return the backbone classifier with the parameters ($\mathbf{\theta}$) that minimized the hold-out validation ISNet loss.
  \end{algorithmic}
\end{algorithm}

We introduce an efficient implementation of Layer-wise Relevance Propagation in PyTorch, dubbed LRP Block, which produces differentiable LRP heatmaps in parallel. The LRP Block allows automatic backpropagation through LRP. Thus, the heatmap loss gradient can be backpropagated from the loss output until the backbone classifier's parameters, allowing the minimization of the ISNet loss (Algorithm \ref{BRMAlgo}, step 2e). The block is presented in detail in Appendix \ref{layers}, which explains its LRP implementation for multiple types of classifier layers. The structure can be deactivated or removed after the training procedure, representing no run-time computational cost. Alternatively, it can explain the trained ISNet's decisions with LRP: minimal background relevance indicates the success of background relevance minimization. Since LRP can be applied to virtually any DNN\cite{LRPBook}, the ISNet can accept virtually any classifier backbone architecture.

We need to produce one heatmap for each possible class, starting the LRP relevance propagation at the output neuron that classifies it (Algorithm \ref{BRMAlgo}, step 2c). We cannot minimize background LRP relevance for a single class (e.g., the winning one). Imagine that we have a bias in the image background, associated with class C, and we minimize only the background LRP relevance for class C. In this case, the classifier can negatively associate all other classes with the bias, using it to lower their outputs, making the class C output neuron the winning one. This negative association is expressable as negative relevance in the other classes' heatmaps. Consequently, the penalization of positive and negative background relevance in all maps is a solution to the problem. Accordingly, the ISNet training time increases with the number of categories in the classification task (K). For efficiency, the LRP Block propagates relevance in batches, producing multiple independent heatmaps in parallel. If a memory limit is reached, heatmaps will need to be produced in series, making training time linearly increase with K. Future ISNet implementations may reduce this drawback. However, as heatmaps are not necessary after training, the run-time ISNet will be as fast and efficient as its backbone classifier. Thus, the ISNet exchanges training time for run-time performance, especially when compared to the benchmark segmentation-classification pipeline (Appendix \ref{speed}). This trade-off may be very profitable, given that DNNs can be trained with powerful computers, then later deployed in less expensive or portable devices. 

\subsection{ISNet Loss Function}

The ISNet loss, $L_{IS}$, is the function minimized during ISNet training (Algorithm \ref{BRMAlgo}, step 2d). It is a linear combination of two terms (Equation \ref{LIS}): $L_{C}$, a standard classification loss (e.g., cross-entropy), and $L_{LRP}$, the heatmap loss. Their influence over the loss gradient is balanced by a hyper-parameter P. The heatmap loss, $L_{LRP}$, is also a linear combination of two functions, the background ($L_{1}$) and the foreground ($L_{2}$) losses (Equation \ref{LLRP}). The combination utilizes two hyper-parameters, $w_{1}$ and $w_{2}$. Both $L_{1}$ and $L_{2}$ depend on the LRP heatmaps and the foreground segmentation masks.

\begin{gather}
\label{LIS}
L_{IS}=(1-P).L_{C}+P.L_{LRP}, \mbox{ where } 0 \le P \le 1\\
\label{LLRP}
L_{LRP}=w_{1}.L_{1}+w_{2}.L_{2}, \mbox{ where } 0 < w_{1} \mbox{ and } 0 < w_{2}
\end{gather}

Algorithms \ref{LbkgAlgo} and \ref{LforgAlgo} describe the calculation of $L_{1}$ and $L_{2}$, respectively. The background loss quantifies and penalizes background relevance in LRP heatmaps. Meanwhile, the foreground loss is an auxiliary term, which ensures the stability of the LRP heatmaps during ISNet training, avoiding zero maps or exploding LRP relevance. Appendix \ref{loss} presents a more detailed description of the functions. Appendix \ref{hyperParameterTuning} details which hyper-parameters require a fine search, a coarse search, or no search. It also explains our hyper-parameter tuning strategy. The ISNet loss is differentiable, and PyTorch can perform automatic gradient backpropagation through it.

\begin{algorithm}
  \caption{The Background Loss, $L_{1}$\label{LbkgAlgo}}
  \hspace*{\algorithmicindent} Input: $B\times K$ heatmaps, where B is the mini-batch size and K the number of classes in the classification task.\\
  \hspace*{\algorithmicindent} A corresponding foreground segmentation mask, $\mathbf{M_{bk}}$, for each heatmap $\mathbf{H_{bk}}$.\\
  \hspace*{\algorithmicindent} Output: $L_{1}$, the scalar background loss for the mini-batch.  
  \begin{algorithmic}[1]
    \State Absolute heatmaps: take the absolute value of the LRP heatmaps (element-wise), abs($\cdot$). This step ensures that the background loss equally penalizes positive and negative LRP relevance. $\mathbf{H_{bk}}$ is the LRP heatmap explaining the classifier output (logit) for class k, considering the mini-batch image b as the DNN's input.\\
    $\mathrm{abs}(\mathbf{H_{bk}})$
    \State Normalized absolute heatmaps: normalize the absolute heatmaps, dividing each map by the absolute average relevance in its foreground region. Use foreground masks, $\mathbf{M_{bk}}$, to define such regions. This step makes the background loss relative. I.e., the loss minimization makes the influence of background features on the classifier progressively smaller than the influence of foreground features. Moreover, because $L_{1}$ is relative, an overall reduction of both background and foreground LRP relevance (which does not represent foreground focus) cannot minimize the loss. Sum($\cdot$) adds all elements in a tensor, and $\odot$ represents element-wise multiplication.\\
    $\mathbf{H'_{bk}}=\mathrm{abs}(\mathbf{H_{bk}})\div\{[\mathrm{Sum}(\mathrm{abs}(\mathbf{H_{bk}}) \odot \mathbf{M_{bk}})/(\mathrm{Sum}(\mathbf{M_{bk}})+e)] +e\}, \mbox{ where } 0 < e << 1$
    \State Segmented heatmaps: in the normalized absolute heatmaps, set all foreground relevance to zero, by element-wise multiplying the maps and inverted foreground masks (i.e., figures valued one in the background and 0 in the foreground, $\mathbf{1}-\mathbf{M_{bk}}$). This step ensures the loss penalizes only background LRP relevance.\\
    $\mathbf{UH'_{bk}}=(\mathbf{1}-\mathbf{M_{bk}}) \odot \mathbf{H'_{bk}}$
    \State Raw background attention scores: use Global Weighted Ranked Pooling\cite{GWRP} (GWRP) over the segmented heatmaps, obtaining one scalar score per map channel. GWRP can be seen as a hybrid between average pooling and max pooling. It is governed by a hyper-parameter d ($0 \le d \le 1$). If $d=0$, GWRP matches max pooling. If d=1, GWRP matches average pooing. The lower the d, the more the background loss penalizes the existence of small background regions with strong influence over the classifier. Thus, lowering d increases the ISNet resistance to background bias. However, smaller values can decrease training stability. GWRP outputs one scalar, $r_{bkc}$, per channel (c) in a segmented heatmap $\mathbf{UH'_{bk}}$. $\mathbf{UH'_{bk}}$ has 3 channels, $\mathbf{UH'_{bkc}}$, when the ISNet classifies RGB images.\\
    $r_{bkc}=\mathrm{GWRP}(\mathbf{UH'_{bkc}})$
    \State Activated scores: pass the raw scores, $r_{bkc}$, through the non-linear function $f(r_{bkc})=r_{bkc}/(r_{bkc}+E)$, where E is a constant hyper-parameter, normally set as 1. The activated scores are naturally limited between 0 and 1, being an adequate input for cross-entropy.
    \State Background attention loss: calculate the cross-entropy between the activated scores and a zero target, $\mathrm{CE}(\cdot)$. We utilize a zero objective in cross-entropy, because lower activated scores represent weaker background attention.\\
    $\mathrm{CE}(f(r_{bkc}))=-\mathrm{ln}(1-f(r_{bkc}))$
    \State $L_{1}$: calculate the average background attention loss for all heatmaps in the training mini-batch:\\
    $L_{1}=\frac{1}{B.K.C}\sum_{b=1}^{B}\sum_{k=1}^{K}\sum_{c=1}^{C}\mathrm{CE}(f(r_{bkc}))$
  \end{algorithmic}
\end{algorithm}

\begin{algorithm}
  \caption{The Foreground Loss, $L_{2}$\label{LforgAlgo}}
  \hspace*{\algorithmicindent} Input: $B\times K$ heatmaps, where B is the mini-batch size and K the number of classes in the classification task.\\
  \hspace*{\algorithmicindent} A corresponding foreground segmentation mask, $\mathbf{M_{bk}}$, for each heatmap $\mathbf{H_{bk}}$.\\
  \hspace*{\algorithmicindent} Output: $L_{1}$, the scalar background loss for the mini-batch.  
  \begin{algorithmic}[1]
    \State Absolute heatmaps: take the absolute value of the LRP heatmaps (element-wise), abs($\cdot$). $\mathbf{H_{bk}}$ is the LRP heatmap explaining the classifier output (logit) for class k, considering the mini-batch image b as the DNN's input.\\
    $\mathrm{abs}(\mathbf{H_{bk}})$
    \State Segmented heatmaps: in the absolute heatmaps, set all background relevance to zero, by element-wise multiplying them by the foreground masks. This steps avoids a direct interference of the foreground loss on the minimization of LRP background relevance, caused by the background loss ($L_{1}$) optimization.\\
    $\mathrm{abs}(\mathbf{H_{bk}}) \odot \mathbf{M_{bk}}$
    \State Absolute foreground relevance: sum all elements in the segmented maps, Sum($\cdot$), obtaining the total absolute foreground relevance per-heatmap. \\
    $g(\mathbf{H_{bk}})=\mathrm{Sum}(\mathrm{abs}(\mathbf{H_{bk}}) \odot \mathbf{M_{bk}})$
    \State Square losses: if a heatmap's absolute foreground relevance, $g(\mathbf{H_{bk}})$, is smaller than a hyper-parameter, $C_{1}$, the corresponding square loss is $L_{2}^{bk}=(C_{1}-g(\mathbf{H_{bk}}))^{2}/C_{1}^{2}$. If it is larger than $C_{2}$ ($C_{2}>C_{1}>0$), we have $L_{2}^{bk}=(C_{2}-g(\mathbf{H_{bk}}))^{2}/C_{1}^{2}$. If $C_{1}>g(\mathbf{H_{bk}})>C_{2}$, we define $L_{2}^{bk}=0$. Accordingly, the foreground loss, $L_{2}^{bk}$, and its gradient are zero when the heatmap's absolute foreground relevance ($g(\mathbf{H_{bk}})$) is within a pre-defined range, $[C_{1},C_{2}]$. However, the loss raises quadratically when the absolute foreground relevance exits the range. The $C_{1}$ and $C_{2}$ hyper-parameters are set to represent a natural range of absolute relevance (Appendix \ref{hyperParameterTuning}).\\
    $
    L_{2}^{bk}=
    \begin{cases}
    \frac{(C_{1}-g(\mathbf{H_{bk}}))^{2}}{C_{1}^{2}}, \mbox{ if } g(\mathbf{H_{bk}}) < C_{1} \\
    0, \mbox{ if } C_{1} \le g(\mathbf{H_{bk}}) \le C_{2} \\
    \frac{(g(\mathbf{H_{bk}})-C_{2})^{2}}{C_{1}^{2}}, \mbox{ if } g(\mathbf{H_{bk}}) > C_{2}
    \end{cases}$
    \State $L_{2}$: take the average of all square losses, considering all heatmaps in the mini-batch. $L_{2}$ will be zero if the LRP relevance stays within normal values, but it quickly raises if it exits this natural range. Thus, $L_{2}$ avoids zero or exploding LRP heatmaps during ISNet training.\\
    $L_{2}=\frac{1}{B.K}\sum_{b=1}^{B}\sum_{k=1}^{K}L_{2}^{bk}$
  \end{algorithmic}
\end{algorithm}

\subsection{ISNet Theoretical Fundamentals}
\label{mathBackground}

We could not find other works optimizing LRP explanations to improve classifiers' behavior. Layer-wise relevance propagation sits among the most robust, interpretable, and high-resolution explanations to date\cite{LRPRobustness}\textsuperscript{,}\cite{LRPVsAttention}. While the relevance signal is propagated through the DNN (from output to input), it extracts context and high level features from late DNN layers, and captures precise spatial information from earlier layers. Accordingly, the resulting explanation heatmap carries both high resolution and high level of abstraction. Therefore, LRP optimization can teach the ISNet to precisely identify the images' foreground features, and to form decision rules based on them (Appendix \ref{precision}).

In this Section, we theoretically justify the empirically verified (Section \ref{results}) ISNet capacity of avoiding background attention, and the shortcut learning caused by background bias. We begin by summarizing the mathematical fundamentals behind LRP, elucidating its relationship with Taylor expansions. Afterwards, we discuss how LRP optimization makes the ISNet robust to background bias. 

\subsubsection{LRP Mathematical Fundamentals}

A DNN implements a function of its input, $y=f(\mathbf{X})$, where $y$ is a network's logit of interest. A first-order Taylor expansion can decompose the logit into a summation of one term per input dimension ($x_{j}$), plus an additional zero-order term ($f(\mathbf{\tilde{X}})$), and an approximation error ($\rho$) with respect to higher-order Taylor expansions (Equation \ref{taylorRelevance}). To minimize the approximation error (i.e., the Taylor residuum, $\rho$), the Taylor reference ($\mathbf{\tilde{X}}$, with elements $\tilde{x}_{j}$) should be close to the data point $\mathbf{X}$. Furthermore, by setting the reference point as a root of the function $f(\mathbf{X})$, we remove the zero-order element ($f(\mathbf{\tilde{X}})=0$). Thus, with a nearby root as the Taylor reference, the logit is decomposed as the summation of the first-order terms, with one term per input dimension (Equation \ref{Taylor2}). In this case, the terms explain the differential contribution of each input element to the logit, with respect to the logit's state of maximal uncertainty, when it is zero (50\% class probability for a sigmoid function, and the ReLU function hinge)\cite{LRP}. I.e., the terms explain how each input element contributed to making the logit different from zero. Accordingly, the terms can form a heatmap, explaining and decomposing the network's output, $y$. For the explanations to be more meaningful, the Taylor reference should reside in the classification problem data manifold\cite{LRP}. 

\begin{gather}
\label{taylorRelevance}
    f(\mathbf{X}) = f(\mathbf{\tilde{X}}) + \sum_{j}(x_{j}-\tilde{x}_{j})\at{\frac{\partial f}{\partial x_{j}}}{\mathbf{\tilde{X}}} + \rho \\
    \label{Taylor2}
    f(\mathbf{X}) \approx \sum_{j}(x_{j}-\tilde{x}_{j})\at{\frac{\partial f}{\partial x_{j}}}{\mathbf{\tilde{X}}}
\end{gather}

Although principled, adequate Taylor explanations are difficult to compute for deep neural networks. The function $f(\mathbf{X})$ is complicated, highly non-linear, and can have noisy gradients. Finding a root $\mathbf{\tilde{X}}$ that satisfies all aforementioned requirements is a complex and analytically intractable problem\cite{LRPZb}. However, a DNN function, $f(\mathbf{X})$, is defined as a structure of simpler sub-functions, learned at each neuron. Finding Taylor references for the simpler functions is easier, and approximate analytical solutions can be found for the corresponding Taylor expansions\cite{LRPZb}.

The LRP relevance propagation rules we employ for the ISNet (LRP-$\varepsilon$ and LRP-z$^{\mathrm{B}}$) are justified by the Deep Taylor Decomposition (DTD) framework \cite{LRPZb}\textsuperscript{,}\cite{LRPBook}. DTD propagates relevance, from the classifier's logit to its inputs, according to local Taylor expansions performed at each neuron. The neuron output relevance ($R_{k}$, received from the subsequent layer) is viewed as a function of the neuron's inputs, $\mathbf{a}$ (composed of elements $a_{j}$). Accordingly, Equation \ref{relevanceDecompose} displays a first-order Taylor expansion of $R_{k}(\mathbf{a})$, considering a reference point $\mathbf{\tilde{a}}$.

\begin{equation}
\label{relevanceDecompose}
    R_{k}(\mathbf{a}) = R_{k}(\mathbf{\tilde{a}}) + \sum_{j}(a_{j}-\tilde{a}_{j})\at{\frac{\partial R_{k}}{\partial a_{j}}}{\mathbf{\tilde{a}}}+ \rho 
\end{equation}

The relevance $R_{k}(\mathbf{a})$ is redistributed to the neuron's inputs $a_{j}$ according to the first-order terms in the expansion (summed terms in Equation \ref{relevanceDecompose})\cite{LRPBook}. Ideally, we choose a reference point ($\mathbf{\tilde{a}}$) that minimizes the zero-order term $R_{k}(\mathbf{\tilde{a}})$. We also want the reference to be close to the data point $\mathbf{a}$, as the proximity reduces the Taylor residuum ($\rho$). Finding this reference point is still not simple, nor computationally inexpensive, considering the complexity of the function $R_{k}(\mathbf{a})$\cite{LRP}. Therefore, to obtain a closed-form solution for the Taylor expansion, LRP considers an approximation of $R_{k}(\mathbf{a})$ (dubbed approximate relevance model, $\hat{R}_{k}(\mathbf{a})$), and standardized choices of the reference $\mathbf{\tilde{a}}$. These choices are justified by the approximate relevance model and the neuron's input domain\cite{LRPBook}. For a neuron with ReLU activation (Equation \ref{neuron}), a modulated ReLU activation is the most common relevance model ($\hat{R}_{k}(\mathbf{a})$). It is defined in Equation \ref{relevanceModel}, where $c_{k}$ is a constant chosen to force the model to match the true relevance at the data point ($R_{k}(\mathbf{a})=\hat{R}_{k}(\mathbf{a})$). For further explanation and theoretical justification of the relevance model, please refer to \cite{LRPBook}.

\begin{equation}
\label{neuron}
a_{k}=\mathrm{max}(0,z_{k})=\mathrm{max}(0,\sum_{j} w_{jk}a_{j})
\end{equation}

\begin{equation}
\label{relevanceModel}
\hat{R}_{k}(\mathbf{a})=\mathrm{max}(0,\sum_{j} w_{jk}a_{j})c_{k}
\end{equation}

The LRP-$\varepsilon$ and LRP-0 rules are derived by selecting different reference points ($\mathbf{\tilde{a}}$) for the approximate local Taylor expansions\cite{LRPBook}. All these points satisfy $\tilde{a}_{j} \geq 0$, thus residing in the neuron's input domain (considering that it follows other neurons with ReLU activations). For LRP-0, we have $\mathbf{\tilde{a}}=\mathbf{0}$. For LRP-$\varepsilon$ the definition is: $\mathbf{\tilde{a}}=\frac{\varepsilon}{a_{k}+\varepsilon}\mathbf{a}$. The Euclidean distance between the LRP-$\varepsilon$ reference point and the actual data point, $\mathbf{a}$, is much smaller than the distance between $\mathbf{a}$ and the LRP-0 reference point\cite{LRPBook}. Therefore, in comparison to LRP-0, LRP-$\varepsilon$ reduces the Taylor residuum ($\rho$) in relation to higher order Taylor expansions, creating heatmaps that are more faithful, less noisy, and more contextualized\cite{LRPBook}.

\subsubsection{Optimizing for Background Bias Resistance: Why LRP?}

There are some key qualities we expect the optimized explanation strategy to have: it must be differentiable, consider both positive and negative evidence (Section \ref{IsNetSec}), and be computationally efficient. Moreover, we search for an explanation methodology that can fundamentally justify why its optimization leads to resistance to background bias. LRP-$\varepsilon$\cite{LRP} satisfies these requirements. As we show in the LRP block, the strategy is differentiable. Second, it is fast, constructing an LRP heatmap requires a single backpropagation of relevance through the neural network. Efficiency and differentiability are essential for explanations that must be created during training and optimized. Third, LRP-$\varepsilon$ considers both positive and negative relevance. Finally, the technique is principled, because it approximates the sequential application of local Taylor expansions (per-neuron) in deep neural networks with ReLU activations. Interestingly, we had no success in preliminary experiments with the optimization of LRP rules that do not consider negative evidence (e.g., LRP-$\gamma$\cite{LRPBook}), or that are not justified by the deep Taylor framework (e.g., LRP-$\alpha \beta$\cite{LRPBook}). They could not produce background bias resistance.

A previously explained, the terms of a first-order Taylor expansion, considering an adequate Taylor reference (a nearby root in the data manifold), indicate the contribution of a function's input elements to its output variation, when the input moves from the Taylor reference to the actual data point. Thus, the minimization of terms associated with bias should minimize the bias contribution to the function output variation, when its input moves from a point of maximal uncertainty (root) to the current data point. I.e., it minimizes the bias influence over the function's output. However, the creation of adequate Taylor explanations of a DNN is a complex and analytically intractable problem\cite{LRP}. Thus, such explanations violate our requirement of computational efficiency. 

However, LRP-$\varepsilon$ is a fast procedure, which explains a DNN output by approximating a sequence of local Taylor expansions. For this reason, LRP-$\varepsilon$ optimization is an efficient and justifiable alternative to minimize the influence of biased input elements on the network's logits. Recapitulating, LRP-$\varepsilon$ relevance propagation starts with the classifier logit we are explaining and uses an approximate Taylor expansion to decompose its value, and redistribute it to the inputs of the last DNN layer, according to their contribution to the logit. Repeating this procedure, the decomposition results (relevance) are further decomposed and redistributed multiples times through the DNN. I.e., an approximate Taylor expansion at each DNN neuron decomposes its output relevance and redistributes it to the neuron's inputs, according to how much they contributed to the neuron's output relevance (Equation \ref{relevanceDecompose}). The procedure ends at the input layer, forming the LRP heatmap. 

For the background bias to influence the logit, it needs to influence neurons (or convolutional activations) throughout the entire DNN, until its last layer. The layer L neurons that carry and process the bias information must influence neurons in layer L+1, or the bias information will not reach the DNN output. This influence will be captured by the local Taylor expansions performed at the layer L+1 neurons, affecting the LRP relevance flow. When background bias influences the classifier's decisions, it produces a flow of influence from the bias to the logit, encompassing the neurons and connections that carry and process the bias information. This influence flow will cause a corresponding flow of LRP relevance, bringing relevance from the logit to the background region of the LRP heatmap. The ISNet's background relevance minimization procedure optimizes DNN parameters to constrict and ultimately stop the relevance flow from the logits to the LRP heatmaps' background, thus minimizing the corresponding influence and information flow from the background bias to the logit. Accordingly, the ISNet hinders the background bias influence on the classifier's outputs.

We selected the LRP-$\varepsilon$ rule to propagate relevance through all DNN, except for its first layer. The input domain of the first DNN layer is diverse. In a network with ReLU activations and analyzing images, the first layer inputs range from 0 to 1 (standardized pixels), while other layers' inputs assume values in $\mathbf{R^{+}}$. While LRP-$\varepsilon$ is adequate for the remaining layers, the LRP-z$^{\mathrm{B}}$ rule represents a more accurate choice of Taylor reference for the first layer\cite{LRPZb}. According to preliminary tests, the ISNet works with the LRP-$\varepsilon$ in layer one, but LRP-z$^{\mathrm{B}}$ produced an accuracy improvement. 

Meanwhile, LRP-0 explanations are highly noisy, and less interpretable than LRP-$\varepsilon$. In the LRP-$\varepsilon$ propagation rule (Equation \ref{LRPeEquation}), the $\varepsilon$ term not only avoids division by zero, but it absorbs some of the relevance that would have been propagated to the lower layer. The absorption reduces the influence of neurons with small activations ($z_{k}$ in Equation \ref{LRPeEquation}) over the relevance signal being propagated. Therefore, it reduces noise and contradiction in the resulting heatmaps\cite{LRPBook}. This improvement is justified by the DTD framework. As previously noticed, in relation to LRP-0, LRP-$\varepsilon$ represents a deep Taylor decomposition using Taylor references that are closer to the data points. Thus, LRP-$\varepsilon$ reduces the Taylor residuum, and explanations become more contextualized and coherent with the network's behavior\cite{LRPBook}.

Past studies showed that, in DNNs based on ReLU activations, LRP-0 is equivalent to Gradient*Input explanations (assuming no division by zero or numerical instabilities in LRP-0)\cite{LRPBook}. The equality demonstrates that the advantages of LRP-$\varepsilon$ over LRP-0 directly apply when comparing LRP-$\varepsilon$ to Gradient*Input. Gradient*Input is an explanation technique proposed to improve the sharpness of input gradients\cite{GradInput}, by multiplying them (element-wise) with the DNN input itself. Input gradients (or saliency maps) are another explanation technique\cite{saliency}, representing the gradient of a DNN logit with respect to the model's input. To create input gradients and Gradient*Input explanations, we backpropagate the gradient of the logit corresponding to the class we want the heatmap to explain. The ISNet Grad*Input is an ablation study, where we substituted the ISNet LRP heatmaps by Gradient*Input. Right for the Right Reasons (RRR) is a model that optimizes input gradients (alongside a standard classification loss), minimizing their background values, which are identified by ground-truth foreground masks\cite{RRR}. Essentially, DNN optimizers treat the networks' inputs as constants. Consequently, background minimization in Gradient*Input minimizes the input gradient's background elements. Therefore, Gradient*Input optimization shares the solid fundamental from RRR: the minimization of input gradients' backgrounds make classifier's outputs locally invariant to changes in images' backgrounds. The learned local invariance can be generalizable: during testing, RRR also based its decisions on foreground features, instead of background bias\cite{RRR}. Both input gradients and Gradient*Input are computationally efficient (similar to LRP, refer to Appendix \ref{speed}), differentiable, and show positive and negative evidence.

However, both input gradients and Gradient*Input are noisy for large DNNs analyzing high-resolution images\cite{LRPvsGrad}. RRR was originally tested in networks with few hidden layers\cite{RRR}. This study considers deeper classifier backbones (DenseNet121, with 121 layers, and VGG-19, with 19 layers) and 224x224 images. Accordingly, when employing the VGG-19, the ISNet's generalization and background bias resistance significantly surpassed RRR, and slightly surpassed the ISNet Grad*Input (with confidence interval overlap, Table \ref{synth}). However, with the deeper backbone, the ISNet significantly surpassed both models, without overlaps (Table \ref{synth}). In summary, our empirical findings (Tables \ref{synth} to \ref{tb}) indicate that the LRP-$\varepsilon$ theoretical advantages over LRP-0 or Gradient*Input (lower noise, higher coherence and better contextualization) allow the ISNet to more effectively and stably minimize background attention, better hindering the shortcut learning caused by background bias and improving o.o.d. generalization. 

Appendix \ref{SOTACompare} thoroughly compares the ISNet to RRR and the ISNet Grad*Input, and it explains the alternative models in more detail. Appendix \ref{lrpVsGradInput} formally demonstrates the equivalence between LRP-0 and Gradient*Input. Afterwards, it provides an alternative view on LRP-$\varepsilon$, which makes its denoising quality clear. Finally, Appendix \ref{GAINComparison} displays the advantages of LRP optimization over the optimization of another popular explanation technique, Grad-CAM\cite{GradCAM}.

\section*{Data Availability}
Source data are provided with this paper. The X-ray data from healthy and/or pneumonia positive subjects used in this study are available in the Montgomery and Shenzen databases\cite{ChineseDataset1}, \url{http://archive.nlm.nih.gov/repos/chestImages.php}; in ChestX-ray14\cite{chex14}, \url{https://paperswithcode.com/dataset/chestx-ray14}; and in CheXpert\cite{irvin2019chexpert}, \url{https://stanfordmlgroup.github.io/competitions/chexpert/}. The COVID-19 radiography data used in this study are available in The BrixIA COVID-19 project\cite{BrixiaSet}, \url{https://brixia.github.io/}; and in the BIMCV-COVID19+ database\cite{BimcvSet}, \url{https://bimcv.cipf.es/bimcv-projects/bimcv-covid19/}. The tuberculosis-positive X-rays used in this study are available in NIAID TB Portals\cite{TBPortals}, \url{https://tbportals.niaid.nih.gov/download-data}. The Images for facial attribute estimation (along with their segmentation masks) used in this study are available in the Large-scale CelebFaces Attributes (CelebA) Dataset\cite{celebA}, \url{https://mmlab.ie.cuhk.edu.hk/projects/CelebA.html}. The MIMIC-CXR-JPG database (v2.0.0)\cite{MIMIC}\textsuperscript{,}\cite{MIMIC1}\textsuperscript{,}\cite{PhysioNet} used in this study is available at \url{https://physionet.org/content/mimic-cxr-jpg/2.0.0/}. The Stanford Dogs\cite{StanfordDogs} dataset used in this study is available at \url{http://vision.stanford.edu/aditya86/ImageNetDogs/}. X-rays with lesions marked by the radiologist and the corresponding ISNet Layer-wise Relevance Propagation heatmaps (Supplementary Data 1), generated in this study, have been deposited in \url{https://doi.org/10.6084/m9.figshare.24243895.v2}\cite{SupData}. All data supporting the findings described in this manuscript are available in the article and in the Supplementary Information and from the corresponding author upon request.

\section*{Code Availability}
The code containing the ISNet PyTorch implementation is available at \url{https://github.com/PedroRASB/ISNet}\cite{CodeGit}. It also presents the implementations for the benchmark deep neural network architectures.

\section*{Acknowledgments}
P.R.A.S.B. thanks the funding from the Center for Biomolecular Nanotechnologies, Istituto Italiano di Tecnologia (73010, Arnesano, LE, Italy). A.C. thanks the funding from the Istituto Italiano di Tecnologia (6163, Genova, GE, Italy). We gratefully acknowledge the HPC infrastructure and the Support Team at Fondazione Istituto Italiano di Tecnologia. Data were obtained from the TB Portals (https://tbportals.niaid.nih.gov), which is an
open-access TB data resource supported by the National Institute of Allergy and
Infectious Diseases (NIAID) Office of Cyber Infrastructure and Computational Biology
(OCICB) in Bethesda, MD. These data were collected and submitted by members of the
TB Portals Consortium (https://tbportals.niaid.nih.gov/Partners). Investigators and other
data contributors that originally submitted the data to the TB Portals did not participate
in the design or analysis of this study.

\section*{Author Contributions Statement}
P. R. A. S. B. developed the study's concept, implemented the neural networks, and analyzed the results. S. S. J. D. annotated lesions in the X-rays. A. C. supervised and reviewed the work.

\section*{Competing Interests Statement}
The authors declare no competing interests.

\section*{Corresponding Author}
Correspondence to Pedro R.A.S. Bassi and Andrea Cavalli.


\bibliography{sample}

\appendix

\section{Details about the ISNet Loss Function}
\label{loss}

Two terms comprise the ISNet loss ($L_{IS}$), a classification loss ($L_{C}$), which penalizes the classifier's classification outputs considering the labels, and a term that minimizes the amount of background attention present in the heatmaps, created by the LRP Block. We call this term 'heatmap loss' or $L_{LRP}$.

\begin{equation}
L_{IS}=(1-P).L_{C}+P.L_{LRP}, \mbox{ where } 0 \le P \le 1
\end{equation}

The hyper-parameter $P$ in the above equation balances the influence of the two losses in the gradient and parameters update. It must be valued between 0 and 1, with larger values increasing the strength of the heatmap loss. If P is too small, the network will not minimize $L_{LRP}$ effectively, and background attention will be high. Training datasets with stronger biases may require higher P values. However, if P is too high, training classification accuracy will decrease very slowly. To tune P, we advise a search through different values, as the ideal choice may change for different backbones and tasks. With adequate values of P, the two losses will converge. Please refer to Appendix \ref{hyperParameterTuning} for details about the hyper-parameter tuning procedure.

$L_{C}$ is a standard loss function for classification. Here, we used cross-entropy or binary cross-entropy (for a single-label or multi-label task, respectively), averaged across the mini-batch samples. We will now explain the calculation of the second term, $L_{LRP}$. First, some definitions are needed: we denote as $\mathbf{H}$ a tensor containing the heatmaps for all classes, created for a single input image or for a mini-batch of B images (in this paper we use bold letters to indicate tensors). For a multi-class classification problem with K classes and input images of shape (C,Y,X), where C is the number of channels, Y the image height, and X its width, $\mathbf{H}$ assumes the following shape: (B,K,C,Y,X).The second dimension (class) indicates at which output neuron the relevance propagation began. Thus, in the k-th heatmap, the positive relevance is associated with the class k. A single LRP heatmap (for image b and class k), $\mathbf{H_{bk}}$, has the same shape as an input image, (C,Y,X). The mini-batch and classes dimensions were merged inside the LRP block, but we separate them before loss calculation. $\mathbf{M}$ is a tensor containing the segmentation targets (masks) for each figure in the mini-batch. A mask separates the image foreground and background. It is defined as a single-channel image with shape (Y,X), valued 1 in the regions of interest and 0 in areas of undesired attention. Having B masks, one for each mini-batch image, we just repeat them in the channels and classes dimensions for $\mathbf{M}$ to match the shape of $\mathbf{H}$. We denote the mask for a single image as $\mathbf{M_{bk}}$. It is possible to use different masks for different classes in the image, or to repeat the same mask for all classes.

The heatmap loss is composed of two terms, the first is the heatmap background loss, $L_{1}$, and it penalizes background attention. Its calculation starts by the computing of the element-wise absolute value of the heatmaps, $\mathrm{abs}(\mathbf{H_{bk}})$, because the training procedure must minimize both the positive and the negative undesired relevance. Afterwards, the absolute maps are normalized, as all their elements are divided by the mean value inside their region of interest (RoI). To compute this mean, we perform an element-wise product ($\odot$) between the absolute map and the image's segmentation target, $\mathbf{M_{bk}}$, sum the elements of the resulting tensor, $\mathrm{Sum}(\cdot)$, and divide the summation result by the number of elements in the region of interest, $\mathrm{Sum}(\mathbf{M_{bk}})$. In all divisions we add a small $e$ value ($10^{-10}$) to the denominator, enforcing the relationship $0/0=0$, and avoiding indeterminate results. We call the normalized absolute heatmaps $\mathbf{H'_{bk}}$. If an image region produced small relevances (less than 1) in its $\mathbf{H'_{bk}}$ maps, the attention paid to the area was necessarily smaller than the average attention paid to the region of interest. Notice that our normalization procedure does not shift the zero value in the heatmaps. Thus, minimizing relevance in $\mathbf{H'_{bk}}$ also minimizes it in the original LRP heatmaps.

\begin{equation}
\mathbf{H'_{bk}}=\frac{\mathrm{abs}(\mathbf{H_{bk}})}{[\mathrm{Sum}(\mathrm{abs}(\mathbf{H_{bk}}) \odot \mathbf{M_{bk}})/(\mathrm{Sum}(\mathbf{M_{bk}})+e)] +e}
\label{eqabs}
\end{equation}

$\mathbf{H'}$ is a tensor containing all normalized absolute heatmaps $\mathbf{H'_{bk}}$, with the same shape as $\mathbf{H}$. The next step in the $L_{1}$ loss calculation is zeroing the relevance inside the regions of interest in the $\mathbf{H'_{bk}}$ heatmaps. This is because background relevance minimization should not affect the classifier's ability to analyze these regions. With $\mathbf{1}$ being a tensor whose elements are 1, with the same shape as $\mathbf{H}$ and $\mathbf{M}$, we create the inverted masks tensor, $\mathbf{1}-\mathbf{M}$, whose values are 1 in the region of undesired attention and 0 in the important image areas. Thus, an element-wise multiplication of $\mathbf{H'}$ with the inverted masks generates a new tensor, $\mathbf{UH'}$, containing heatmaps that show only the undesired relevance:

\begin{equation}
\mathbf{UH'}=(\mathbf{1}-\mathbf{M}) \odot \mathbf{H'}
\end{equation}

Afterwards, we apply a Global Weighted Ranked Pooling\cite{GWRP} (GWRP) operation to $\mathbf{UH'}$. GWRP is a weighted arithmetic mean of $\mathbf{UH'}$ in the two spatial dimensions (y and x), outputting a tensor $\mathbf{R}$ (with elements $r_{bkc}$), which has the remaining three dimensions (batch, class, and channel). GWRP begins by ranking the elements in each (Y,X) matrix inside the tensor $\mathbf{UH'}$ in descending order, creating vectors $\mathbf{V_{bkc}}$, whose elements we will call $V_{i}^{bkc}$, $i=\{1,2,...,X.Y\}$. Naturally, we will have one vector for each batch element (b), class (k), and channel (c). Each vector is reduced to a scalar, which we will call $r_{bkc}$, by taking the weighted arithmetic mean of its elements. The weights in the mean decrease exponentially, such that high relevance elements in the heatmap's background strongly increase $r_{bkc}$. A constant scalar, $d$, is the hyper-parameter controlling the rate of the exponential decay (Equation \ref{gwrpEquation}). Notice that, when $d=0$, GWRP is the same as max pooling, and, when $d=1$, it is the same as average pooling. Training is more stable with higher d, but higher values may require higher P to avoid background attention, which can reduce training speed. We suggest testing d values like 0.9, 0.996 and 1 (refer to Appendix \ref{hyperParameterTuning} for our hyper-parameter tuning procedure). Alternatively, the parameter can be set to 1 to simplify hyper-parameter tuning and be reduced only if more bias resistance is needed even for high P (e.g., for $P>0.8$). The purpose of using GWRP is to allow a strong penalization of minute high attention regions in the image background.

\begin{equation}
\label{gwrpEquation}
r_{bkc}=\frac{\sum_{i}d^{i}.V_{i}^{bkc}}{\sum_{i}d^{i}}, \mbox{ where } 0 \le d \le 1
\end{equation}

The scalars $r_{bkc}$ are real positive values, which measure of the background relevance in channel c of the heatmap for the mini-batch image b, starting the LRP relevance propagation at class k.

Cross-entropy is the most common error function for image segmentation. We want the classifier to distinguish the image foreground and background, to ignore the latter. Therefore, cross-entropy seemed a sensible choice for the $r_{bkc}$ penalization. As the cross-entropy's target for the undesired relevance, the natural choice is zero, which represents a model not considering background features during classification. Under this specific condition, the cross-entropy function, $\mathrm{CE}(\cdot)$, for a scalar, $x$, can be expressed as:

\begin{equation}
\mathrm{CE}(x)=-\mathrm{ln}(1-x)
\end{equation}

Which evaluates to 0 when $x=0$, and to infinity when $x=1$. Consequently, we must choose a function, $f(\cdot)$, to map $r_{bkc}$ to the interval $[0,1[$ before applying cross-entropy. Some design requirements for $f(\cdot)$, considering positive arguments ($r_{bkc}\geq 0$), are: being monotonically increasing, differentiable, limited between 0 and 1, and to map 0 to 0. The sigmoid function cannot be used, since it evaluates to 0.5 for a null input. After practical tests with some candidates, we decided to use the following function:

\begin{equation}
f(x)=\frac{x}{x+E}
\end{equation}

Where $E$ is a hyper-parameter controlling the function's slope and how fast it saturates to 1. We set this value to 1 in all experiments, as the ISNet does not appear to be overly sensitive to the parameter (Appendix \ref{hyperParameterTuning}). Smaller values lead to a stronger penalization of background attention, and earlier saturation of the loss function, which may decelerate the beginning of the optimization process.

Now, we can calculate $\mathrm{CE}(f(r_{bkc}))$, the cross-entropy between $f(r_{bkc})$ and 0, and average it across the channels (c), classes (k) and batch (b) dimensions, resulting in the scalar heatmap background loss, $L_{1}$. Considering B mini-batch images, K classes, and C image channels, we have:

\begin{gather}
\mathrm{CE}(f(r_{bkc}))=-\mathrm{ln}(1-\frac{r_{bkc}}{r_{bkc}+E})\\
L_{1}=\frac{1}{B.K.C}\sum_{b=1}^{B}\sum_{k=1}^{K}\sum_{c=1}^{C}\mathrm{CE}(f(r_{bkc}))
\end{gather}

The loss function monotonically grows with the undesired relevance in a heatmap, $r_{bkc}$, presenting a minimum at 0, when there is an absence of focus on the image background. As $\mathrm{CE}(f(r_{bkc}))$ diverges if $f(r_{bkc})=1$, we clamp $f(r_{bkc})$ between 0 and $1-10^{-7}$ for numerical stability.

Notice that $L_{1}$ evaluates to zero if all values in the LRP heatmaps are zero. Moreover, due to the heatmap normalization, the loss could be reduced with an unnatural growth of the heatmap ($\mathbf{H_{bk}}$) relevance inside the region of interest, instead of a reduction of the background relevance, which is our objective. To avoid these two undesired solutions to $L_{1}$, we included a second term in the heatmap loss, $L_{2}$. We called it heatmap foreground loss. Given an unnormalized heatmap $\mathbf{H_{bk}}$, $L_{2}^{bk}$ evaluates the absolute values of the elements inside the heatmap region of interest. If the sum of these values, $g(\mathbf{H_{bk}})$ (the total absolute foreground relevance), falls within an expected range, $[C_{1},C_{2}]$, the loss and its gradient are both zero. However, the error function increases quadratically if the sum moves below $C_{1}$, or above $C_{2}$. We set the hyper-parameters $C_{1}$ and $C_{2}$ with an analysis of the LRP heatmaps of a non-ISNet classifier (refer to Appendix \ref{hyperParameterTuning}), making $[C_{1},C_{2}]$ reflect a natural range of total absolute relevance. The loss $L_{2}^{bk}$ is calculated per heatmap, and the scalar $L_{2}$ is its mean across all classes and mini-batch elements.

\begin{gather}
\label{gLoss}
g(\mathbf{H_{bk}})=\mathrm{Sum}(\mathrm{abs}(\mathbf{H_{bk}}) \odot \mathbf{M_{bk}}) \\
\label{L2}
L_{2}^{bk}=
\begin{cases}
\frac{(C_{1}-g(\mathbf{H_{bk}}))^{2}}{C_{1}^{2}}, \mbox{ if } g(\mathbf{H_{bk}}) < C_{1} \\
0, \mbox{ if } C_{1} \le g(\mathbf{H_{bk}}) \le C_{2} \\
\frac{(g(\mathbf{H_{bk}})-C_{2})^{2}}{C_{1}^{2}}, \mbox{ if } g(\mathbf{H_{bk}}) > C_{2}
\end{cases} \\
L_{2}=\frac{1}{B.K}\sum_{b=1}^{B}\sum_{k=1}^{K}L_{2}^{bk}
\end{gather}

The division by $C_{1}^{2}$ in Equation \ref{L2} ensures that zero valued heatmaps have high loss even when $C_{1}$ is small. Finally, we define the heatmap loss as a linear combination of the background and the foreground losses. We did not need to fine-tune the combination weights (Appendix \ref{hyperParameterTuning}). Our total loss function, $L_{IS}$, can be defined as:

\begin{equation}
L_{IS}=(1-P).L_{C}+P.L_{LRP}, \mbox{ where }L_{LRP}=w_{1}.L_{1}+w_{2}.L_{2}
\end{equation}

The main idea behind $L_{2}$ is to avoid unnatural (too low or too high) attention to the region of interest, and to make the heatmap loss ($L_{LRP}$) have minimal effect over the region when its LRP relevance is within a natural range ($[C_{1},C_{2}]$). Thus, $L_{LRP}$ minimizes background focus, while the attention pattern inside the RoI is naturally guided by the classification loss ($L_{C}$) optimization.

\section{Detailed Comparisons to the State-of-the-Art}
\label{SOTACompare}

\subsection{The Alternative Models: U-Net Followed by Classifier, Standard Classifier, Multi-task U-Net, Attention Gated Sononet, Vision Transformer, Guided Attention Inference Network, and Right for the Right Reasons}
\label{BaselineImplementations}

The ISNet PyTorch and PyTorch Lightning (for simple multi-GPU training) implementations are publicly available at \url{https://github.com/PedroRASB/ISNet}\cite{CodeGit}, along with the implementations for the multiple state-of-the-art benchmark DNNs. Our first baseline DNN is a standard classifier, without any special mechanism to avoid attention to background bias. We refer to it as standard classifier, it consists of a standalone VGG-19 for Stanford Dogs, and a DenseNet121 in the remaining tasks.

Attention mechanisms were introduced to the field of computer vision to make neural networks focus on salient image features, and pay less attention to irrelevant aspects of the figure. Especially, the paradigm of spatial attention for classification aims to create DNNs that focus on image regions that are the most relevant for the classification task\cite{attentionSurvey}. Examples of spatial attention mechanisms are recurrent attention models (RAM)\cite{RAM}, spatial transformer networks (STN)\cite{STN} and Attention Gated Networks\cite{AGNet}. A recent review\cite{attentionSurvey} extensively covered many spatial attention mechanisms employed in computer vision. Although the strategies vary greatly, there is a common characteristic in the mechanisms employed for image classification: they rely solely on the images and their classification labels to learn where the DNN shall place its attention. I.e., the networks learn that an image region is important if paying attention to it improves classification performance. They do not learn from semantic segmentation masks. Because focusing on bias can increase training accuracy, we argue that most state-of-the-art attention mechanisms are not adequate to prevent shortcut learning.

Exemplifying attention mechanisms that do not rely on semantic segmentation, we trained an attention gated Sononet\cite{AGNet} as a benchmark model. It was specifically chosen for two reasons: it is an attention mechanism for convolutional neural networks (CNNs), and the tested ISNet is also a CNN; and the mechanism was designed as an efficient substitute for the segmentation-classification pipeline commonly used to classify medical images\cite{AGNet}, an architecture that we also employ as baseline. Attention gates are a visual attention mechanism designed to make CNNs automatically focus on important image regions and ignore irrelevant features. They present minimal computational overhead and were effective for the classification of fetal ultrasound screenings, which are noisy and hard to interpret images\cite{AGNet}. In summary, an attention gate multiplies (element-wise) a convolutional layer feature map with attention coefficients. They aim to reinforce layer activations inside important regions, and attenuate activations in irrelevant areas. The attention gate computes the coefficients based on the gated layer feature map, and on a feature map from a later convolutional layer (e.g., the last). This map contains more context information, but is coarse, limiting the mechanism's resolution. In this study we implemented the AG-Sononet, an attention gated CNN based on the VGG-16 architecture, which also bases the U-Net in the alternative segmentation-classification pipeline and the multi-task U-Net. We utilize the model PyTorch implementation developed by the method's authors. Its detailed description can be found in their paper\cite{AGNet}.

The Vision Transformer\cite{VisionTransformer} is another benchmark model, as it represents an increasingly popular choice of attention mechanism, which also does not learn from segmentation masks. The model is based on the transformer, a highly successful structure in natural language processing. In summary, the transformer implements a self-attention mechanism, which relates diverse positions in its input sequence, producing a new representation of the sequence\cite{transformer}. Initially, the sequence elements are based on linear projections of patches in the input image, which are processed by a series of transformer encoder layers. Unlike all other tested models, the Vision Transformer is not a convolutional neural network. Specifically, we employed the ViT-base (ViT-B/16) architecture\cite{VisionTransformer}, which has 12 layers, breaks the input image into 16x16 patches, and contains 86M parameters. The specific model was chosen because its parameter count scale more closely matches the other neural networks in this study. Moreover, since we do not employ pretraining in this work (a decision justified in Appendix \ref{TransferLearning}), utilizing a model with a more conservative size can reduce the risk of overfitting. The network's implementation is the standard PyTorch Vision Transformer model.

To avoid classifier focus on background bias, attention mechanisms based on semantic segmentation are required. Image segmentation consists of semantically partitioning a figure, assigning a category to each of its pixels. The process allows a differentiation between objects, and background identification. The correlation between background features and image classes is harmless to a segmenter, since these features do not contribute to the optimization of the segmentation loss, which does not rely on the classification labels. Therefore, attention mechanisms based on semantic segmentation can force a classifier to disregard the image background. As a general definition, we can characterize a spatial attention mechanism for a classifier as a strategy that dynamically selects where a DNN pays attention to. Accordingly, the most common segmentation-based attention mechanism is the segmentation-classification pipeline. Other strategies involve using segmentation outputs/masks to weight classifier convolutional feature maps\cite{SSG}. In both cases, prior to the classification procedure we need to run a segmenter and obtain its output. The necessity to run two deep neural networks makes these models computationally expensive. 

In this study, we compare the ISNet to a segmentation-classification pipeline consisting of a U-Net (segmenter) followed by a classifier. The classification model is the same one used in the ISNet: VGG-19 in dog breed classification, and DenseNet121 in all other Tasks. The U-Net uses its original architecture proposal, which can be seen in Figure 1 of\cite{unet}. We trained the segmenter beforehand, using the same dataset that would be later used for classification, and employing the segmentation masks as targets. The same targets were used in the ISNet training. Analyzing the U-Net validation performance, we found the best threshold to binarize its outputs. For COVID-19 detection the optimal value was 0.4 when trained without synthetic bias, and 0.3 with it; for facial attribute estimation, 0.5 without synthetic bias, and 0.4 with it; for Stanford Dogs we used a threshold of 0.3.

The U-Net architecture\cite{unet} and its variants are a common model choice for biomedical image segmentation, including lung segmentation\cite{bassi2021covid19}\textsuperscript{,}\cite{covidSegmentation}\textsuperscript{,}\cite{TuberculosisUNet}. It was created with this type of task in mind and so is designed to obtain strong performances even in small datasets. The U-Net is a fully convolutional deep neural network with two parts. First, a contracting path, created with convolutions and pooling layers, captures the image's context information. Then, an expanding path uses transposed convolutions to enlarge the feature maps and perform precise localization. Skip connections between the paths allow the expanding path access to earlier high-resolution feature maps (before down-sampling), which carry accurate positional information. Its combination with the context information in the later feature maps results in the U-Net's ability to conduct precise image segmentation.

As is commonly done, the U-Net parameters are kept frozen when training the pipeline's classifier. Algorithm \ref{algoPipeline} summarizes the sequential model.

\begin{algorithm}
  \caption{Segmentation-classification Pipeline}
  \label{algoPipeline}
  \begin{algorithmic}[1]
    \State The U-Net processes the image (e.g., X-ray) and segments the foreground (e.g., lung region).

    \State We threshold the U-Net result and create a binary mask, where the foreground pixels are represented as 1 and the remaining image parts as 0.

    \State We use an element-wise multiplication between the image and the mask to erase the figure's background.
    
    \State The DenseNet121/VGG-19 classifies the segmented image. Therefore, the classifier will not have access to background features.
    
  \end{algorithmic}
\end{algorithm}

As flexible models, DNNs are capable of performing image classification and semantic segmentation at the same time, using a single network. Multi-task learning\cite{MultiTaskOriginal} allows the simultaneous optimization of multiple loss terms, normally by minimizing their linear combination. Accordingly, it is possible to combine a classification and a segmentation loss. Past studies concluded that such multi-task models can be advantageous for both the classification and the segmentation tasks and improve the analysis of noisy images\cite{MultiTask1}\textsuperscript{,}\cite{MultiTask2}. Moreover, the utilization of shared representations may promote similar attention foci for the two tasks\cite{MultiTask1}. However, we maintain that the utilization of this strategy does not guarantee a similar attention pattern for segmentation and classification, especially in the presence of background bias. Thus, we pose that using the multi-task DNN will not reliably prevent shortcut learning.

As another baseline, we created a multi-task model. It consists of a U-Net with an additional classification head, connected to the U-Net bottleneck feature representation (i.e., the end of its contracting path). The model is trained with multi-task learning for the tasks of classification and semantic segmentation. Past studies have shown that such models are useful when analyzing images with cluttered backgrounds, because one task benefits the other, allowing the DNN to achieve better results on both\cite{MultiTask1}\textsuperscript{,}\cite{MultiTask2}. As the U-Net contracting path is based on the VGG architecture\cite{vggOriginal}, we employed a classification head that is similar to the VGG last layers. It comprises an adaptive average pooling layer, with output size of 7x7 and 1024 channels, leading to a sequence of 3 dense layers. The first two have 4096 neurons and ReLU activation. The output layer has one neuron accounting for each possible class, and it uses a softmax or sigmoid activation function (in multi-class or multi-label problems, respectively). To minimize overfitting, the first two dense layers are followed by dropouts of 50\%. Moreover, to accelerate and stabilize the training procedure, we added batch normalization after each convolutional layer in the U-Net, before their ReLU activations. We experimented with different classification heads (e.g., two convolutional layers followed by global average pooling and a dense output layer), but we found no significant advantage in relation to the chosen structure. We trained the multi-task U-Net to perform segmentation and classification simultaneously, by minimizing a linear combination of a classification loss and a semantic segmentation loss (both being cross-entropy functions). The first error function compares the classification head outputs with classification labels. Meanwhile, the second compares the U-Net expanding path outputs to segmentation masks. Therefore, the trained multi-task U-Net produces classification scores with its classification head, and segmentation outputs with the expanding path.

As the multi-task model, the ISNet is trained with a linear combination of two loss functions. However, unlike a standard multi-task DNN, which aims to solve a segmentation and a classification task, both ISNet loss functions have the same objective: to improve classification performance. Besides having a different objective, the ISNet also presents drastic methodological differences in relation to multi-task learning for simultaneous classification and semantic segmentation. The second loss function in the multi-task network optimizes a standard DNN output, created by a dedicated neural network branch. It makes this output match segmentation targets, solving a semantic segmentation task. This optimization procedure does not restrain the model's classifier attention or explanation, as our experiments prove. Thus, the methodology cannot be regarded as an attention mechanism, nor can it reliably minimize the influence of background bias over the classification scores. Meanwhile, the ISNet creates no segmentation output. Unlike the multi-task model, it does not have a dedicated branch for producing standard semantic segmentation masks. The LRP block is a temporary structure, without independent parameters, which uses LRP to explain the classifier's decisions with heatmaps. The second loss term for the ISNet takes as input the classifier's LRP heatmap, and the cost function minimization directly constrains the classifier's attention pattern, giving rise to a classifier spatial attention mechanism. Therefore, unlike a multi-task DNN, the ISNet training procedure optimizes classifier explanations (and attention) instead of a standard DNN segmentation output.

Besides the ISNet, four benchmark DNNs in this study control classifier attention by optimizing its explanations, with the aid of foreground masks: GAIN, HAM, RRR and the ISNet Grad*Input ablation. They are important to empirically demonstrate the advantages of LRP optimization (ISNet). In the models, we employ the same classifier backbones as in the ISNet, i.e., VGG-19 for dog breed classification, and DenseNet121 for the remaining tasks. Guided Attention Inference Networks\cite{GAIN} (GAIN) train classifiers optimizing their Grad-CAM\cite{GAIN} (a simpler explanation technique) heatmaps. The main objective of the DNN is to produce Grad-CAM heatmaps that will serve as priors for training weakly-supervised semantic segmenters (i.e., segmenters trained on datasets with little to no available ground-truth segmentation targets)\cite{SEC}. A classifier's heatmaps may serve as localization cues for training such models. Standard classifiers normally do not pay attention to all parts of the object of interest, creating heatmaps that partially highlight the image areas related to the classes. To produce more complete Grad-CAM explanations, GAIN classifiers are trained by minimizing a linear combination of a standard classification loss, the attention mining loss, and, optionally, the external supervision loss\cite{GAIN}. The last two functions are based on the optimization of Grad-CAM heatmaps, the model using the three terms can also be called extended GAIN. The attention mining's objective is to make the DNN pay attention to all image regions correlated with the sample's classes. Meanwhile, the external supervision loss optimizes the Grad-CAM heatmaps to match the available ground-truth segmentation targets. Although GAIN's main purpose is the creation of priors for weakly-supervised semantic segmentation, its authors claim that, as an additional benefit, the GAIN classifiers are more robust to background bias. Supporting this statement, they trained DNNs on databases with background bias, and showed that GAIN's o.o.d. test performance surpassed a standard classifier's, especially when using the three loss functions\cite{GAIN}.

Basically, GAIN uses Grad-CAM to locate the classifier attention for a given class (present in the image), then it removes the high attention areas from the figure (masking) and classifies it again. The class prediction score the DNN gives for the masked image constitutes the attention mining loss. When the image presents multiple ground-truth classes, multiple masked images are created (running Grad-CAM for the different classes) and classified. In this case, the attention mining loss is the arithmetic mean of the prediction scores for all present classes. If Grad-CAM attention covered all image regions related to the present classes, the DNN prediction scores for the masked images would be low, resulting in low attention mining loss. Thus, the minimization of the loss function extends the attention in Grad-CAM. The external supervision loss is the mean squared loss between the Grad-CAM heatmap for a present class (interpolated to the DNN input size) and the ground-truth foreground segmentation target for the class. When multiple classes are present in one image, the losses for the diverse Grad-CAMs and targets are averaged.

Our GAIN implementation was based on a PyTorch version of the network\cite{GAINTorch}, which we modified to include the external supervision loss, and to also deal with multi-class single-label problems. Our modifications followed the guidelines in the paper presenting GAIN\cite{GAIN}. Grad-CAMs considered the DenseNet's feature map just before global average pooling, matching the original GAIN paper's choice of using the output of the classifier's last convolutional layer for Grad-CAM. Accordingly, in the VGG-19 Grad-CAM utilizes the last convolutional layer's output. This is the standard choice for producing Grad-CAM explanations in DenseNets and VGG. Classification loss was cross-entropy for the single-label problems, and binary cross-entropy for multi-label tasks. A second Grad-CAM-based attention mechanism, Hierarchical Attention Mining (HAM)\cite{HAM}, is considered as a baseline in Appendix \ref{CheXpertClassification}.

LRP and Grad-CAM are fundamentally diverse techniques: Grad-CAM is based on a linear combination of a DNN layer's output feature maps, while LRP back-propagates a signal through the entire DNN. We hypothesize that Grad-CAM contains fundamental weaknesses, which make it difficult for attention mechanisms optimizing Grad-CAM heatmaps to securely avoid background bias attention and shortcut learning. We analyze these weaknesses in Appendix \ref{GAINComparison}. As a more robust and theoretically founded explanation technique\cite{LRP}\textsuperscript{,}\cite{LRPRobustness} (main article section ``ISNet theoretical fundamentals''), LRP does not have these flaws. Therefore, it more reliably identifies background bias, making it a more trustworthy foundation for attention mechanisms that hinder shortcut learning. To support this claim, we compare the ISNet to the extended GAIN (employing segmentation targets for all training samples) and prove our proposed LRP-based attention mechanism is more reliable, and more effective in reducing attention to background bias, thus improving o.o.d. test performances.

The network from the study titled Right for the Right Reasons\cite{RRR}, which we will refer to as RRR, optimizes input gradients to avoid a classifier's attention to its input background. Input gradients (gradients of the DNN outputs, with respect to its input), or saliency maps, are a simple and efficient explanation technique. In RRR, the authors use a special loss function term, dubbed right reasons, to minimize the input gradient background elements. In essence, the loss masks out the gradient foreground elements (identified by ground-truth foreground masks), and follows by applying a square loss, with targets zero. Utilizing a multilayer perceptron with two hidden layers as classifier, RRR was very successful in avoiding background attention and shortcut learning produced by background bias\cite{RRR}. However, input gradients (saliency maps) are known to be very noisy when considering deep classifier architectures\cite{LRPvsGrad}\textsuperscript{,}\cite{LRP}. Thus, here we test RRR with deep backbones (VGG-19 and DenseNet121) and large images (224x224) in multiple experiments, comparing it to the ISNet. To justify the observed empirical results, we theoretically analyze why LRP-$\varepsilon$ produces more coherent and less noisy explanations of deep networks, thus improving the ISNet optimization convergence, and leading to superior resistance to background attention and better overall accuracy (main article section ``ISNet theoretical fundamentals'').

The element-wise multiplication of input gradients and the input itself is said to be an improvement over simple input gradients, increasing sharpness\cite{GradInput}\textsuperscript{,}\cite{LRPvsGrad}. These heatmaps are known as Gradient*Input, and we use them for an ISNet ablation study. The ISNet Grad*Input is an ISNet, where the LRP heatmaps are substituted by Gradient*Input. Therefore, the ISNet Grad*Input uses the ISNet loss function, which is much more restrictive than the RRR loss, thus changing the optimization dynamics. In summary, RRR's right reasons loss can be reduced by an overall reduction in the magnitude of all the heatmap's elements. Meanwhile, due to its relative nature, the ISNet loss is only small when the background attention is small in relation to the foreground attention. Moreover, the loss foreground component avoids an overall change in the heatmap scale (Appendix \ref{loss}). The ISNet Grad*Input represents a controlled experiment, where only the optimized explanation is changed. It should indicate if the optimization of LRP is advantageous over that of Gradient*Input. For an in-depth comparison of LRP, input gradients and Gradient*Input, please refer to main article section ``ISNet theoretical fundamentals'' and Appendix \ref{lrpVsGradInput}.

\subsection{Analysis of the Benchmark Models' Results and Comparison to the ISNet}
\label{baselineComparisons}

\subsubsection{Experiments with Synthetic Background Bias}
Here, we analyze the behavior of the multiple alternative benchmark models, and explain their disadvantages in relation to the ISNet, considering the quantitative results in main article Table 1. First, the standard segmentation-classification pipeline was able to efficiently avoid attention to the synthetic background bias, and hinder shortcut learning. In COVID-19 detection and facial attribute estimation, we observe no F1-Score drop when comparing the biased dataset evaluation to the unbiased and deceiving tests (main article Table 1). A tiny maF1 reduction can be observed when the synthetic bias is substituted by confounding bias in the Stanford Dogs test dataset. The U-Net in the pipeline produced a test intersection over union (IoU) of 0.61 with the ground-truth segmentation targets in Stanford Dogs, 0.875 in COVID-19 detection, and 0.924 in facial attribute estimation. The worse segmentation performance in Stanford Dogs may indicate that the U-Net could not perfectly remove bias from all images in the task, explaining the small accuracy drop in main article Table 1. Overall, the pipeline's high resistance to shortcut learning is expected. Since the geometrical shapes do not affect the U-Net training procedure, their addition to the images have negligible effect over the segmenter's test performance. Indeed, when trained without the synthetic bias for lung segmentation (in COVID-19 detection) or face segmentation, the U-Net achieved 0.893 and 0.925 IoU, respectively. The results are remarkably similar to what was observed with the artificial bias. Thus, assuming that the segmenter is able to properly erase the background bias, the classifier receives unbiased images. Like the ISNet, the pipeline's results for COVID-19 detection and facial attribute estimation with synthetic bias (main article Table 1) is very similar to the model's performance when trained without the artificial shapes: 0.645 +/-0.009 maF1 in COVID-19 detection, and 0.806 +/-0.027 maF1 in facial attribute estimation. The ISNet sightly surpassed the pipeline in facial attribute estimation and Stanford Dogs (with confidence interval overlap), and it significantly surpassed the pipeline in COVID-19 detection. The ISNet's unexpected advantage over the pipeline in COVID-19 detection will be analyzed in detail in Appendix \ref{heatmapAnalysis}. Due to the combination of a dedicated segmenter and a classifier, two large deep neural networks, the run-time alternative pipeline was much slower and more memory consuming than the ISNet (Appendix \ref{speed}).

The ISNet Grad*Input represents an ablation experiment, where the ISNet's Layer-wise Relevance Propagation heatmaps were substituted by simpler explanations, Gradient*Input\cite{GradInput}. In Stanford Dogs, the model presented a tiny maF1 drop when the geometrical shapes were removed or substituted by deceiving bias. It performed well on the task, minimizing shortcut learning and practically matching the original ISNet. However, in the two other tasks, we could not get the models' losses to properly converge. Even though we tried multiple hyper-parameter configurations (changing momentum, learning rate, and loss hyper-parameters), the ISNet Grad*Input heatmap loss always remained high and unstable. Consequently, resistance to bias was not guaranteed (main article Table 1), the model's maF1 was the worst in the facial attribute estimation standard and biased tests, and it was overshadowed by the original ISNet's in COVID-19 detection. The reason for the low performance and unsatisfactory loss convergence may lie on the different classification backbones utilized in the three experiments: VGG-19 in Stanford Dogs, and DenseNet121 in the other tasks. Indeed, in preliminary tests using a DenseNet121 as backbone in Stanford Dogs, we could not make the ISNet Grad*Input heatmap loss properly converge. Input gradients are known to be highly noisy for deep models and are normally used to explain only simpler neural networks\cite{LRPvsGrad}. Therefore, noisy Gradient*Input heatmaps in the very deep DenseNet121 may prevent the proper minimization of the explanations' background attention. This is a substantial drawback of optimizing Gradient*Input instead of LRP, as very deep classifier architectures are required for state-of-the-art performance in multiple computer vision applications. For example, DenseNet121 is among the best performing models for disease classification in chest X-rays\cite{chexnet}. As we could not obtain proper loss convergence with the deep backbone, we did not implement the ISNet Grad*Input in the remaining applications in this study. For an in-depth theoretical analysis, and comparison between the optimization of LRP and Gradient*Input, please refer to main article section ``ISNet theoretical fundamentals'' and Appendix \ref{lrpVsGradInput}.

Instead of optimizing the multiplication of the input and the input gradient (Gradient*Input), the RRR neural network (Right for the Right Reasons) optimizes just the input gradient\cite{RRR}. The model showed more resistance to shortcut learning than the standard classifier, with smaller maF1 reduction across the three test settings (the biased, standard, and deceiving). However, the geometrical shapes could affect the network's decisions and cause shortcut learning, as shown by the performance drops across the columns in main article Table 1. Thus, in relation to the ISNet (which displays no performance drop), RRR was less resistant to the effect of background bias. There are a few reasons for the ISNet surpassing RRR. First, like Gradient*Input, input gradients are nosier than the LRP heatmaps that the ISNet optimizes when considering deep classifiers. Second, from an optimization standpoint, input gradients are similar to Gradient*Input. In the Deep Taylor Decomposition framework, LRP-$\varepsilon$ represents a more faithful decomposition of the network output, and a more accurate representation of its behavior in relation to Gradient*Input. Please refer to main article section ``ISNet theoretical fundamentals'' and Appendix \ref{lrpVsGradInput} for a detailed analysis of the advantages of LRP optimization. Finally, the ISNet heatmap loss formulation is more restrictive than the RRR loss (Appendix \ref{loss}), better ensuring that reduced loss values correspond to the smaller influence of background features on the classifier decisions.

Although Gradient*Input explanations are similar to input gradients\cite{LRPvsGrad}, main article Table 1 displays differences between RRR and the ISNet Grad*Input. A main reason for the discrepancies may be the diverse loss functions that the two models utilize. The RRR heatmap loss (dubbed right reasons\cite{RRR}) is, in essence, a square loss comparing the input gradients' background region and zero targets. Meanwhile, the ISNet Grad*Input employs the restrictive ISNet heatmap loss. The RRR right reasons loss just requires the explanations' background components to be small (in magnitude). Conversely, the ISNet loss is relative, demanding the background components to be much smaller, in magnitude, than the foreground elements. Moreover, it has an additional term, which forces the sum of the absolute values in the heatmap foreground to stay within a pre-defined range. Therefore, the RRR loss can drop due to an overall reduction of the input gradients, affecting both their foreground and background elements. However, the relative nature of the ISNet heatmap loss makes the classifier pay progressively more attention to the images' foreground, in relation to the background. Meanwhile, the additional loss term keeps the heatmaps' average value stable. Empirically, we observed that, when increasing the weight of RRR's right reasons loss ($\lambda_{1}$ parameter\cite{RRR}), the loss value decreased, but this effect was not necessarily accompanied by a significant reduction in shortcut learning. Indeed, we could not get the same maF1 in the three testing scenarios (the biased, standard, and deceiving), unless $\lambda_{1}$ was set so high that the DNN essentially resorted to random guessing (we tested $\lambda_{1}$ values up to  $10^{20}$). On the other hand, with the VGG-19 backbone, the ISNet Grad*Input was able to mostly avoid the accuracy drop that characterizes shortcut learning. However, unlike RRR, its could not properly converge with the DenseNet121 backbone. Consequently, we hypothesize that the utilization of a less restrictive loss allowed RRR to converge more easily, considering the noisiness of both input gradients and Gradient*Input. Notice that the ISNet loss had no difficulty converging when LRP, a less noisy explanation (Appendix \ref{lrpVsGradInput}), was optimized in the original ISNet.

Grad-CAM\cite{GradCAM} is possibly the most popular explanation technique in computer vision, known to be simple and easily interpretable\cite{GradCAM}. The Extended GAIN\cite{GAIN} is the benchmark neural network that optimizes such explanations. As RRR, GAIN was more resistant to background bias than the baseline classifiers. However, it could not match the ISNet capacity of avoiding background attention: the geometrical shapes influenced the DNN's decisions, as shown by its accuracy drop across the columns in main article Table 1. We tried increasing the weights of GAIN's attention mining and/or external supervision losses\cite{GAIN},
up to an external supervision weight (w) of 10000. Even such settings could not effectively reduce GAIN’s maF1 gap. Unlike input gradients and Gradient*Input, Grad-CAM is not a very noisy explanation for large DNNs. However, we discovered that the optimization of Grad-CAM commonly leads to a phenomenon we named spurious mapping. It is defined by a classifier learning to produce Grad-CAM explanations that display no attention to the background, minimizing Grad-CAM-based losses, while the model's decisions do rely on background bias, and shortcut learning is evident. I.e., spurious mapping is when the deep neural network learns to produce unreliable Grad-CAM heatmaps, which hide its background attention. The phenomenon was verified for the three GAIN models in main article Table 1. The possibility of spurious mapping is intrinsic to the Grad-CAM formulation, which is fundamentally diverse from LRP. Please refer to Appendix \ref{GAINComparison} for an in-depth analysis of the issue. Examples of the spurious Grad-CAM heatmaps are available in Figure \ref{spurious}. In summary, the propensity for spurious mapping undermines the capacity of Grad-CAM optimization to minimize attention to background bias, or to hinder shortcut learning, leading to it being surpassed by the ISNet in main article Table 1.

The multi-task U-Net could not prevent shortcut learning in any of the experiments in main article Table 1, as shown by a significant maF1 drop when comparing the biased evaluation to the standard and deceiving tests. Such as the multi-task U-Net, most multi-task models initiate as a sequence of shared layers, which later forks, creating two or more branches. One path can produce segmentation outputs while the other generates classification scores. However, a deep neural network is a very flexible model, and the shared layers can learn a representation that contains enough information to satisfactorily optimize both tasks. The model's objective is to minimize both the classification and the segmentation losses. The segmentation-exclusive path receives the shared layers' output and learns decision rules that ignore information related to the background bias, which is useless for improving segmentation performance. The classifier branch has the same input as the segmentation one, but it focuses on background bias, because it improves the minimization of the training classification loss. Even multi-task architectures that promote interactions between the segmentation and classification branches\cite{MultiTask1} allow them to process task-specific features, permitting diverse foci. We observed that the multi-task model achieved a good segmentation performance even when the images had the artificial background bias, on which the classification decisions strongly focused, as shown in main article Figures 1 and 2. Indeed, the presence of the geometrical shapes on the background did not influence the segmentation performance. In COVID-19 detection, the DNN trained and tested on the original dataset (without geometrical shapes) achieved 0.893 IoU, the one trained with the artificially biased dataset and tested on the original test database, 0.887, and the DNN trained and tested with the artificially biased images, 0.886. In facial attribute estimation, the IoU values in these three scenarios were 0.924, 0.926 and 0.926, respectively. Thus, we observe that the segmenter in a multi-task model can find the region of interest regardless of background bias, ignoring it. On the other hand, significant classification performance reductions when the synthetic bias was removed or substituted by deceiving bias (main article Table 1) quantitatively prove that the segmenter strongly focuses on the image's backgrounds. We did not find a correlation between good segmentation and lack of background attention for classification, as the tasks' foci diverged due to background classification bias.

Not being able to effectively constrain the classifier attention to the region of interest, a multi-task model is not able to replace an ISNet for the purpose hindering background bias attention and avoiding shortcut learning. A possible strategy to use a standard multi-task model (trained for classification and semantic segmentation) to avoid shortcut learning would include running it twice. Firstly, one can run the DNN, store its segmentation output, and use it to find and remove the image background (creating a segmented image). Afterwards, one would need to rerun the DNN, now analyzing the segmented input image, and obtain the classification output. This is similar to the strategy employed in the alternative segmentation-classification pipeline but considering that the classifier and segmenter would be parts of one multi-task model (e.g., multi-task U-Net). Since the U-Net has more parameters than the DenseNet121, this strategy shows only a small advantage in memory consumption. As the model needs to run twice, it does not present a compelling speed improvement. For these reasons, we did not find the strategy much beneficial in relation to the state-of-the-art, nor worth implementing.

One may argue that it is possible to use a multi-task model segmentation output to selectively weigh the feature map produced by the DNN shared layers block, reducing background activations. This strategy would represent a time economy in relation to the previous one, as we would need to run the model shared layers just once, then sequentially run the segmentation branch (getting its output), and the classification one. However, modern DNNs have an exceptionally large receptive field in later convolutional layers. For example, the VGG-16 has a receptive field length of 212 pixels in its last convolution, while deeper models (e.g., ResNet101) can have more than 1000, effectively the whole input image\cite{ReceptiveField}. Consequently, using a segmentation mask to gate the shared layers' last feature map would not be effective to mitigate shortcut learning. Due to the large receptive field, a non-suppressed feature map element would be able to carry information from the background. Moreover, reducing this receptive field means strongly limiting the number of shared layers, or gating an earlier shared layer (making it necessary to run twice all shared layers after the gated feature map). Thus, both strategies negate the advantages of a multi-task model.

All neural networks previously discussed in this Section, except for the standard classifier, rely on semantic segmentation targets to optimize a classifier attention pattern. Conversely, the last two benchmark models, AG-Sononet and Vision Transformer, do not learn from semantic segmentation masks. Their attention profile is only guided by the classification loss optimization. Attention to background bias can help the DNN classify the training samples, reducing training classification loss. Without the information contained in segmentation targets, the classifier cannot effectively distinguish between background bias and foreground features, because both are correlated to classification labels. Consequently, the classifier may deem any of them, or both, relevant.

The AG-Sononet and most spatial attention mechanisms are designed to focus on relevant image features and learn to ignore irrelevant regions \cite{AGNet}. The background geometrical shapes in our experiments represent an easy to learn feature, which is correlated with the image classes. For this reason, the AG-Sononet attention mechanism learns to consider the background bias a relevant region and focuses on it. Accordingly, the model consistently showed the strongest discrepancy between results on the biased, standard, and deceiving evaluations, for all experiments in main article Table 1. Indeed, the maF1 drop was larger than what we observed for the standard classifier, showing a stronger focus on the geometrical shapes, in detriment of the foreground features (faces, lungs, or dogs). We observe that, although useful for dealing with background clutter\cite{AGNet} (irrelevant background features, uncorrelated with the images' classes), attention gated networks are not adequate for avoiding shortcut learning in datasets presenting background bias. Appendix \ref{heatmapAnalysis} will provide further evidence for this conclusion.

Finally, we analyze the Vision Transformer\cite{VisionTransformer}. Unlike all other trained models, it is not a convolutional neural network. It relies on the transformer, a structure that produced state-of-the-art performance in natural language processing tasks, which is built upon the concept of self-attention\cite{transformer}. Self-attention mechanisms relate diverse positions in a sequence (its input), to produce a new representation of the sequence\cite{transformer}. In relation to the standard classifier (CNN), the Vision Transformer showed smaller maF1 gaps when comparing the three testing scenarios, indicating less shortcut learning. The model significantly surpassed the standard classifier in all deceiving bias tests, and on the standard test for COVID-19 detection. Such results indicate a more diffuse attention profile: although the Vision Transformer considers the geometrical shapes in its decisions, their influence does not overpower the impact of the foreground features as much as we observed in the standard classifier. However, not learning from segmentation targets, the Vision Transformer cannot effectively comprehend that it should not consider the geometrical shapes. For this reason, the network's capacity of avoiding background attention and shortcut learning lagged significantly behind the ISNet's, as demonstrated by all main article Table 1 experiments.

\subsubsection{COVID-19 Detection}

Comparing the external test metrics' macro-averages in main article Tables 2 and 3 (COVID-19 detection without synthetic background bias) for the five models containing DenseNet121s (ISNet, DenseNet121, segmentation-classification pipeline, Extended GAIN and RRR), we observe that the ISNet obtained the highest results, followed by the alternative segmentation-classification pipeline, RRR, the model without segmentation, and finally GAIN. The multi-task U-Net and the AG-Sononet, two models based on the VGG-16 architecture, presented similar and poor metrics in main article Tables 2 and 3, being strongly surpassed by the standalone DenseNet121. During training, the two models converged, with training and validation losses similar to the DenseNet121-based DNNs. However, their poor evaluation performances reveal a failure in generalizing to the o.o.d. test dataset. The same can be said for the Vision Transformer, whose performance could not surpass the standard DenseNet121.

We confirmed that a standard classifier (DenseNet121 analyzing whole images) failed to properly generalize to an external, o.o.d. COVID-19 detection dataset (images from hospitals not seen in training). It achieved only 0.546 +/-0.01 macro-average F1-Score, and its heatmaps revealed strong background attention (main article Figure 2). The results are consistent with shortcut learning, a known tendency for mixed COVID-19 datasets\cite{NatureCovidBias}\textsuperscript{,}\cite{ShortcutCovid}. Besides the ISNet and the segmentation-classification pipeline, main article Tables 2 and 3 shows that none of the other benchmark models' average performance scores could significantly surpass the standard classifier (DenseNet121). To overcome the standard DNN's o.o.d. test performance, the benchmark models should be able to efficiently hinder the shortcut learning while not losing overall accuracy in the process. The quantitative results in the table indicate a failure to achieve this goal. This finding corroborates the results from the synthetic bias experiments, which quantitatively showed that the ISNet and the pipeline were the models that could best avoid the influence of background bias on the classification outputs. Thus, the results from the experiment with non-synthetic background bias support the analysis in Appendix \ref{baselineComparisons}.

The U-Net in the alternative segmentation-classification pipeline achieved a test intersection over union (IoU) of 0.893 segmenting the lungs, and the resulting segmentation masks seemed adequate upon visual inspection. However, the ISNet could considerably surpass the alternative segmentation-classification pipeline's classification performance in COVID-19 detection. This result will be analyzed in Appendix \ref{heatmapAnalysis}. 

\subsubsection{Tuberculosis Detection}

Regarding lung segmentation performances, the multi-task U-Net had IoU of 0.941 on the o.o.d. tuberculosis test dataset, and the segmenter inside the segmentation-classification pipeline 0.82, according to the automatically generated segmentation masks that (Appendix \ref{dataset}). Considering the o.o.d. dataset's manually created segmentation targets\cite{ChineseDataset1}\textsuperscript{,}\cite{ShenMasks}, the two DNNs achieved 0.904 and 0.908 IoU, respectively. The produced segmentation outputs seemed adequate upon visual inspection.

In tuberculosis (TB) detection (main article Table 5), we applied a threshold of 0.3 to the model's U-Net outputs. This value was chosen because it maximized the segmenter's validation IoU (considering that the training and validation datasets are i.i.d.). However, a larger threshold, 0.5, would optimize the pipeline's o.o.d. evaluation result: it would achieve 0.631 +/-0.048 maF1, becoming again one of the best generalizing DNNs (main article Table 5). In this task the segmentation-classification pipeline is sensible to changes in the U-Net output threshold, which is difficult to optimize without access to the o.o.d. test database. The same effect was not seen with the other applications in this study: they showed no significant maF1 benefit by optimizing the segmentation threshold beyond the value obtained with validation. Interestingly, even in TB-detection the threshold change from 0.3 to 0.5 did not improve IoU with the o.o.d. dataset's manual segmentation masks (it slightly decreased, from 0.904 to 0.896). As a final note, in preliminary tests training with unthresholded U-Net outputs, the pipeline was not able to ignore the background.

In the tuberculosis experiment, RRR was more promising than in COVID-19 detection, as its maF1 and AUC significantly surpassed the baseline DenseNet121 (main article Table 5). Moreover, the model showed a reduced gap between the i.i.d. and o.o.d. performance, a result consistent with a reduction in shortcut learning. However, the RRR's o.o.d. performance could not match the ISNet, possibly indicating that the network traded overall accuracy (o.o.d. and i.i.d.) for improved resistance to shortcut learning. Indeed, we needed to set the right reasons loss weight\cite{RRR} to a remarkably high value, $\lambda_{1}=10^{8}$, to maximize RRR's o.o.d. performance. We suspect the possible overall accuracy reduction to be caused by the noisiness of the input gradients the model optimizes\cite{LRPRobustness}, and by the fact that, in relation to LRP, input gradients are related to a less accurate decomposition of the classifier's outputs (refer to main article section ``ISNet theoretical fundamentals''). These disadvantages could have undermined RRR's joint minimization of the classification loss and the right reasons loss.

The remaining benchmark DNNs (GAIN, Vision Transformer, multi-task U-Net and AG-Sononet) displayed unimpressive generalization capability, with mean maF1 scores in the 50\% to early 60\% range (main article Table 5). Meanwhile, they had strong results on the i.i.d. test dataset. The discrepancy between performances on the i.i.d and o.o.d. test databases is evidence that the classification problem is prone to shortcut learning. As in COVID-19 detection, multi-task learning, attention gates, the extended GAIN and the Vision Transformer did not strongly improve generalization; the four model's maF1 95\% confidence intervals overlap with the DenseNet121's. Again, this result agrees with the findings from the synthetic bias experiments, and it corroborates with the discussion about the benchmark models' weaknesses, presented in Appendix \ref{baselineComparisons}.

\subsection{Heatmap Analysis}
\label{heatmapAnalysis}

Besides being necessary for the calculation of the heatmap loss, the heatmaps created by our LRP block can be analyzed, improving the explainability of our models. In this section we compare heatmaps for the different DNN architectures, evaluating their attention to background bias. Please refer to Appendix \ref{radiologistAnalysis} for a comparison between ISNet X-ray heatmaps and lung lesions segmented by a radiologist. Main article Figure 1 shows heatmaps for the artificially biased test datasets, according to the implemented DNNs, all trained with databases containing white geometrical shapes correlated with the image's labels in the image corners. The samples displayed in the figure are from the biased evaluation scenario, thus containing the same background bias that was present during training. In the LRP heatmaps, red (positive relevance) indicates areas more associated with the image true class (indicated on the top of the figure), and blue (negative relevance) are regions that reduced the DNN confidence for the class (e.g., areas more associated to the other possible classes). For better visualization, we normalized the LRP maps in the image, making the blue and red channels range between 0 and 255. The face photograph has positive labels for the three attributes. Thus, it has three geometrical shapes, the square indicates rosy cheeks, while the circle represents high cheekbones, and the triangle smiling. Other images come from single-label datasets. Thus, just one geometrical shape is present, which is correlated with the class stated above the images. The image last row displays heatmaps for a standard classifier trained and tested without the artificial bias, allowing the comparison of the ISNet attention pattern to the natural attention profile, which emerged without the influence of the geometrical shapes. Notice that the images' foreground (region of interest), defined in the ground-truth segmentation targets, are: the lungs, excluding the region overlapping the heart, the dogs' entire bodies, and the faces, excluding ears and hair, but including facial hair, eyes and mouth.

The Vision Transformer heatmaps in main article Figure 1 are not created by LRP, they are generated with attention rollout, the standard methodology to visualize the model's attention\cite{VisionTransformer}. It does not differentiate positive and negative relevances, nor is it class selective. Although LRP can be used with Vision Transformers, the method usually requires custom model implementations, while attention rollout worked directly with the official PyTorch Vision Transformer code, which we utilized. Moreover, attention rollout served the purpose of showing that the model has noticeable background and bias attention.

The ISNet clearly shows no attention to the geometrical shapes (main article Figure 1). This result is confirmed by the quantitative scores in main article Table 1, which demonstrate that the synthetic bias could not affect the ISNet's decisions; its removal from the test dataset, or even its substitution by confounding bias, could not reduce the ISNet's maF1 score. Moreover, even comparing the ISNet heatmaps with the explanations for a standard classifier trained without synthetic bias, we observe that the ISNet could better focus on the region of interest. In the dogs classification, positive relevance (red) can be clearly seen especially in the dogs' faces. In facial attribute estimation, for the smile class there is positive attention in the mouth region. Finally, in lung disease classification, a correlation between the ISNet attention and radiological findings is clear (refer to Appendix \ref{radiologistAnalysis}).

Heatmaps for the ISNet Grad*Input show some attention over the synthetic bias. Thus, the explanations and the quantitative results in main article Table 1 indicate that the model's capacity to avoid background attention was surpassed by the ISNet, showcasing the advantages of LRP optimization (refer to main article section ``ISNet theoretical fundamentals'' and Appendix \ref{lrpVsGradInput}). The segmentation-classification pipeline has no discernible attention over the synthetic bias, which could not influence its decisions (main article Table 1). All the remaining DNNs' heatmaps (standard classifier, multi-task U-Net, AG-Sononet, GAIN, RRR, and Vision Transformer) consistently reveal significant attention over the geometrical shapes (main article Figure 1). Thus, the LRP heatmaps explain the results in main article Table 1: these models' explanations indicate that the geometrical shapes influenced their decisions, and they had a significant accuracy drop when the synthetic bias was removed from the test dataset or substituted by deceiving bias.


Main article Figure 2 presents test images and the associated heatmaps for the classification tasks with non-synthetic background bias (caused by dataset mixing): COVID-19 detection and tuberculosis detection. All X-rays are from the o.o.d. test databases. High resolution versions of the X-rays and heatmaps can be found in the Supplementary Data 1\cite{SupData}. Figure \ref{Gmaps2} presents the ISNet Grad-CAM heatmaps for the X-rays in main article Figure 2. Grad-CAM was implemented with the library PyTorch Grad-CAM (1.4.6)\cite{GradCAMTorch}. The figure also presents the Grad-CAM explanations for the extended GAIN, which is based on the optimization of Grad-CAM heatmaps. Grad-CAMs for the synthetic bias experiments will be analyzed in Appendix \ref{GAINComparison}, as they motivate an in-depth comparison of LRP optimization and Grad-CAM optimization.

In both the LRP and Grad-CAM heatmaps, we observe that the ISNet successfully kept the classifier attention inside the foreground (the lungs), with little to no background focus. LRP produced explanations of much higher resolution. Moreover, some details visible with LRP were lost with Grad-CAM. For example, the lesions in the left lungs of the pneumonia and COVID-19 X-rays can only be seen with LRP. Such lesions match with radiologist findings (see Appendix \ref{radiologistAnalysis}). Some areas of negative relevance (blue) in the LRP heatmaps were highlighted by Grad-CAM. However, LRP red regions have a stronger correlation with the radiologist's annotations (available in Supplementary Data 1\cite{SupData}). As an illustration, the top tuberculosis X-ray in Figure \ref{Gmaps2} presents a lesion in the left lung's upper lobe, identified by the radiologist and the LRP heatmap (red), but not by Grad-CAM. Except for resolution, details, and class selectivity (telling apart positive and negative relevance), the ISNet's LRP and Grad-CAM explanations are consistent, with Grad-CAM focusing on regions of high relevance in the LRP maps. Overall, the heatmap analysis shows that the ISNet capability of avoiding background attention was not matched by any other DNN.

Except for the ISNet and the model combining a segmenter and a classifier, main article Figure 2 LRP heatmaps reveal significant background attention for all X-ray classifiers. X-ray classification with mixed databases is a task prone to shortcut learning\cite{ShortcutCovid}, as is confirmed by the amount of background attention seen in most heatmaps. The lack of background focus in the ISNet explanations confirms the model's superior capability of reducing shortcut learning and explains its better performances on the external (o.o.d.) X-ray databases (main article Tables 2 to 5).

A very visible example of background bias can be found in the pneumonia X-ray, as markings over the right shoulder. The ISNet and the segmentation-classification pipeline are the only DNNs that did not pay any attention to it. The same can be said for an R in the neck area of the left TB X-ray, and an L in the corner of the other tuberculosis X-ray (over the shoulder). In the pneumonia sample, we observe that the multi-task U-Net and the AG-Sononet were the models that focused the most on the bias. Accordingly, the two DNNs had the worst performances on the o.o.d. test database for pneumonia and COVID-19 detection (main article Tables 2 to 5). To create the multi-task model heatmaps, we considered LRP propagation from the classification outputs, since we aim to analyze the input features that influenced the classification decisions. Besides focusing on markings, we can observe that the X-ray classifiers (except for the ISNet and segmentation-classification pipeline) also paid attention to parts of the body outside the lungs. In the left extended GAIN LRP heatmap for tuberculosis we can even notice attention to the body contour. Such undesired sources of focus can also be seen as background bias and signs of shortcut learning.

The heatmaps lead us to the same conclusion we had when analyzing the multi-task DNN o.o.d. F1-Scores and its results in the synthetic bias experiments: a multi-task network learning segmentation and classification can show good segmentation outputs, while its classifier concentrates on background bias. The multi-task DNN heatmaps confirm that its classification decision rules strongly consider areas outside of the lungs, creating an attention pattern similar to the DenseNet121's. Therefore, multi-task classification and segmentation is ineffective for preventing focus on background bias and avoiding shortcut learning.

An attention gated DNN focuses on features that improve classification performance, and, since background bias reduces the classification training loss, the model learns to focus on it. In fact, in the main article Figure 2 X-ray heatmaps we observe that the AG-Sononet is the DNN that paid the most attention to background markings. It is noticeable that the DNN concentrates its attention on fewer image features (in relation to a standard classifier, like the DenseNet), but the selected features can be in the background and constitute bias. In conclusion, models that do not employ an attention mechanism based on semantic segmentation cannot reliably differentiate relevant image features from background features that are correlated with the images' classes. This conclusion is also valid for the Vision Transformer. The model has a more diffuse attention distribution, and its o.o.d. performance scores surpassed the AG-Sononet in COVID-19 and tuberculosis detection, but it could not overcome a standard classifier (DenseNet121). Indeed, the diffuse attention profile indicates that the Vision Transformer's decisions are influenced by both foreground features and by background bias, explaining why the model could not hinder shortcut learning (as demonstrated in main article Table 1).

The LRP heatmaps (main article Figures 1 and 2) for RRR show that the model could not effectively concentrate attention inside the regions of interest. Background bias relevance is visible, corroborating with the quantitative results in main article Table 1. This finding indicates that, in our experiments, the observed low values in the networks right reasons loss\cite{RRR} were not associated with a true focus on the images' foregrounds. We utilized deep networks and high image resolution, which increase the noise in input gradients. We suspect the noise to be detrimental for the achievement of true foreground focus during the training process. Thus, the minimization of the right reason loss can mostly reflect an overall reduction in the magnitude of the classifier input gradients, instead of a selective reduction of background attention. Please refer to main article section ``ISNet theoretical fundamentals'' for a more profound analysis.

In COVID-19 detection, there is significant relevance alongside the lungs' borders in the alternative segmentation-classification pipeline heatmaps. This effect is even more pronounced in the model's heatmaps for tuberculosis detection. It was not noticeable in the dogs classification task, as the dogs' segmenter cannot accurately follow the animal's contours. In TB classification we see attention in the background of the pipeline's heatmaps, as exemplified in main article Figure 2. Upon visual inspection, we verified that the model's segmentation outputs for the images were adequate, and the background was properly erased. Indeed, the model presents high IoU (over 0.9) even with manually created lung segmentation masks from the o.o.d. TB database. Thus, the classifier is paying attention to the dark (erased) background, as is confirmed by the homogeneous and undetailed aspect of this attention. The classifier inside the alternative segmentation-classification pipeline receives a segmented image, where the borders between the region of interest and the background are easily identifiable, as is the lungs' placement, size and shape. Therefore, the classifier may have learned to analyze the shape of the lungs' borders and the background format, using this information to distinguish between the classes in the training dataset (even though we employed translation and rotation as data augmentation). Being unrelated to the diseases, decision rules based on the aforementioned aspects shall hinder generalization capability, being considered another form of shortcut learning. This phenomenon may explain why the alternative pipeline performance in the o.o.d. COVID-19 and TB test datasets was worse than the ISNet's, which does not show significant border or background attention. In preliminary tests we applied different thresholding techniques to the pipeline segmenter's output, allowing a range of values and thresholding only quantities outside of this range. The methodology produced blurry lung borders, but it did not improve generalization performances.

\subsection{GAIN and Reliability Comparison Between LRP and Grad-CAM}
\label{GAINComparison}

\begin{figure}[!h]
     \centering
     \begin{subfigure}[b]{0.69\textwidth}
         \centering
         \includegraphics[width=\textwidth]{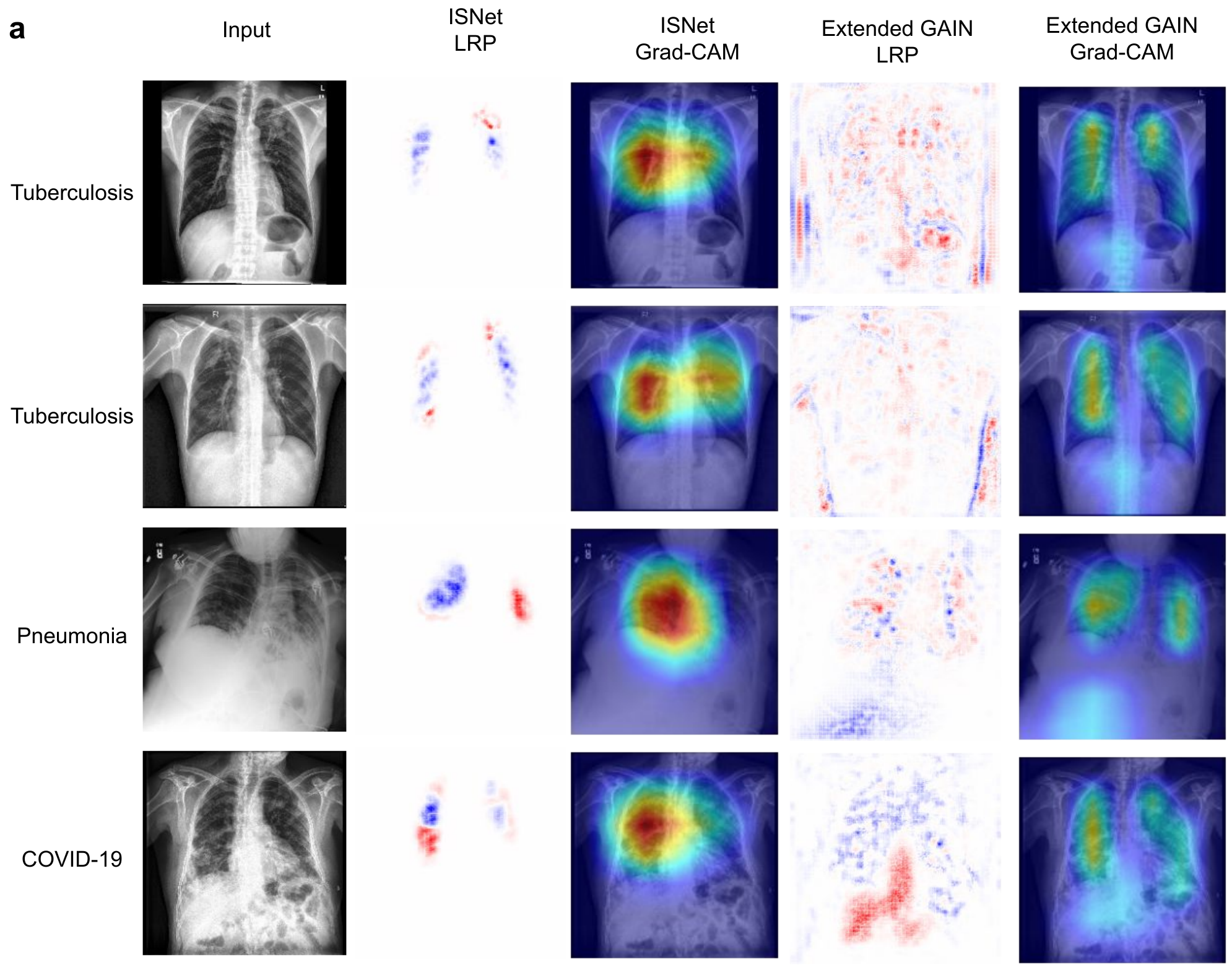}
         \caption{}
         \label{Gmaps2}
     \end{subfigure}
     \hfill
     \begin{subfigure}[b]{0.69\textwidth}
         \centering
         \includegraphics[width=\textwidth]{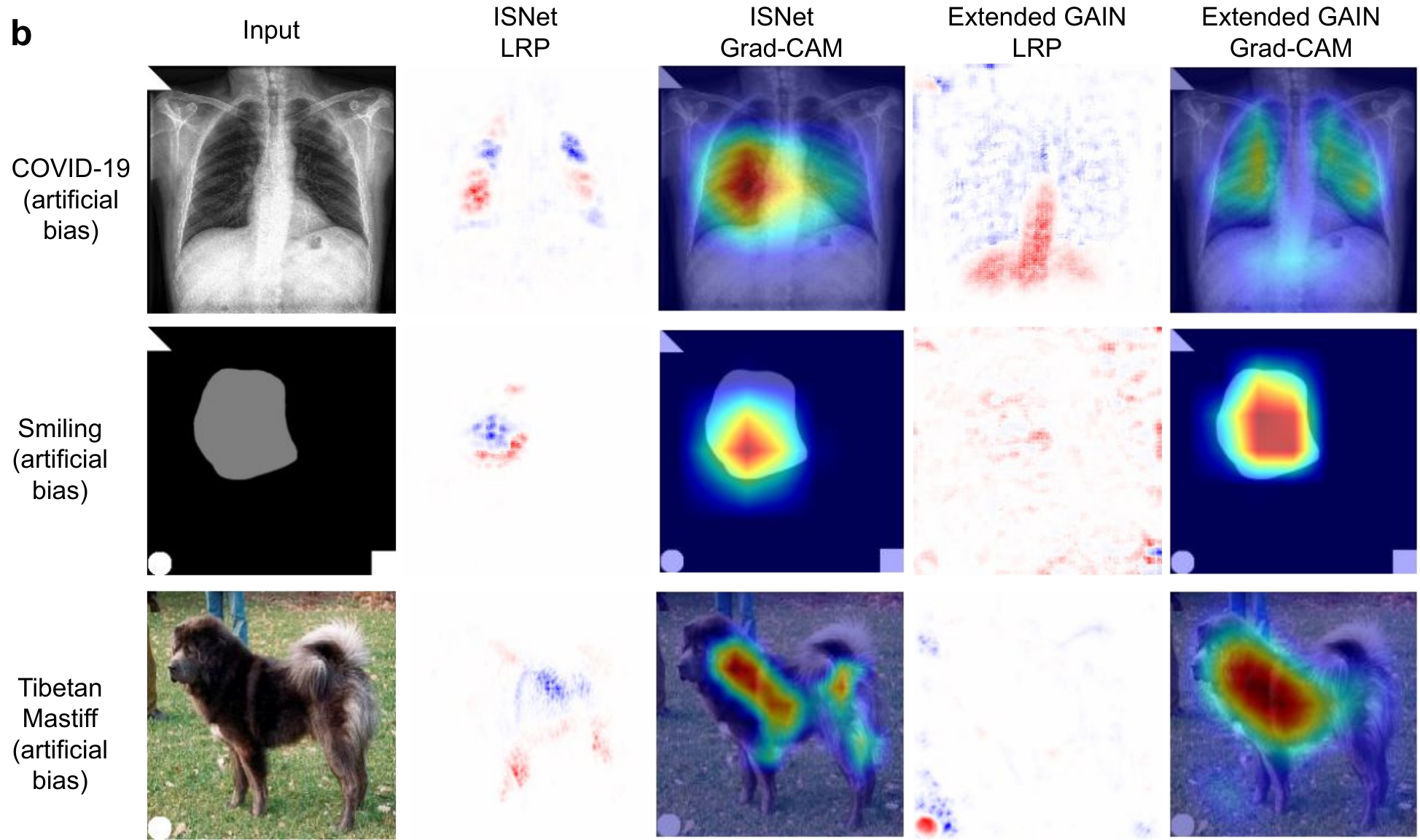}
         \caption{}
         \label{spurious}
     \end{subfigure}
        \caption{\textbf{LRP (Layer-wise Relevance Propagation) heatmaps and Grad-CAM (Gradient-weighted Class Activation Mapping) for X-rays and photographs.} Image classes presented on the left, heatmap type and classifier on the top. In the Grad-CAMs, the redder the area, the more attention was paid to it. Red colors in the LRP maps indicate areas the DNN associated to the ground truth classes. Blue colors in LRP heatmaps are areas that reduced the network confidence for the classes. White indicates little DNN attention. \textbf{a} Images from the COVID-19 and Tuberculosis detection datasets, without synthetic bias, and heatmaps from classifiers trained on data without synthetic bias. \textbf{b} Images extracted from the artificially biased test datasets, and classifiers trained on synthetically biased data. The triangle (background bias) indicates the presence of the classes COVID-19, smiling or Pug. The circle indicates pneumonia, high cheekbones and Tibetan Mastiff. For privacy, the face picture was substituted by a representation of the face (gray) and bias (white) locations, but classifiers received the real picture}
\label{Gmaps}
\end{figure}

In Figure \ref{Gmaps2}, the ISNet Grad-CAM explanations mostly focus on parts of the region of interest (lungs). This is an expected and acceptable behavior, as classifiers naturally focus more on the parts of the image foreground that they deem more relevant\cite{GAIN} (e.g., the mouth region when classifying a smile). Meanwhile, the GAIN Grad-CAM heatmaps successfully achieved the model's losses' main objective\cite{GAIN}: to highlight the region of interest (lungs) in a more complete manner. Unlike the ISNet, GAIN's main purpose is to generate Grad-CAM heatmaps that will be used to train weakly supervised semantic segmenters, which benefit from heatmaps highlighting the region of interest entirely\cite{GAIN}.

Besides the standard classification loss, extended GAIN utilizes the attention mining and the external supervision losses\cite{GAIN}. The first has the objective of expanding the DNN attention to all image regions related to the classes. The second should make the Grad-CAM explanations match the ground-truth segmentation masks. Analyzing the model's LRP heatmaps, we observe that it focuses both on the region of interest and on the background. For example, main article Figure 2 shows that the AG-Sononet learned to mostly focus on bias for X-ray classification, while GAIN pays attention to bias and to the lungs. Moreover, GAIN's o.o.d. test results for COVID-19 and TB detection could not significantly surpass the standard classifier (main article Tables 2 and 5). Such results indicate that the model also suffered strong shortcut learning, corroborating the background focus shown in its LRP heatmaps.

In the LRP explanations for the artificially biased images (exemplified in Figure \ref{spurious} and main article Figure 1), we observed an analogous situation. GAIN heatmaps indicate focus on the geometrical shapes, and on the region of interest. Once more, GAIN's performance scores confirm the validity of the LRP explanations: maF1 decreased when the synthetic bias was removed from testing or substituted by deceiving bias (main article Table 1). Overall, LRP heatmaps and numerical results point out that, although GAIN could reduce the concentration of attention on background bias in relation to a standard classifier, it could not effectively prevent the attention to background bias. Possibly, the observed distribution of attention across most of the image is a consequence of the attention mining loss, while the external supervision loss could not effectively hinder the background influence on the classifier's decisions.

Contrary to its LRP maps, GAIN's Grad-CAM heatmaps indicate little to no background attention, as can be seen in Figure \ref{Gmaps2}. For example, the markings in the pneumonia and tuberculosis X-rays can be seen in the GAIN LRP heatmaps, but not in Grad-CAM. Moreover, GAIN's Grad-CAM explanations show no attention over the artificial biases, again disagreeing with GAIN's LRP heatmaps, as clearly demonstrated in Figure \ref{spurious}. Therefore, unlike LRP, GAIN's Grad-CAM cannot explain the model's quantitative results, especially how the inclusion of the synthetic background bias improved F1-Scores, and how deceiving bias reduced it (main article Table 1). Accordingly, we conclude that the extended GAIN classifier learned to produce Grad-CAM heatmaps that hide background focus, inaccurately representing the DNN's attention, while the classifier's decision rules consider background features.

The GAIN performance scores and LRP heatmaps evidentiate that Grad-CAM has fundamental flaws, which allow the creation of heatmaps that do not reflect the neural network's real attention pattern. In essence and in its most common implementation (including GAIN's), Grad-CAM heatmaps are based on a linear combination of the different channels (feature maps) in the output of a chosen DNN layer (normally, the last convolution)\cite{GAIN}\textsuperscript{,}\cite{GradCAM}. The combination weight of a channel is given by the average value of the gradient in the channel, which is back-propagated from a DNN output neuron until the chosen layer. Afterwards, a ReLU function is applied to the result of the linear combination, removing its negative values. In the end, the resulting low-resolution heatmap is interpolated to the original image size. Below, we analyze two shortcomings of this procedure.

First, by deleting negative values, the ReLU operation can ignore negative evidence, i.e., feature map areas that reduced the DNN confidence for the class considered in the Grad-CAM heatmap. Negative evidence can cause classifier bias. As an illustration, if the class COVID-19 is highly correlated with a triangle, instead of increasing the COVID-19 prediction score when the geometrical shape is present, the classifier can reduce the confidence for the other classes when it sees a triangle. For this reason, Grad-CAM's ReLU function can make GAIN losses less effective in hindering background attention related to negative evidence. LRP indicates both the positively and the negatively relevant input features. To avoid both types of background attention, the ISNet's heatmap loss (Appendix \ref{loss}) must minimize positive and negative relevance. Moreover, to fully solve the problem of negative evidence, the ISNet minimizes background relevance in the heatmaps of all possible image classes. Otherwise, negative background relevance in the losing classes can produce bias (main article section ``Background relevance minimization and ISNet'').

This first Grad-CAM weakness could be alleviated by substituting the ReLU in Grad-CAM by an alternative function, such as the absolute value, and GAIN can be easily adapted to consider heatmaps for all classes in the external supervision loss. We experimented with these modifications, but had no success improving GAIN's o.o.d. generalization, and the DNN still paid attention to background bias. Such results point to a more fundamental problem in Grad-CAM. The technique's heatmaps are produced from a linear combination of a DNN layer's feature maps. Normally, the output from the last convolutional layer is chosen, as it contains a higher level of abstraction, better capturing the DNN behavior\cite{GradCAM}\textsuperscript{,}\cite{GAIN}. However, modern neural networks are very deep, and their latter convolutional layers have exceptionally large receptive fields, commonly covering the entire input. When layers' feature maps are interpolated to the input size, we expect their activations to align with the input features that caused such activations. This assumption is fundamentally necessary for Grad-CAM to produce faithful explanations, and it seems to normally hold for standard classifiers, and for DNNs that do not directly optimize Grad-CAM heatmaps. However, due to the large receptive fields, there is no guarantee that a DNN will not learn to map the input's background features to late layers' activations that align with the image foreground. For example, an activation in the corner of the first layer's feature map may cause an activation in the center of a feature map from the last convolutional layer. In such cases, Grad-CAM will not show background attention, but background features will be able to alter the outputs of the deep layers, influencing the classifier decisions. Thus, Grad-CAM heatmaps will become misleading explanations of the DNN behavior. We dubbed this phenomenon spurious mapping.

Spurious mapping does not seem common for classifiers trained to minimize a standard classification loss, nor for the ISNet, whose Grad-CAM explanations are consistent with LRP heatmaps (Figures \ref{Gmaps2} and \ref{spurious}). However, the GAIN classifiers may learn spurious mapping to satisfy the GAIN's Grad-CAM-based external supervision loss, while the hidden attention to background bias allows the DNN to better optimize the training classification loss. When comparing spurious mapping to the desired solution for the classification problem (a classifier with high accuracy and not influenced by the background bias), spurious mapping could represent a lower overall loss minimum. The dataset, task, initialization point, and training hyper-parameters may influence whether Grad-CAM optimization finds a spurious mapping solution or the desired solution. The experiments in this study indicate that spurious mapping is more easily achieved in many settings.

Unlike the problems caused by the ReLU function, the possibility of spurious mapping is intrinsic to the nature of deep neural networks. Furthermore, the fundamental definition of Grad-CAM allows such mapping strategies to generate unreliable heatmaps. Limiting receptive field to avoid spurious mapping would restrict the DNN design, limiting its capacity and depth. Using Grad-CAM heatmaps created with earlier layer activations (which have smaller receptive fields) is also not ideal. Firstly, they would not effectively capture high-level semantics. Moreover, Grad-CAM-based losses (e.g., GAIN) using such heatmaps would prompt the DNN to filter out the influence of background features at an early stage in the network. To ignore the background, a DNN must learn to identify the images' foreground features, an operation that may be complex and represent a high level of abstraction. The utilization of early Grad-CAMs would limit the number of layers that the DNN can use to implement this operation, restricting its complexity. In conclusion, due to Grad-CAM weaknesses, we advocate that Grad-CAM heatmaps are not reliable enough to be directly optimized with the objective of controlling classifier attention when bias is present.

LRP and Grad-CAM create explanation heatmaps with drastically diverse procedures. Instead using a linear combination of one layer's feature maps, LRP backpropagates a signal through the entire neural network, capturing information from all parameters and activations to produce a high-definition heatmap, which incorporates high level semantics and precise localization\cite{LRPBook}. Consider a case of spurious mapping, where the classification was influenced by a late layer's activation aligned with the region of interest, but the activation was actually caused by a background feature in the input. In this case, due to its backpropagation rules, LRP will map the late activation's relevance back to the background region in the DNN's input (main article section ``Layer-wise relevance propagation'' and ``Optimizing for background bias resistance: why LRP?''). Thus, LRP will reveal background attention, but Grad-CAM will not (as exemplified in Figure \ref{Gmaps} in the GAIN heatmaps). Consequently, LRP-based attention mechanisms such as the ISNet are not vulnerable to a solution based on spurious mapping.

In summary, past studies have shown that LRP is more robust than Grad-CAM\cite{LRPRobustness}, the technique has a more consistent mathematical background\cite{LRP}\textsuperscript{,}\cite{LRPBook}, its optimization is more theoretically founded (main article section ``ISNet theoretical fundamentals''), and it does not present Grad-CAM's aforementioned flaws (e.g., vulnerability to spurious mapping). Accordingly, in this study's experiments, our LRP-based attention mechanism surpassed the Grad-CAM based GAIN, demonstrating less background attention and improved o.o.d. test performance. The ISNet's o.o.d. generalization performance also surpassed another Grad-CAM-based architecture, Hierarchical Attention Mining (HAM)\cite{HAM}, when background bias was present during training. Like GAIN, HAM also displayed spurious mapping and deceiving Grad-CAM explanations (Appendix \ref{CheXpertClassification}). Therefore, we conclude that Layer-wise Relevance Propagation is more reliable than Grad-CAM for the construction of attention mechanisms and the reduction of shortcut learning.

\subsection{Concluding Thoughts}

With multiple synthetic bias experiments, considering different tasks, dataset sizes and classifier backbones, we systematically demonstrated the ISNet's superior capacity to avoid background attention and the shortcut learning induced by background bias. Quantitative results show that only two of the multiple tested neural networks could consistently hinder background bias attention, having their decisions not affected by the synthetic bias: the ISNet and the alternative segmentation-classification pipeline. However, the ISNet surpassed the large pipeline in all synthetic bias experiments, and in all the experiments with naturally occurring background bias (COVID-19 and TB detection). Furthermore, at run-time, the proposed model is much more memory efficient and faster (Appendix \ref{speed}).

All other state-of-the-art neural networks we tested failed to avoid shortcut learning, as they considered the synthetic bias in their decisions (standard classifier, multi-task DNN, AG-Sononet, GAIN, RRR, Vision Transformer, and HAM in Appendix \ref{CheXpertClassification}). Thus, they demonstrate significant performance degradation when the synthetic bias is removed or replaced by deceiving bias (main article Table 1). When analyzing the LRP heatmaps for the models, attention to the synthetic bias is apparent, unlike in the case of the ISNet.

The quantitative results in COVID-19 and TB detection confirm the findings from the synthetic bias experiments: the ISNet reduced attention to background bias and achieved the highest generalization performance in both tasks. Like in the synthetic bias experiments, except for the segmentation-classification pipeline and the ISNet, all other neural networks displayed significant background attention in their LRP heatmaps. This finding explains why the ISNet surpassed their o.o.d. test performances. As past studies suggest\cite{bassi2021covid19}\textsuperscript{,}\cite{covidSegmentation}, the utilization of lung segmentation to remove the X-ray background before classification reduced shortcut learning in COVID-19 detection. Accordingly, and corroborating with the findings from the synthetic bias experiments, the segmentation-classification pipeline strongly surpassed the other benchmark DNNs on an external COVID-19 dataset (achieving maF1, with standard deviation, of 0.645 +/-0.009), and its LRP heatmaps showed focus on the lungs. However, as best demonstrated in TB detection, the alternative pipeline can pay attention to lung borders and to the erased background. This possibly reveals decision rules that consider the region of interest shape and location, hindering generalization. This attention profile was more apparent in TB detection. Unsurprisingly, the pipeline's generalization was worse in the task (0.576 +/-0.05 maF1 with o.o.d. evaluation). The ISNet did not display this problem, and it surpassed the alternative model in both TB and COVID-19 detection (with 0.773 +/-0.009 maF1 in the o.o.d. COVID-19 test dataset, and 0.738 +/-0.044 maF1 in the o.o.d. TB dataset). 

An attention gated network and a Vision Transformer could not attenuate shortcut learning, they paid attention to sources of background bias and generalized poorly. This result suggests that attention mechanisms that do not learn from semantic segmentation targets are not useful to avoid shortcut learning. Without an external guide indicating what should be ignored, the models learn to focus on the image features with high potential to improve classification loss during training, like background bias. Using a multi-task learning model, which performed image segmentation and classification simultaneously, was also ineffective in dealing with background bias. The model could correctly segment lungs, faces or dogs, regardless of background features. However, the classification procedure focused on the background and showed poor generalization. Therefore, in the presence of background bias, the multi-task DNN attention profiles for segmentation and classification diverged.

Besides the ISNet, GAIN\cite{GAIN} and RRR\cite{RRR} also optimize explanation heatmaps to control a classifier's attention. However, instead of LRP, GAIN's heatmaps are built with Grad-CAM\cite{GradCAM}, and RRR's with input gradients\cite{saliency}. In relation to the ISNet, the models' capability of avoiding background attention is not up to par. Unlike the LRP-based ISNet, they could not avoid the artificial biases' influence on the classifier decisions and F1-Scores, as demonstrated by significant accuracy drops when the synthetic bias was removed or substituted by deceiving bias. Accordingly, the models' LRP explanation heatmaps (main article Figures 1 and 2) indicate attention to the natural and artificial sources of background bias. Thus, the models' o.o.d. performances were strongly surpassed by the ISNet. Like GAIN, Hierarchical Attention Mining\cite{HAM}, another Grad-CAM-based attention mechanism, was also unable to avoid background bias attention (Appendix \ref{CheXpertClassification}). In an ablation experiment (the ISNet Grad*Input), we substituted the ISNet LRP heatmaps by Gradient*Input explanations, which can be seen as an improvement over input gradients\cite{GradInput}. The model was tested with the three synthetic bias experiments. With a VGG-19 backbone, it displayed a small performance degradation when background bias was removed from the test data. However, the model could not converge properly when utilizing a large classifier backbone (DenseNet121), leading to small overall accuracy. In summary, we empirically demonstrate the higher reliability of LRP optimization in deep neural networks. When utilizing very deep backbones, the ISNet accuracy and resistance to background bias significantly surpassed models optimizing the more common explanation methodologies of Grad-CAM, input gradients and Gradient*Input.

We justify these empirical results with an in-depth theoretical analysis (main article section ``ISNet theoretical fundamentals'' and Appendix \ref{lrpVsGradInput}). Briefly, LRP-$\varepsilon$ has a solid mathematical background, with roots on the application of Taylor expansions at the DNN's neurons\cite{LRPBook}. These mathematical fundamentals allow the classifier resulting from LRP optimization (ISNet) to have superior bias resistance, without losing overall accuracy. Meanwhile, Grad-CAM optimization led to classifiers learning to produce spurious Grad-CAM explanations, which showed no background bias attention, while quantitative performance scores (main article Table 1) proved that the models' decisions were being influenced by the background bias. LRP heatmaps revealed the attention to background bias for such models. On the other hand, input gradients, optimized by RRR, are a powerful tool to optimize shallower classifiers, as shown by the study presenting the method\cite{RRR}. However, input gradients are commonly not seen as adequate explanations for large DNNs analyzing high-resolution images, especially due to their noisiness\cite{LRPvsGrad}. We hypothesize that such characteristic hinders the optimization of RRR when considering large classifier backbones and high-resolution images, similarly to Gradient*Input optimization. We formally demonstrate how LRP-$\varepsilon$ reduces explanation noise in comparison to input gradients and Gradient*Input (Appendix \ref{lrpVsGradInput}). Moreover, we show that, from a Deep Taylor Decomposition standpoint, LRP-$\varepsilon$ represents more contextualized and coherent explanations of the classifier behavior. In summary, LRP's theoretical fundamentals justify why its optimization leads to robust and accurate classifiers (main article section ``ISNet theoretical fundamentals''), explaining its superior bias resistance, empirically verified in diverse experiments with background bias (main article Tables 1 to 5).

\section{Comparing ISNet X-ray Heatmaps to Radiologist's Annotations}
\label{radiologistAnalysis}

With the objective of evaluating the correlation between lung lesions and the ISNet's focus during X-ray classification, we compared the DNN's LRP heatmaps and the corresponding X-rays, with lesions localized by a radiologist. The specialist, Dr. Dertkigil, is a professor of radiology at the Department of Radiology of the University of Campinas (UNICAMP), Campinas, SP, Brazil. He has been practicing medicine at the Clinics Hospital of the University of Campinas since 2009 and is the Director of the hospital's Radiology Service.

We randomly selected 30 X-rays from the external (o.o.d.) test databases, consisting of 10 pneumonia images, 10 COVID-19 images, and 10 tuberculosis samples. We selected only images that were correctly classified by the ISNets (true positives), because we aim to understand if the fundamentals for a correct DNN decision are similar to the reasoning of a human specialist. The radiologist was asked to mark lung lesions caused by the disease present in the X-rays. He had no access to the DNN outputs or heatmaps before completing the requested task. Afterwards, the annotated X-rays were superposed over the corresponding LRP heatmaps, produced by the ISNets, and we analyzed the resulting images. Annotations are presented in magenta (lines); X-rays were equalized (histogram equalization), heatmaps were resized to the original X-ray shape and normalized (with red and blue values ranging between 0 and 255). LRP employed the LRP-$\varepsilon$ rule, with LRP-z$^{\mathrm{B}}$ for the first DNN layer, and the softmax-based LRP rule\cite{LRPBook} for the output layer (ensuring more class-specific heatmaps).

Although we utilized annotated X-rays to evaluate the clinical soundness of the ISNets' LRP heatmaps, the classifiers were not trained to segment lesions in their heatmaps, nor were they informed of the lesions' locations during training. The ISNet architecture was created to hinder attention outside a region of interest (lungs), but it does not control the attention pattern inside this region (refer to Appendix \ref{loss}). Thus, the configuration of attention inside the lungs is a natural result of the classification loss optimization. For the precise localization of lesions, object detectors or semantic segmenters, trained with lesion masks, may be preferred in relation to classifiers.

When classifying images, classifiers can focus mostly on parts of the object of interest. Thus, explanation heatmaps generally highlight the objects partially\cite{SEC}. Accordingly, our DNNs may correctly classify a disease by focusing on just some of the lung lesions, or only parts of some lesions. We must also consider that classification accuracies were not perfect. Test performance metrics for the DNNs can be found in main article Tables 2 and 3 for the model classifying healthy, pneumonia and COVID-19, and in main article Table 5 for the network classifying healthy and tuberculosis. Class F1-Scores (and standard deviations) for the classification of COVID-19 and pneumonia are 0.907 +/-0.006 and 0.858 +/-0.007, respectively. For tuberculosis, the F1-Score (and 95\% confidence interval) is 0.748 +/-0.043. For these reasons, we did not expect a perfect correlation between the radiologist lesion annotations and the DNN LRP heatmaps. However, the existence of a correlation was clear, and its quality was superior for COVID-19, the class with the best F1-Score.

\begin{figure}[!h]
\includegraphics[width=0.7\textwidth]{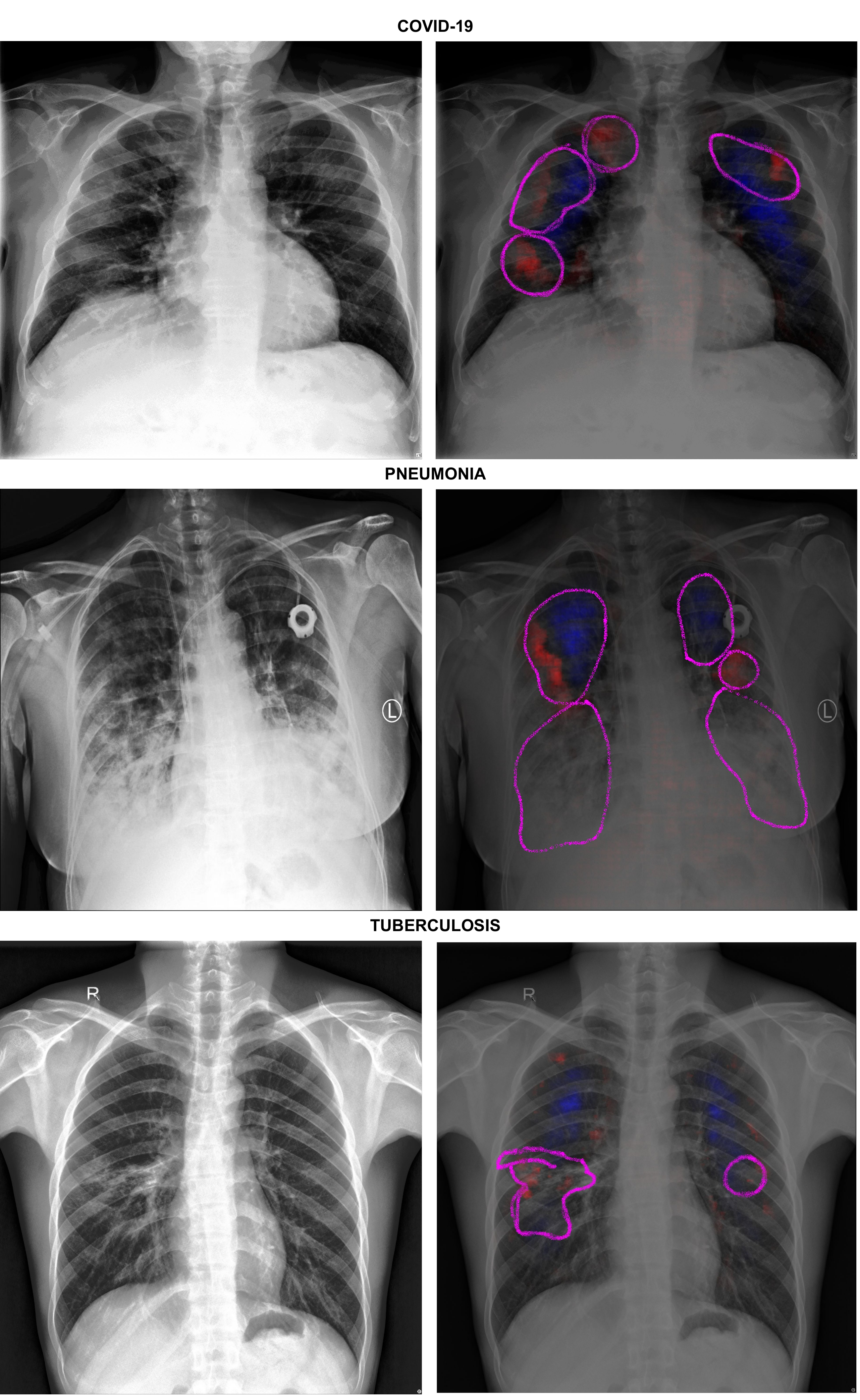}
\centering
\caption{\textbf{X-rays and their LRP (Layer-wise Relevance Propagation) heatmaps (blue and red), superposed over the image annotated by the radiologist (magenta lines).} Red colors indicate areas that the DNN associated to the true class (indicated over the X-rays), blue colors are areas that decreased the network confidence for the class}
\label{rad}
\end{figure}

Figure \ref{rad} presents a few examples of X-rays and their heatmaps, superposed over the annotated X-rays. Red colors indicate regions that the DNN associated with the classified disease, and blue represents areas that reduced the DNN confidence for the disease. In the COVID-19 heatmap, blue areas can be associated to the pneumonia and healthy classes; in the pneumonia image, to COVID-19 and healthy; in the TB sample, to healthy. The magenta lines are the radiologist's annotations, surrounding lesions caused by the conditions. As expected, high relevance for the diseases (red colors) partially covered the annotations, and not all annotations contained relevance. In the tuberculosis heatmap, some positive relevance (red) can be seen outside of the annotations. If we recreate the COVID-19 heatmap by executing LRP for pneumonia instead of COVID-19, blue regions encompassed by the magenta lines will become red. Moreover, when running LRP for the COVID-19 class with the pneumonia-positive image, blue regions inside annotations also become red. Thus, we observe a bit of confusion between damage produced by COVID-19 pneumonia and non-COVID-19 pneumonia, which can be explained by the similarity of two conditions' lesions. In the three Figure \ref{rad} X-rays we see annotations encompassing red regions, sometimes almost perfectly. Therefore, it clearly shows a correlation between the ISNet heatmaps and the radiologist's analysis.

We individually compared all annotated X-rays with their heatmaps. Radiologist's annotations, heatmaps, and individual analyses are available in the Supplementary Data 1\cite{SupData}. With a qualitative evaluation, we concluded that the correlation between annotations and red regions in the heatmaps is clearly observable. As exemplified in Figure \ref{rad}, the DNN generally focused on some of the annotated lesions and paid more attention to parts of them. We noticed that the ISNets tend to avoid paying attention to lung regions that are overly opaque. When there was a very opaque lesion, the DNN preferred to focus on its borders or on other lesions. This behavior may be caused by our ground-truth lung segmentation targets (used to train the ISNets), which could not always identify strongly opaque regions as part of the lungs. Accordingly, we believe that the ISNet attention pattern could be improved if manual lung segmentation masks existed for all training X-rays. Many images had some relevance outside of annotations, but the majority had a large part of the red regions covered by annotations, including the reddest heatmap area. Comparing the three classes, the COVID-19 heatmaps presented the largest number of annotations encompassing red regions, and the smallest quantity of red outside of annotations. Accordingly, the correlation between the human specialist annotations and the disease positive relevance (red) in AI (Artificial Intelligence) heatmaps is stronger for the class with the highest test performance, COVID-19.

From a quantitative point of view, in 9 of the 10 COVID-19-positive images, at least half of the COVID-19 lesions marked by the radiologist encompassed red regions (i.e., areas that the DNN associated with COVID-19). All figures presented at least one annotation covering red areas. In 8 of the 10 X-rays, most of the COVID-19 relevance (red) was inside the radiologist's annotations. Moreover, in 8 of the 10 images the reddest heatmap area (i.e., the ISNet strongest focus for the COVID-19 class) was covered by an annotation. As in the Figure \ref{rad} example, a few annotations covered blue regions, which were normally more associated with pneumonia than with COVID-19 by the DNN. Only one of the 10 X-ray heatmaps showed a significant amount of relevance outside of the lungs. This is the only image in the 30 annotated samples with significant attention outside of the lungs, demonstrating the success of the ISNet attention mechanism. The undesired relevance was around the armpit, a region that some ground-truth lung segmentation targets falsely identified as lungs. Thus, the problem could be solved by training the ISNet with more accurate lung masks. Overall, the quantitative results indicate that the ISNet COVID-19 heatmaps and the radiologist's analysis are strongly correlated.

Of the 10 pneumonia-positive images, 7 presented at least 50\% of the radiologist's annotations covering pneumonia relevance (red). Furthermore, 9 had at least one annotation containing red areas. The only exception was an image with a single annotation, located in the lung region that overlaps the heart. However, our ground-truth segmentation targets trained the ISNet not to consider this area as region of interest. Of the 10 images, 4 had the reddest area inside an annotation. They also had most of the pneumonia relevance contained by the annotations. In line with the ISNet's F1-Scores, the quantitative results show that the correlation between the pneumonia heatmaps and the radiologist's analysis is not as strong as for the COVID-19 images. However, the results clearly indicate a correlation.

Finally, 9 of the 10 TB heatmaps have at least half of their annotations encompassing red regions (TB relevance). All images had at least one annotation covering a red area. Six samples had most of the TB relevance contained by annotations. Moreover, these 6 X-rays also had the reddest heatmap area inside an annotation. Like the TB classification test F1-Score, these quantitative results did not match those found for COVID-19. However, a correlation is observable once more.

The qualitative and quantitative comparison between the ISNet heatmaps and the radiologist's annotations shows the existence of a correlation, which is the strongest for the class with the highest test performance. Even though the ISNet was not trained to segment lesions, the correlation indicates a consistency between a human specialist's reasoning and the decision rules naturally learned by the model. In line with the ISNet superior o.o.d. generalization performance (main article Tables 2 and 5), we observe that its attention mechanism diverted focus from background bias to the lesions.

\section{LRP Block}
\label{layers}

The LRP block defines each LRP relevance propagation rule as a special DNN layer. Thus, it consists on a stack of layers, each one performing LRP propagation through a corresponding classifier layer. The block propagates relevance from the classifier logits until its input, producing the heatmaps. Unlike actual layers (e.g., convolutions or dense layers), a layer in the LRP Block does not have independent trainable parameters. It only implements an LRP relevance propagation rule through a classifier layer. As exemplified in the main article Equation 3, such rules require access to the classifier layer's parameters and input. Accordingly, the LRP block layer shares parameters with the classifier layer, and a skip-connection carries the classifier layer's input to the LRP block layer. Due to parameter sharing, after gradient backpropagation through the LRP block, the classifier's parameters will be automatically optimized to minimize the heatmap loss. By defining the LRP relevance propagation process as a sequence of special neural network layers, optimizers in standard deep learning libraries can automatically backpropagate the heatmap loss gradient through the LRP Block. Therefore, we avoid the task of manually defining LRP gradient backpropagation rules. The LRP block is an implementation of Layer-wise Relevance Propagation, which generates the same heatmaps as those created with traditional LRP implementations. Our implementation, created in PyTorch, aims at improving computational efficiency in the LRP relevance propagation and the subsequent gradient backpropagation, making it feasible to produce and optimize multiple differentiable LRP heatmaps in parallel during ISNet training. 

\begin{figure}[h]
    \includegraphics[width=0.5\textwidth]{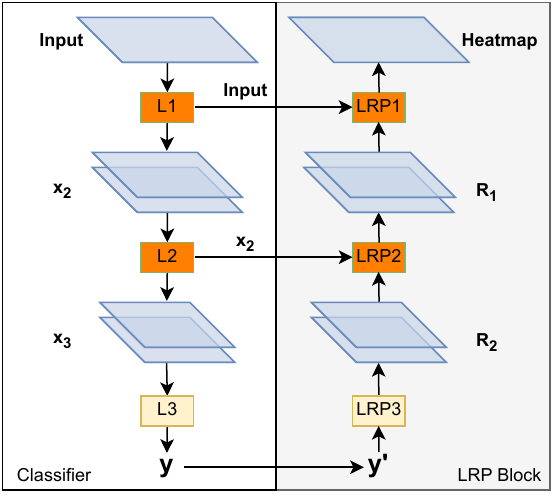}
    \centering
    \caption{\textbf{Representation of an ISNet for a classifier containing 3 layers.} The classifier is on the left, and the corresponding LRP (Layer-wise Relevance Propagation) block on the right. L1 and L2 represent convolutional layers, L3 represents a fully-connected layer. $\mathbf{x_{i}}$ indicates the input of classifier layer Li. With LRPi being the layer in the LRP block responsible for performing the relevance propagation through Li, its output, $\mathbf{R_{i-1}}$, is the relevance at the input of layer Li, and at the output of layer Li-1. $\mathbf{y}$ is the classifier's output, and $\mathbf{y'}$ is the output for one of the classes (with all other logits set to zero), which will be used to generate the heatmap associated with the class. $\mathbf{Input}$ is the classifier's input and $\mathbf{Heatmap}$ its LRP heatmap. Convolutional layers and their LRP counterparts are displayed in orange, while fully-connected layers and their LRP counterparts are in yellow}
    \label{SimpleISNet}
\end{figure}

Figure \ref{SimpleISNet} presents a simple ISNet example, whose classifier comprises two convolutional layers, L1 and L2, followed by a linear layer, L3. LRPi represents the LRP block layer performing relevance propagation for layer Li. To perform the propagation, LRPi shares parameters with Li.

Besides defining the LRP relevance propagation rules as special neural network layers, we also modify their implementation to train an ISNet. For each layer in the classifier, a corresponding LRP layer is added to the LRP block to perform the relevance propagation through it. All LRP layers are based on an efficient LRP implementation in four steps. Considering the propagation of relevance through a classifier convolutional or dense layer L with ReLU activation, the efficient procedure is defined by Algorithm \ref{StandardLRP}\cite{LRPBook}:

\begin{algorithm}
  \caption{LRP Standard Procedure for Fully-connected or Convolutional Layers}
  \label{StandardLRP}
  \begin{algorithmic}[1]
    \State Forward pass the layer L input, $\mathbf{x_{L}}$, through layer L, generating its output tensor $\mathbf{z}$ (without activation).
    \State For LRP-$\varepsilon$, modify each $\mathbf{z}$ tensor element ($z_{k}$) by adding $\mathrm{sign}(z_{k})\varepsilon$ to it. Defining the layer L output relevance as $\mathbf{R_{L}}$, perform its element-wise division by $\mathbf{z}$: $\mathbf{s}=\mathbf{R_{L}}/\mathbf{z}$.
    \State Backward pass: backpropagate the quantity $\mathbf{s}$ through the layer L, generating the tensor $\mathbf{c}$.
    \State Obtain the relevance at the input of layer L by performing an element-wise product between $\mathbf{c}$ and the layer input values, $\mathbf{x_{L}}$ (i.e., the output of layer L-1): $\mathbf{R_{L-1}}=\mathbf{x_{L}} \odot \mathbf{c}$.
  \end{algorithmic}
\end{algorithm}

In the ISNet, we apply the following changes to Algorithm \ref{StandardLRP}:

In Step 1, we perform a forward pass through a copy of layer L, whose weights are shared with its clone (i.e., they have the same parameters). The input for the operation is layer L's original input, $\mathbf{x_{L}}$ (i.e., the output of layer L-1), captured by a skip connection between the LRP block and the classifier. Although the bias parameters are also shared, we do not directly optimize them for the $L_{LRP}$ loss minimization (in PyTorch we accomplish this with the detach function on the LRP layer shared bias). Knowing that the biases are responsible for relevance absorption during its propagation\cite{LRP}, this choice prevents the training process from increasing them to enforce an overall reduction of relevance. As in the original method, we do not use an activation function in this step.

Step 2 is unchanged. We note that LRP-0 can be unstable for ISNet training. We based the LRP layers on LRP-$\varepsilon$, having chosen $\varepsilon$ as $10^{-2}$. LRP-z$^{\mathrm{B}}$ is used for the first DNN layer (Appendix \ref{lrpZb}). We do not employ LRP-$\gamma$ or LRP-$\alpha\beta$, because they did not result in background focus minimization (main article section ``ISNet theoretical fundamentals'').

For Step 3, we propose a fundamental modification because using a backward pass during the forward propagation through the LRP block could complicate the subsequent backpropagation of the heatmap loss gradient in some deep learning libraries. Consequently, we substitute the backward pass in Step 3 with an equivalent operation: the forward propagation of $\mathbf{s}$ through a transposed version of layer L. If L is fully-connected, its transposed counterpart is another linear layer, whose weights are the transpose of the original ones ($\mathbf{W}$). Thus, Step 3 becomes: $\mathbf{c}=\mathbf{W^{T}} \cdot \mathbf{s}$. Similarly, convolutional layers have transposed convolutional layers as their counterpart, using the same padding, stride, and kernel size. As in Step 1, the transposed layer share weights with layer L, but this time it does not use bias parameters.

Step 4 is unchanged and, as in Step 1, $\mathbf{x_{L}}$ is carried by the skip connection. The obtained relevance value, $\mathbf{R_{L-1}}$, is forwarded to the next LRP block layer, which will perform the relevance propagation for the classifier layer L-1. 

Using italics to emphasize our changes to the original algorithm, we summarize the relevance propagation through the LRP layer corresponding to a convolutional or fully-connected classifier layer L, which uses ReLU activation, in Algorithm \ref{LRPAlgoConvLin}.

\begin{algorithm}
  \caption{LRP Layer for Fully-connected or Convolutional Layers}
  \label{LRPAlgoConvLin}
  \begin{algorithmic}[1]
    \State Forward pass the layer L input, $\mathbf{x_{L}}$, through \textit{a copy of} layer L, generating its output tensor $\mathbf{z}$ (without activation). \textit{Use parameter sharing, use the detach() function on the biases. Get $\mathbf{x_{L}}$ via a skip connection with layer L.}
    \State For LRP-$\varepsilon$, modify each $\mathbf{z}$ tensor element ($z_{k}$) by adding $\mathrm{sign}(z_{k})\varepsilon$ to it. Defining the layer L output relevance as $\mathbf{R_{L}}$, perform its element-wise division by $\mathbf{z}$: $\mathbf{s}=\mathbf{R_{L}}/\mathbf{z}$.
    \State \textit{Forward pass the quantity $\mathbf{s}$ through a transposed version of the layer L (linear layer with transposed weights or transposed convolution), generating the tensor $\mathbf{c}$. Use parameter sharing but set biases to 0.}
    \State Obtain the relevance at the input of layer L by performing an element-wise product between $\mathbf{c}$ and the layer input values, $\mathbf{x_{L}}$ (i.e., the output of layer L-1): $\mathbf{R_{L-1}}=\mathbf{x_{L}} \odot \mathbf{c}$.
  \end{algorithmic}
\end{algorithm}

The above-defined convolutional and fully-connected LRP layers are equivalent to the traditional LRP-0/LRP-$\varepsilon$ propagation rules. Therefore, the proposed execution methodology is not detrimental to the explanational ability of LRP. In the following subsections, we explain the LRP layers for other common operations in DNNs, namely, pooling layers and batch normalization. Finally, we explain the implementation of the LRP-z$^{\mathrm{B}}$ rule for the first convolutional or fully-connected layer in the classifier.

\subsection{LRP Layer for Pooling Layers}
\label{pool}

As linear operations, sum pooling and average pooling can be treated like convolutional layers, whose corresponding LRP layer was defined in Algorithm \ref{LRPAlgoConvLin}. However, since pooling operations do not have trainable parameters, no weight sharing is needed. Furthermore, the $\mathbf{z}$ tensor in Step 1 of Algorithm \ref{LRPAlgoConvLin} can be explicitly defined as the pooling output (obtained via a skip connection). We represent convolutional kernels by a tensor $\mathbf{K}$ of shape (C1,C2,H,W), where C1 is the number of input channels, C2 of output ones, W the kernel width, and H its height. Then, for a convolution to be equivalent to pooling, C1 and C2 become the pooling number of channels, while W and H are defined by its kernel size. Naturally, the two layers use the same padding and stride parameters. To represent sum pooling, the convolutional kernel elements, $k_{c1,c2,h,w}$, are defined as constants:
 
\begin{equation}
k_{c1,c2,h,w}=
\begin{cases}
1,& \text{if }c1=c2\\
0,&  \text{otherwise}
\end{cases}
\label{weights}
\end{equation}

And considering average pooling with kernel size of (H,W), we have:

\begin{equation}
k_{c1,c2,h,w}=
\begin{cases}
\frac{1}{H.W},& \text{if }c1=c2\\
0,&  \text{otherwise}
\label{avgPoolWeights}
\end{cases}
\end{equation}

In summary, Algorithm \ref{LRPAlgoPool} explains the LRP relevance propagation for average or sum pooling.

\begin{algorithm}
  \caption{LRP Layer for Average Pool or Sum Pool}
  \label{LRPAlgoPool}
  \begin{algorithmic}[1]
    \State Get the pooling layer L output, $\mathbf{z}$, via a skip connection with layer L.
    \State For LRP-$\varepsilon$, modify each $\mathbf{z}$ tensor element ($z_{k}$) by adding $\mathrm{sign}(z_{k})\varepsilon$ to it. Defining the layer L output relevance as $\mathbf{R_{L}}$, perform its element-wise division by $\mathbf{z}$: $\mathbf{s}=\mathbf{R_{L}}/\mathbf{z}$.
    \State Find the equivalent convolution for pooling. Define its kernel according to Equation \ref{weights} for sum pooling or \ref{avgPoolWeights} for average pooling. Set biases to 0. Forward pass the quantity $\mathbf{s}$ through a transposed version of the equivalent convolution, generating the tensor $\mathbf{c}$.
    \State Obtain the relevance at the input of layer L by performing an element-wise product between $\mathbf{c}$ and the layer input values, $\mathbf{x_{L}}$ (i.e., the output of layer L-1): $\mathbf{R_{L-1}}=\mathbf{x_{L}} \odot \mathbf{c}$.
  \end{algorithmic}
\end{algorithm}

For max pooling, the corresponding LRP layer adopts a winner-takes-all strategy, distributing all the relevance to the layer inputs that were chosen and propagated by the pooling operation. Therefore, to obtain the relevance at the input of the max pooling layer, we propagate its output relevance through an equivalent MaxUnpool layer. This operation, available in the PyTorch library, uses the indices of the maximal values propagated by max pooling to calculate its partial inverse, which sets all non-maximal inputs to zero. As in average/sum pooling, no parameter sharing is needed, and indices can be directly obtained via a skip connection. Algorithm \ref{LRPMaxPool} defines the procedure.

\begin{algorithm}
  \caption{LRP Layer for Max Pool}
  \label{LRPMaxPool}
  \begin{algorithmic}[1]
    \State Using a skip connection with max pooling layer L, obtain the indices of the layer's inputs that were propagated by the pooling operation, $\mathbf{indices_{L}}$.
    \State Defining the layer L output relevance as $\mathbf{R_{L}}$, propagate it with a MaxUnpool operation, which sends all output relevance to the layer L's input elements that were chosen by max pooling.\\
     $\mathbf{R_{L-1}}=\mathrm{MaxUnpool}(\mathbf{R_{L}};\mathbf{indices_{L}})$
  \end{algorithmic}
\end{algorithm}

\subsection{LRP Layer for Batch Normalization}
\label{BN}
Batch normalization (BN) layers are linear operations that can be fused with an adjacent convolutional or linear layer during relevance propagation, creating a single equivalent convolutional/fully-connected layer. Thus, we calculate the parameters of the equivalent layer and create its corresponding LRP layer with the methodology presented in Algorithm \ref{LRPAlgoConvLin}.

A study\cite{BNLRP} presented equations to fuse batch normalization with an adjacent convolutional or fully-connected layer. To analyze a convolution, followed by BN, and then by a ReLU activation function (a configuration used in DenseNets), we define: $\mathbf{K}$ of shape (C1,C2,H,W) as the convolution weights/kernels; $\mathbf{B}$ as the convolutional bias, of shape (C2); $\bm{\gamma}$ as batch normalization weights, of shape (C2); $\bm{\beta}$ as the BN bias, also of size (C2); $\bm{\mu}$ the per-channel (C2) mean of the batch-normalization inputs, and $\bm{\sigma}$ its standard deviation, defined as the square root of the input's variance plus a small value for numerical stability (a parameter of the BN layer). After replicating the tensors $\bm{\gamma}$ and $\bm{\sigma}$ in the C1, W, and H dimensions to match the shape of $\mathbf{K}$, we can define the equivalent convolutional layer weights, $\mathbf{K'}$, with element-wise divisions and multiplications:

\begin{equation}
\label{BN1}
\mathbf{K'}=\frac{\mathbf{K} \odot \bm{\gamma}}{\bm{\sigma}}
\end{equation}

While the equivalent convolution bias, $\mathbf{B'}$, is given by the following element-wise operations:

\begin{equation}
\label{BN2}
\mathbf{B'}=\bm{\beta}+\bm{\gamma} \odot \frac{\mathbf{B}-\bm{\mu}}{\bm{\sigma}}
\end{equation}

Dense neural networks also present BN layers between pooling and ReLU layers, thus an adjacent convolutional or fully-connected layer is not available. Observing that the preceding pooling operation is also a linear operation (not followed by any activation function), we can fuse it with batch normalization. If we have average or sum pooling, we simply calculate its equivalent convolution, as explained in Appendix \ref{pool}, and perform the fusion according to Equations \ref{BN1} and \ref{BN2}. 

In the case of max pooling, we start by imagining a convolutional layer performing identity mapping before the batch normalization. This imaginary layer has no bias or padding, its stride is unitary, and its weights, $\mathbf{K}$, have shape (C,C,1,1), where C is the max pooling number of channels. An element $k_{c1,c2,h,w}$ is then given by Equation \ref{weights}. Thus, we can fuse the batch normalization with the identity convolution according to the Equations \ref{BN1} and \ref{BN2}, creating a BN equivalent convolution. The LRP layer for relevance propagation through the sequence of Max Pool, BN, and ReLU is then defined by the 4-step procedure in Algorithm \ref{LRPAlgoMPBN}.

\begin{algorithm}
  \caption{LRP Layer for the Sequence Max Pool, Batch Normalization, ReLU}
  \label{LRPAlgoMPBN}
  \begin{algorithmic}[1]
    \State Forward pass the Max Pool output through the BN equivalent convolution, generating its output tensor $\mathbf{z}$ (without activation). Use parameter sharing, use the detach() function on the biases, and obtain the Max Pool output and indices ($\mathbf{indices_{L}}$) via a skip connection.
    \State For LRP-$\varepsilon$, modify each $\mathbf{z}$ tensor element ($z_{k}$) by adding $\mathrm{sign}(z_{k})\varepsilon$ to it. Defining the output relevance of the ReLU activation as $\mathbf{R_{L}}$, perform its element-wise division by $\mathbf{z}$: $\mathbf{s}=\mathbf{R_{L}}/\mathbf{z}$. 
    \State  Forward pass the quantity $\mathbf{s}$ through a transposed version of the BN equivalent convolution (transposed convolution), generating a quantity $\mathbf{t}$. Employ parameter sharing but set biases to 0. Then perform a MaxUnpool operation, using $\mathbf{t}$ as input, creating the tensor $\mathbf{c}$.\\
    $\mathbf{c}=\mathrm{MaxUnpool}(\mathbf{t};\mathbf{indices_{L}})$
    \State Obtain the relevance at the input of the Max Pooling layer by performing an element-wise product between $\mathbf{c}$ and the Max Pool input values, $\mathbf{x_{L}}$ (i.e., the output of layer L-1): $\mathbf{R_{L-1}}=\mathbf{x_{L}} \odot \mathbf{c}$.
  \end{algorithmic}
\end{algorithm}

This methodology generates the same result as propagating the relevance first in the BN equivalent convolution and then in the max pooling LRP layer, but it has a lower computational cost.

\subsection{LRP Layer for z$^{B}$ LRP Rule}
\label{lrpZb}

The LRP-z$^{\mathrm{B}}$ rule\cite{LRPZb} considers the DNN input range in its formulation and is used for the first layer in the neural network. For a convolutional or fully-connected layer, it begins with a separation of the positive and negative elements of its weights, $\mathbf{W}$, and bias, $\mathbf{B}$. With $\mathbf{0}$ being a tensor of zeros:

\begin{gather}
\label{w+}
\mathbf{W^{+}}=\mathrm{max}(\mathbf{0},\mathbf{W})\\
\mathbf{B^{+}}=\mathrm{max}(\mathbf{0},\mathbf{B})\\
\mathbf{W^{-}}=\mathrm{min}(\mathbf{0},\mathbf{W})\\
\mathbf{B^{-}}=\mathrm{min}(\mathbf{0},\mathbf{B})\\
\label{b-}
\end{gather}

The parameters define three new layers: weights $l.\mathbf{W^{+}}$ and biases $l.\mathbf{B^{+}}$ define layer L$^{+}$, where $l$ is the lowest pixel value possible (in the common case where $l=0$, L$^{+}$ is ignored); weights $h.\mathbf{W^{-}}$ and biases $h.\mathbf{B^{-}}$ produce layer L$^{-}$, where h is the highest allowed input (normally 1). The original parameters, $\mathbf{W}$ and $\mathbf{B}$, define a copy of the original layer L. The four-step procedure for convolutional/fully-connected layers (defined in Algorithm \ref{LRPAlgoConvLin}) is then changed to the procedure in Algorithm \ref{LRPAlgoZb}.

\begin{algorithm}
  \caption{LRP Layer for Convolutional or Fully-connected Layer Using LRP-z$^{\mathrm{B}}$.}
  \label{LRPAlgoZb}
  \begin{algorithmic}[1]
    \State Forward pass the L layer input, $\mathbf{x_{L}}$, through a copy of layer L, creating $\mathbf{z^{original}}$. Forward an input of ones through layers L$^{+}$ and L$^{-}$, producing $\mathbf{z^{+}}$ and $\mathbf{z^{-}}$, respectively. Do not consider the non-linear activation. The classifier layer L shares parameters with the three instances, but apply the detach function to the bias parameters. Use a skip connection with layer L to obtain $\mathbf{x_{L}}$. Combine the three results in the following manner: $\mathbf{z}=\mathbf{z^{original}}-\mathbf{z^{+}}-\mathbf{z^{-}}$.
    \State Modify each $\mathbf{z}$ tensor element ($z_{k}$) by adding $\mathrm{sign}(z_{k})\varepsilon$ to it. Defining the layer L output relevance as $\mathbf{R_{L}}$, perform its element-wise division by $\mathbf{z}$: $\mathbf{s}=\mathbf{R_{L}}/\mathbf{z}$.
    \State Forward pass the quantity $\mathbf{s}$ through a transposed version of L, L$^{+}$ and L$^{-}$, creating the tensors $\mathbf{c^{original}}$, $\mathbf{c^{+}}$ and $\mathbf{c^{-}}$, respectively. Use parameter sharing but set biases to 0.
    \State Being $\mathbf{x_{L}}$ the layer L input values, obtain the relevance at the input of layer L, $\mathbf{R_{L-1}}$, with: $\mathbf{R_{L-1}}=\mathbf{x_{L}} \odot \mathbf{c^{original}} -\mathbf{c^{+}}-\mathbf{c^{-}}$.
  \end{algorithmic}
\end{algorithm}

If batch normalization follows the first classifier layer, placed before the ReLU nonlinear activation, Algorithm \ref{LRPAlgoZb} can implement the z$^{B}$ rule for the equivalent convolutional/fully-connected layer (the result of fusing the convolutional/linear layer with BN), explained in Appendix \ref{BN}. 

\subsection{Dropout}
The dropout operation can be defined as the random removal of layer input elements during the training procedure, and LRP block layers can automatically deal with it. For a layer L preceded by dropout, its inputs $\mathbf{x_{L}}$ shall have some zero values caused by the operation. According to the element-wise multiplication $\mathbf{R_{L-1}}=\mathbf{x_{L}} \odot \mathbf{c}$ used in the final step of LRP layers (e.g., step 4 in Algorithm \ref{LRPAlgoConvLin}), these null values will also make their respective relevances zero, thus being accounted for in the LRP propagation.

\subsection{Explaining the Implicit Segmentation Precision}
\label{precision}

\begin{figure}[!h]
\includegraphics[width=0.5\textwidth]{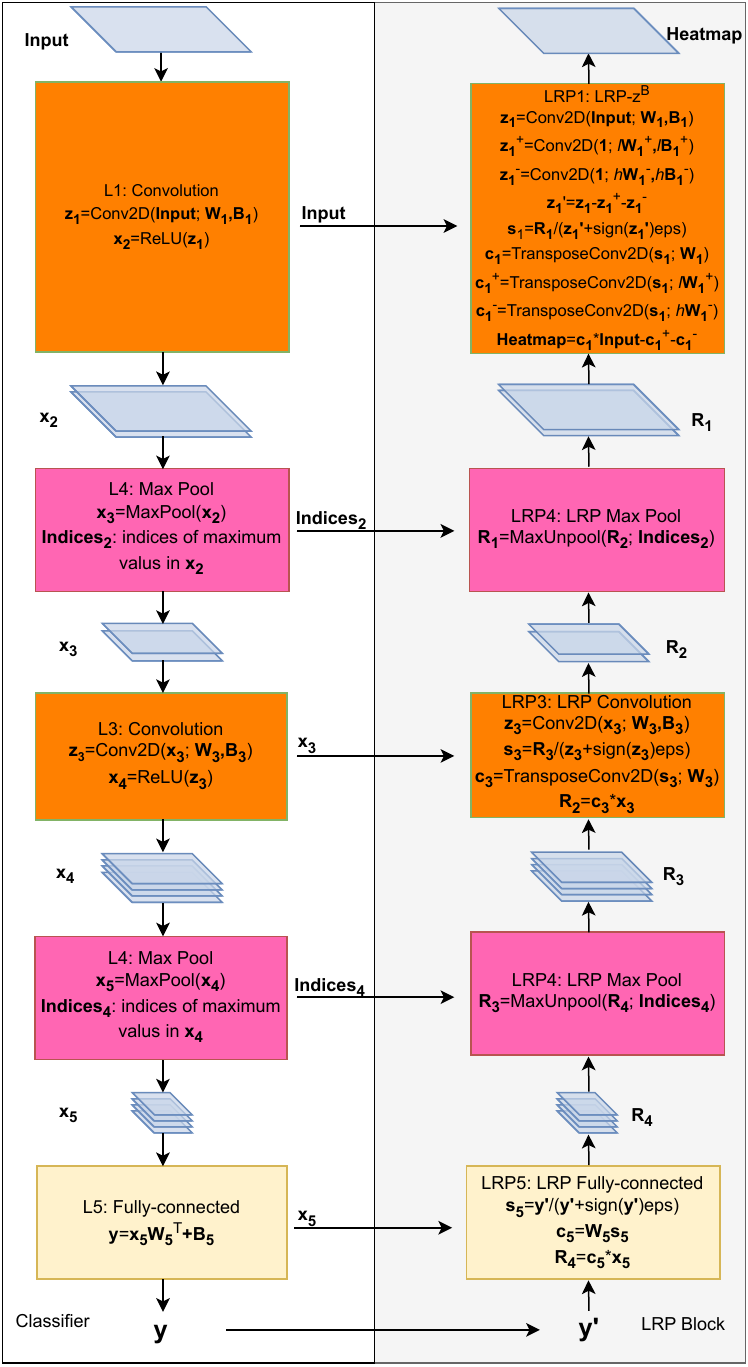}
\centering
\caption{\textbf{Representation of an ISNet.} The classifier is on the left, and the corresponding LRP (Layer-wise Relevance Propagation) block on the right. L1 to L5 represent classifier layers, and LRP1 to LRP5 represent the corresponding LRP block layers. Pooling layers and their LRP counterparts are in pink, convolutional in orange, and linear in yellow. Each layer in the figure presents a summarized description of the sequence of mathematical operations it comprises. Here, we denoted the LRP-$\varepsilon$ stabilizer ($\varepsilon$) as eps, element-wise multiplication was expressed as *, and element-wise division as /. The notation Conv2D(\textbf{x};\textbf{W,B}) indicates a convolution with input \textbf{x}, weight matrix \textbf{W}, and bias vector \textbf{B}. $\mathbf{x_{i}}$ indicates the input of classifier layer Li. $\mathbf{R_{i-1}}$ is the relevance at the input of layer Li, and at the output of layer Li-1. $\mathbf{y}$ is the classifier's output, and $\mathbf{y'}$ is the output for one of the classes (with all other logits set to zero), which will be used to generate the heatmap associated with the class. $\mathbf{Input}$ is the classifier's input and $\mathbf{Heatmap}$ its LRP heatmap. $\mathbf{Indices_{i}}$ indicates the indices of the elements in $\mathbf{x_{i}}$ propagated by the max pooling layer Li. ReLU($\cdot$) is the rectified linear unit activation function, sign($\cdot$) is the sign function, and TransposeConv2D(\textbf{x};\textbf{W}) indicates a transposed convolution with input \textbf{x} and weight matrix \textbf{W}. $h$ and $l$ represent the highest and lowest allowed input values, respectively. $\mathbf{W^{+}}$, $\mathbf{B^{+}}$, $\mathbf{W^{-}}$, and $\mathbf{B^{-}}$ are defined in Equations \ref{w+} to \ref{b-}. For a detailed description of the LRP layers, please refer to Appendix \ref{layers}}
\label{ISNet}
\end{figure}

Having defined the LRP layers of the LRP block, we can visualize the entire ISNet structure. For a common feed-forward classifier without skip connections in its structure (e.g., VGG-19\cite{VGG}), a layer in the middle of the corresponding LRP block, propagating relevance through the classifier layer L, has 3 connections: one with the previous LRP layer, the second with the following LRP layer, and a skip connection with the classifier layer L. The first carries the relevance that can be seen at the layer L output, $\mathbf{R_{L}}$; the second brings the LRP layer result, $\mathbf{R_{L-1}}$ (equivalent to the relevance at the layer L input), to the next layer in the LRP block; and the last carries, from the classifier layer L, the information required for relevance propagation (e.g., L's input, $\mathbf{x_{L}}$, if L is convolutional). Figure \ref{ISNet} exemplifies the ISNet architecture for a simple classifier. The LRP layer propagating relevance through layer Li is named LRPi. In the figure, $\mathbf{y'}$ is the classifier's output $\mathbf{y}$ for one class, with all other outputs made zero. One map shall be created for each class, with each LRP propagation being executed in parallel, in a strategy analogous to mini-batch processing.

As expected, the ISNet structure has a contracting path, defined by the classifier, and an expanding path, the LRP block. The LRP layers for pooling (in pink) increase the size of their relevance inputs. Afterwards, an LRP correspondent for convolution (in orange) reduces the number of channels in the relevance signal and combines it with information from an earlier feature map containing higher resolution, which is brought from the classifier. Because of the LRP rules, we naturally have a structure that combines context information from later classifier feature maps with high resolution from the earlier maps, which suffered less down-sampling. Thus, skip connections between an expanding and a contracting path allow the combination of information from later and earlier feature maps. Unlike common DNNs, our expanding path does not have independent weights, it shares all of its parameters with the contracting path. However, the idea of using skip connections to link a contracting path with an expanding one (combining high resolution and context information) is behind state-of-the art architectures used for image segmentation and object detection, such as the Fully Convolutional Networks for Semantic Segmentation\cite{SegmentationCNN}, U-Net\cite{unet}, and Filter Pyramid Networks\cite{FPN}. In our work, instead of being used for semantic segmentation (like in the U-Net), this concept allows the ISNet to precisely control its attention, containing the LRP relevance inside the image's foreground. Thus, the classifier implicitly and precisely finds the image's region of interest, and constructs decision rules based on its features.

\subsection{DenseNet based ISNet and Classifier Skip Connections}

The Densely Connected Convolutional Network (DenseNet) is characterized by the dense block, a structure where each layer receives, as input, the concatenation of the feature maps produced by every previous layer in the block\cite{DenseNet} (i.e., the block presents skip connections between each layer and all preceding ones). For LRP relevance propagation, all the connections must be considered. Therefore, a mirror image, inside the ISNet LRP block, of a dense block with S skip connections will also have S internal skip connections, now propagating relevance. Naturally, we also have skip connections between the classifier and the LRP block, which carry information from layer L (e.g., its input, $\mathbf{x_{L}}$) to the LRP layer that performs its relevance propagation. In the case of a dense block layer L, $\mathbf{x_{L}}$ is no longer defined as the output of classifier layer L-1, but rather as the concatenated outputs of all layers preceding L in its dense block.

To understand the relevance skip connections, imagine that, in the classifier, a layer L propagates its output to layers L+i, with $i \in \{1,...,N\}$. We define $\mathbf{R_{Inp(L+i),L}}$ as the relevance at the input of layer L+i, considering only its input elements (or channels) connected to L. We define relevance at the input of layer L+i, considering all input channels, as $\mathbf{R_{Inp(L+i)}}$. $\mathbf{R_{Inp(L+i)}}$ (and $\mathbf{R_{Inp(L+i),L}}$) can be obtained with relevance propagation through L+i. Then, to further propagate the relevance through layer L, we set the relevance at L's output as $\mathbf{R_{L}}$, given by Equation \ref{sumRel}. Proceeding with the propagation rules explained in Appendix \ref{layers}, we can obtain $\mathbf{R_{Inp(L)}}$, the relevance at the input of layer L.

\begin{equation}
\mathbf{R_{L}}=\sum_{i=1}^{N}\mathbf{R_{Inp(L+i),L}}
\label{sumRel}
\end{equation}

\begin{figure}[h]
\includegraphics[width=0.5\textwidth]{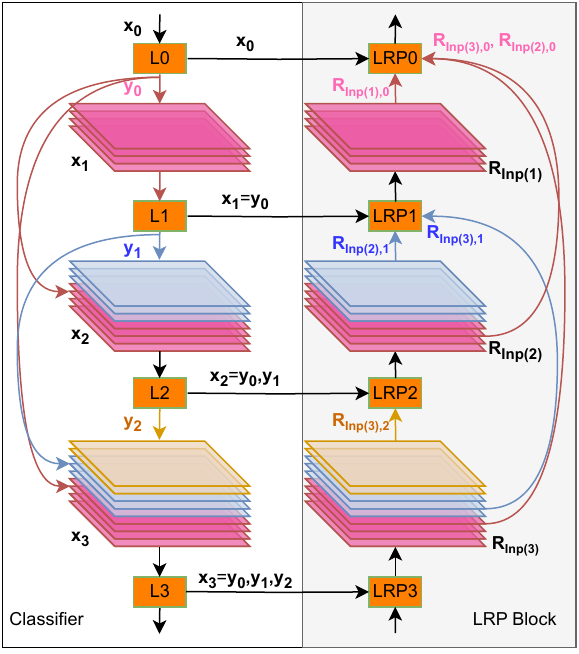}
\centering
\caption{\textbf{Classifier and LRP block for 4 convolutional layers with skip connections in the classifier.}  The classifier is on the left, and the corresponding LRP (Layer-wise Relevance Propagation) block on the right. L0 to L3 represent convolutional layers, and LRP0 to LRP3 represent the corresponding LRP block layers. $\mathbf{x_{i}}$ indicates the input of classifier layer Li, and $\mathbf{y_{i}}$ its output. With LRPi being the layer in the LRP block responsible for performing the relevance propagation through Li, its output, $\mathbf{R_{Inp(i)}}$, is the relevance at the input of layer Li. $\mathbf{R_{Inp(i),j}}$ is the relevance at the input of layer Li, for the channels that originated from layer Lj. The notation $\mathbf{x_{i}=y_{j},y_{k}}$ indicates that $\mathbf{x_{i}}$ is the concatenation of $\mathbf{y_{j}}$ and $\mathbf{y_{k}}$, in the channels dimension. Convolutional layers and their LRP counterparts are displayed in orange}
\label{SimpleDense}
\end{figure}

In a DenseNet, a layer L inside one of its dense blocks is defined as a nonlinear mapping $H_{L}(\cdot)$, which can in turn comprise a standard feed-forward sequence of other layers, e.g., ReLU activation, convolution, and batch normalization. Since there is an absence of skip connections inside one layer L, we propagate relevance through the sequence in the standard manner. Equation \ref{sumRel} can be directly used if $H_{L}(\cdot)$ is defined as sequences in the form $\mathrm{C_{L}(convolution) \rightarrow BN_{L} \rightarrow ReLU_{L}}$, $\mathrm{C_{L} \rightarrow ReLU_{L}}$, $\mathrm{C_{L} \rightarrow Pool_{L} \rightarrow ReLU_{L}}$, or combinations of the previous sequences. Figure \ref{SimpleDense} presents a simple example, considering a sequence of 4 convolutional layers in the classifier, L0, L1, L2, and L3, each defined as $\mathrm{C \rightarrow ReLU}$ and receiving the outputs of all previous layers. The flow of relevance is observable in the LRP block. Note that different connections carry different channels of the $\mathbf{R_{Inp(L+i)}}$ tensor, $\mathbf{R_{Inp(L+i),L}}$. In the Figure, we define the input of layer Li as $\mathbf{x_{i}}$, which is a concatenation (in the channels dimension) of the previous layers' outputs, $\mathbf{y_{j}}$.

The most common proposal for $H_{L}(\cdot)$ in DenseNets is: $\mathrm{BN^{1}_{L} \rightarrow ReLU^{1}_{L} \rightarrow C^{1}_{L} \rightarrow BN^{2}_{L} \rightarrow ReLU^{2}_{L} \rightarrow C^{2}_{L}}$. In Appendix \ref{layers}, we defined LRP layers to propagate relevance in sequences ending with ReLU activation. As such, they cannot be directly applied to $H_{L}(\cdot)$, and nor can we directly rely on Equation \ref{sumRel}. Consider a dense block layer L, followed by N other layers, L+i ($i \in \{1,...,N\}$), which receive L's output. $\mathbf{R_{L+i,L}^{ReLU1}}$ is the relevance at the output of the first ReLU inside layer L+i ($\mathrm{ReLU^{1}_{L+i}}$), considering only channels that came from layer L. Layer L+i starts by processing layer L's output through the sequence $\mathrm{BN^{1}_{L+i} \rightarrow ReLU^{1}_{L+i}}$, producing a tensor that we shall call $\mathbf{y_{L}^{L+i}}$. We can define $\mathbf{y_{L}}$ as the concatenation, in the channels dimension, of the N $\mathbf{y_{L}^{L+i}}$ feature maps, one for each L+i layer. With these definitions, the Algorithm \ref{LRPAlgoDenseNet} calculates the relevance at the output of layer L's first ReLU activation, $\mathbf{R_{L}^{ReLU1}}$, based on the N relevances $\mathbf{R_{L+i,L}^{ReLU1}}$. Thus, Algorithm \ref{LRPAlgoDenseNet} defines a recursion rule to propagate relevance through the LRP block.

\begin{algorithm}
  \caption{LRP Layer for the Dense Block Layer}
  \label{LRPAlgoDenseNet}
  \begin{algorithmic}[1]
    \State Fuse layer L's second convolution, $\mathrm{C^{2}_{L}}$, with the first batch normalization operation in each of the N subsequent layers L+i in the dense block, $\mathrm{BN^{1}_{L+i}}$, obtaining equivalent convolutional kernels and biases according to Equations \ref{BN1} and \ref{BN2}, respectively. Concatenate these parameters in their output channel dimension, creating the kernels and biases that define a single equivalent convolutional layer, which generates $\mathbf{y_{L}}$ from layer L's second ReLU output. Use it to recreate $\mathbf{y_{L}}$ but detach the bias parameters.

    \State Concatenate the N relevances $\mathbf{R_{L+i,L}^{ReLU1}}$ (for each L+i layer) in the channel dimension, producing $\mathbf{R_{L}^{conc}}$. This tensor can be seen as the relevance at the output of the equivalent convolution created in Step 1.

    \State Propagate the relevance $\mathbf{R_{L}^{conc}}$ through equivalent layer from Step 1, using Algorithm \ref{LRPAlgoConvLin}. The operation returns $\mathbf{R_{L}^{ReLU2}}$, the relevance at the output of the second ReLU activation in layer L.
    
    \State Obtain $\mathbf{R_{L}^{ReLU1}}$ by propagating $\mathbf{R_{L}^{ReLU2}}$ through layer L's sequence of operations $\mathrm{C^{1}_{L} \rightarrow BN^{2}_{L} \rightarrow ReLU^{2}_{L}}$, using the procedure in Appendix \ref{BN}. $\mathbf{R_{L}^{ReLU1}}$ can be propagated through earlier layers in the dense block by repeating these four steps.
    
  \end{algorithmic}
\end{algorithm}

A DenseNet transition layer L can be formed by the following sequence: $\mathrm{BN^{1}_{L} \rightarrow ReLU^{1}_{L} \rightarrow C^{1}_{L} \rightarrow ReLU^{2}_{L} \rightarrow AvgPool_{L}}$. $\mathrm{ReLU^{2}_{L}}$ is not part of its original configuration, but we added it to simplify the relevance propagation, as our rules are defined for layers with ReLU activation. It did not seem to have a detrimental effect on the model. The transition layer sits between 2 dense blocks. It receives all outputs from the layers in the first block, B, and propagates its own result to every layer of the next one, B+1. Therefore, a layer in block B naturally considers the transition layer among its consecutive N layers during Steps 1 and 2 of Algorithm \ref{LRPAlgoDenseNet}; the same is true for the $\mathrm{BN \rightarrow ReLU}$ sequence following the last dense block in a DenseNet (every layer in the last block shall consider it in Step 1). The relevance propagation through the transition layer L must take into account all its skip connections with the block B+1, so we use a 4-step procedure like Algorithm \ref{LRPAlgoDenseNet}. There are only two changes. Since L ends in an average pooling instead of a convolution, we modify the fusion process in Step 1 to merge pooling and BN, a technique explained in Appendix \ref{BN}.  Furthermore, in Step 4, we obtain $\mathbf{R_{L}^{ReLU1}}$ after propagating $\mathbf{R_{L}^{ReLU2}}$ through the transition layer $\mathrm{C^{1}_{L}\rightarrow ReLU^{2}_{L}}$ sequence, using the rules defined in Algorithm \ref{LRPAlgoConvLin}. We can treat the max pooling layer in the beginning of the DenseNet similarly, considering its skip connections with the first dense block. The code to automatically generate an ISNet from a DenseNet, in PyTorch, is available at \url{https://github.com/PedroRASB/ISNet}\cite{CodeGit}. 

\section{Datasets}
\label{dataset}
\subsection{COVID-19 Chest X-ray}
\label{COVIDdataset}

In this study we employed the Brixia COVID-19 X-ray dataset\cite{BrixiaSet} as the source of the training and hold-out validation COVID-19 positive samples. It is one of the largest open databases regarding the disease, providing 4644 frontal X-rays showing visible symptoms of COVID-19 (i.e., samples with an assigned COVID-19 severity score, the Brixia Score, higher than 0). All images were collected from the same hospital, ASST Spedali Civili di Brescia, Brescia, Italy. They correspond to all triage and patient monitoring (in sub-intensive and intensive care units) samples collected between March 4th and April 4th 2020, thus reflecting the variability found in a real clinical scenario\cite{BrixiaSet}. The database includes AP (87\%) and PA projections, with CR (62\%) and DX modalities, and the images were obtained from the hospital's RIS-PACs system. Data is available in the DICOM format, and system manufacturers are Carestream and Siemens. All samples are accompanied by a COVID-19 severity score, the Brixia Score. We excluded the 59 images with 0 overall Brixia Score, assuming that they do not present radiological signs of COVID-19 that can be detected. The scores were assigned by the radiologist on shift, who is part of a team composed by about 50 radiologists. They operate in diverse radiology units of the hospital, have different expertise in chest imaging and a wide range of years of experience\cite{BrixiaSet}. Accordingly, our dataset's labeling accuracy reflects the radiologist on shift's. The patients (2315) mean age is 62.8 years, with standard deviation of 14.8 years. They are 64.4\% male. We randomly assigned 75\% of the samples (3483 images) for training, and the remaining for hold-out validation. The two subdivisions were not allowed to have images from the same patients.

The images of healthy and non-COVID-19 pneumonia patients in our training and hold-out validation datasets come from the CheXPert database\cite{irvin2019chexpert}. It is a large collection of chest X-rays (224,316), showing various lung diseases, assembled in the Stanford University Hospital, California, United States of America. Samples correspond to studies performed between October 2002 and July 2017 (before the COVID-19 pandemic, ensuring that there are no COVID-19 samples in the database). The dataset considers both inpatient and outpatient centers. The database's classification labels were created with natural language processing (NLP), utilizing radiological reports, and have an estimated accuracy surpassing 90\%\cite{irvin2019chexpert}. Considering class balance, we randomly gathered 4644 images of healthy patients (corresponding to 4644 different patients), and 4644 of pneumonia patients (from 3899 different patients). CheXPert uncertainty labels were treated as negative. We only considered AP and PA X-rays (excluding lateral views). 64.9\% and 74.3\% of normal and pneumonia X-rays are AP, respectively. We downloaded the images in the jpg format. Samples were also randomly divided with 25\% for hold-out validation and 75\% for training, employing a patient split. Pneumonia patients have a mean age of 62.3 years, with a standard deviation of 18.7 years. They are 57.1\% male. Healthy patients have a mean age of 51.7 years, with standard deviation of 18.2, and are 56.3\% male.

To assess shortcut learning and o.o.d. generalization capacity, an external test database (with dissimilar sources in relation to the training dataset) was assigned. Regarding the COVID-19 class, we selected the dataset BIMCV COVID-19+ (iterations 1+2)\cite{BimcvSet} for evaluation. The database is also among the largest open sources of COVID-19 positive X-rays. The data was gathered from health departments in the Valencian healthcare system, Spain, considering multiple hospitals, mostly in the provinces of Alicante and Castellón\cite{BimcvSet}. Therefore, it is highly unlikely that this database shares patients with the Brixia dataset\cite{BrixiaSet}. Samples were acquired by querying the Laboratory Information System records from the Health Information Systems in the Comunitat Valenciana. They correspond to all consecutive imaging studies from patients with at least one positive PCR, IgM, IgG or IgA test for COVID-19, registered between February 16th and June 8th, 2020. We observe that the BIMCV COVID-19+ data acquisition period includes the Brixia dataset's. Moreover, both databases are representative of the first pandemic wave in Europe, not including more recent COVID-19 variants. The data contains DX and CR X-ray modalities, and we selected only AP or PA views, excluding lateral X-rays. Among the chosen images, 53.2\% were CR, and 58.5\% AP. The dataset authors used natural image processing to label the X-rays, according to their radiological reports. The NLP model reported F1-Scores are above 0.9\cite{BimcvSet}. The model could output 3 COVID-19 related labels, which indicate the disease probability according to the radiological report: COVID-19, COVID-19 uncertain or COVID-19 negative. We obtained 1515 images, corresponding to all COVID-19 positive frontal X-rays, labeled as COVID-19, which had associated DICOM metadata. Therefore, COVID-19 samples in our test dataset were classified as a highly likely COVID-19 case by a radiologist and correspond to a patient with at least one positive PCR or immunological test in the BIMCV COVID-19+ data acquisition period. We chose to include only samples that were labeled as COVID-19 by the NLP model to minimize the number of false positive COVID-19 samples in our test dataset, and not rely solely on PCR or immunological tests. Moreover, our training dataset also only includes samples where radiologists found COVID-19 symptoms. The X-rays were produced with a wide variety of devices, from different manufactures (see Table 5 in the manuscript presenting the dataset\cite{BimcvSet}). The selected test COVID-19 patients (1145) have a mean age of 66 years, with standard deviation of 15.3 years, and they are 59.6\% male. The patients have 6516 reported COVID-19 tests, which are 62\% PCR, 16.5\% IgG, 11.1\% IgM and 10.4\% IgA. 56.6\% of the tests are positive (all subjects have at least one positive test). Considering the positive tests, 51.9\%, 23.4\%, 15\% and 9.8\% are PCR, IgG, IgA and IgM, respectively.

We also performed cross-dataset testing for the pneumonia and normal classes. We chose the ChestX-ray14 database\cite{chex14} as the source of test pneumonia X-rays. Like CheXPert, it is a large dataset, containing 108,948 frontal X-rays, which can display various diseases. The images were mined from the PCS system in the National Institutes of Health Clinical Center, Bethesda, United States of America. They correspond to studies between 1992 and 2015 (prior to the COVID-19 pandemic). The ChestX-ray14 dataset authors\cite{chex14} labeled their database with natural language processing, according to radiological reports (with an estimated accuracy over 90\%). Data is presented in the png format. We utilized 1295 pneumonia images, corresponding to 941 patients, who are over 18 years old. They present a mean age of 48 years, with standard deviation of 15.5 years, and are 58.7\% male. 

Normal test images were extracted from a database assembled in Montgomery County, Maryland, USA (80 images), and Shenzhen, China (326 images)\cite{ChineseDataset1}, published in 2014. The Montgomery images were obtained from the Montgomery County’s Tuberculosis screening program, in collaboration with the local Department of Health and Human Services. X-rays present the AP view, and were created with an Eureka stationary X-ray machine (CR). Meanwhile, the Shenzen X-rays were gathered in Shenzhen No.3 People’s Hospital, Guangdong Medical College (Shenzhen, China). The facility employed a Philips DR Digital Diagnost system, and the X-rays also present the AP view. They were collected for one month, as part of the hospital routine, mainly in September 2012. All images from the Montgomery and Shenzen dataset were acquired in the png format, and the database does not present multiple images per patient. The healthy patients have a mean age of 36.1 years (and standard deviation of 12.3 years) and are 61.9\% male. Because the images in this database were manually labeled by radiologists, we preferred to extract the test normal images from the Montgomery and Shenzhen database, and not from ChestX-ray14. Having an unbalanced test database, we will evaluate our DNNs with macro-average performance metrics.

The Brixia COVID-19 dataset presented only 11 images of patients younger than 20 years old, and the BIMCV COVID-19+ database has a single one. Thus, to avoid bias, patients under 18 were not included in the other classes, which had much higher proportions of pediatric patients in their original datasets. After removing pediatric patients, we observe that the three classes in the training dataset have similar demographics, reducing possible related biases. The utilization of jpg images in the ChexPert database may introduce compression artifacts. However, any related bias should not artificially boost evaluation results, since our test database does not contain jpg figures. Both employed COVID-19 databases were projected to be large and representative of the disease's first wave in Italy or Spain. Moreover, the BIMCV COVID-19+ demographic's is similar to the Brixia dataset's, and it is coherent with demographics of COVID-19+ in Spain at the time of data collection\cite{BimcvSet}. 

The assembled database was chosen to resemble the characteristics of the most popular COVID-19 datasets: the use of dissimilar sources for different classes. Dataset mixing is required to work with some of the largest COVID-19 databases, and is also necessary in other classification problems, where a database does not contain all classes needed for the study. Mixing increases the possibility of background features correlating with classes. Therefore, our mixed COVID-19 dataset is an ideal scenario to test the ISNet ability to focus only on the region of interest (lungs), improving generalization and hindering shortcut learning.

The ISNet and multiple benchmark DNNs require segmentation masks for training. The ISNet utilizes them to calculate the heatmap loss during the training procedure. Thus, we utilized a U-Net, trained for lung segmentation in a previous study\cite{bassi2021covid19}, to create the segmentation targets, which we stored. We applied a threshold of 0.4 on the U-Net output to produce binary masks, which are valued 1 in the lung regions and 0 everywhere else; the chosen threshold maximized the segmenter's validation performance (intersection over union or IoU). The previous study authors\cite{bassi2021covid19} trained the U-Net using 1263 chest X-rays representing four classes: COVID-19 (327 images), healthy (327 images), pneumonia (327 images), and tuberculosis (282 images). They stated that the model achieved an IoU of 0.864 with the ground truth lung masks during testing. Please refer to\cite{bassi2021covid19} for a detailed explanation of the U-Net training procedure. We trained the ISNet and all alternative baselines using the COVID-19 detection datasets described in this section. When they were needed, all the models received the same segmentation targets, to produce a fairer comparison. For this reason, we do not use the U-Net from the previous study\cite{bassi2021covid19} in the alternative segmentation-classification pipeline. Instead, we trained another U-Net, using the same dataset that we employ for classification training. I.e., the segmentation targets that train the U-Net inside the alternative pipeline were created by the U-Net from\cite{bassi2021covid19}. It is possible that directly using the model from the previous study\cite{bassi2021covid19} in the alternative pipeline could boost its results, but, in this case, the alternative model would have been trained with different segmentation targets in relation to the ISNet. Moreover, these targets could be superior to the ones in our dataset, since some of them were manually created\cite{bassi2021covid19}.

Finally, to create an extreme test for the ISNet's attention mechanism, we produced a version of the dataset with synthetic background bias, consisting of white geometrical shapes inserted in the image's upper left corner. The shapes' height and width are 10\% of the image's height and width. All COVID-19 images were marked with a triangle, the normal class was marked with a square, and pneumonia with a circle. See main article Figure 1 for an example. Our data augmentation procedure can sometimes remove the shapes from the training samples (with rotations and translations), which may attenuate the background bias effect. However, the same is true for many natural background features (e.g., text and markers close to the X-ray borders), and the effect of the artificial bias should still be very solid. A common classifier, analyzing unsegmented images, can easily learn to identify the shapes, and to use them for improving classification accuracy. We evaluate the neural networks trained on the biased dataset in three different scenarios: with test images containing the same geometric shapes, with the original test images, and with deceiving bias. Deceiving bias means altering the correlation between the classes and geometrical shapes during testing, which has catastrophic effects on neural networks that focus on the background bias. Specifically, in this scenario we add the circle to the normal images, the triangle to pneumonia, and the square to COVID-19. The strong and controllable synthetic bias, and the three testing scenarios, allow us to better assess the neural networks' resistance to background attention. If the ISNet performs equally in all test scenarios, we can conclude that the proposed attention mechanism is adequate to avoid shortcut learning caused by background bias. The artificial biased was applied to all training and validation samples, simulating a worst-case scenario. 

\subsection{Tuberculosis Detection}
\label{TBDatasetSec}

The NIAID TB Portals\cite{TBPortals} is a multi-national program for multi-domain tuberculosis data sharing. Lead by the National Institute of Allergy and Infectious Diseases (NIAID, National Institutes of Health, USA), the program offers open-access and curated radiological, genomic, clinical, laboratory and geographic data from prospective and retrospective TB cases\cite{TBPortals}. Data was acquired with international collaborations with multiple clinical research sites and academic research organizations in countries that are highly affected by drug-resistant TB. With over 7400 TB patient cases (January 2022 published version), it provides one of the largest open databases of tuberculosis positive chest X-rays. Using the program's API and cohort creation system we downloaded every X-ray from patients diagnosed with tuberculosis of lung, confirmed by sputum microscopy with or without culture (diagnosis code A15.0, 3620 cases), confirmed by culture only (A15.1, 1427 cases) or confirmed histologically (A15.2, 46 cases). TB Portals mostly focus on drug-resistant-tuberculosis. In the downloaded data, 29.1\% present MDR non XDR resistance, 21.3\% are drug-sensitive, 9.5\% XDR resistant, 4.64\% Mono DR, 1.87\% Poly DR and 1.47\% Pre-XDR. Regarding outcomes, 38.4\% were cured, 6.55\% died, 6.54\% completed, 5.32\% lost to follow up, 4.16\% are still on treatment, 1.33\% in palliative care, 0.68\% unknown and 0.01\% not reported. Data were collected in the following countries: Georgia (30.5\%), Ukraine (25.5\%), Belarus (14.9\%), Moldova (14.7\%), Romania (5.7\%), Kazakhstan (3.8\%), Azerbaijan (3.2\%), Nigeria (1.3\%) and India (0.3\%). The imaging data is available in the DICOM format and represents both computed radiography and digitalized film X-rays (using flatbed scanners or digital cameras). After downloading the data, we manually reviewed the images, correcting files with wrong color-schemes (inverted DICOM MONOCHROME1 or 2 flag), and removing subpar samples (lateral views, figures with multiple X-rays in a single image or pictures with a small X-ray and large blank spaces). The dataset presented only 50 patients under 18. Thus, we removed underage patients from all databases (the normal class had a much higher percentage of children). We ended up with 6728 images, corresponding to 4227 patients, who are 74\% male, and have a mean age of 43 years with standard deviation of 13.8 years. We randomly divided them into 3 datasets (performing patient split): training (70\%), validation (20\%) and i.i.d. testing (10\%). Accordingly, we ended up with 4800 training images, 1286 validation samples, and 642 test samples.

As the Brixia Dataset for COVID-19 detection, the NIAID TB Portals do not present control cases (TB-negative). To the best of our knowledge, there is no open-source TB dataset with equal size or larger than TB Portals presenting control and TB-positive cases collected from the same sources. Accordingly, as in COVID-19 detection, researchers need to employ dataset mixing to use TB Portals for DNN training, or resort to smaller databases. Accordingly, we randomly gathered 6728 normal X-rays from the CheXPert database (explained in Appendix \ref{COVIDdataset}) and used them as the control cases. Corresponding patients (5562) are 56.9\% male, with mean age of 51.9 years, and standard deviation of 18.1 years. We did not include children. We also randomly divided the normal X-rays into three datasets (employing patient split), with 70\% for training, 20\% for validation and 10\% for i.i.d. testing.

Again, out-of-distribution evaluation is used to assess shortcut learning. We chose the Montgomery and Shenzen database\cite{ChineseDataset1} (described in Appendix \ref{COVIDdataset}) as the o.o.d. test database. It is among the most popular datasets for TB detection\cite{TBReview}. The database is much smaller than the training one (394 TB cases, 800 samples in total), but it presents normal and tuberculosis X-rays, collected from the same sources. From the TB images, 336 came from Shenzen, and 58 from Montgomery. Images were manually classified by the dataset authors, according to the X-rays associated clinical readings\cite{ChineseDataset1}. After removing pediatric patients, the o.o.d. test database ended up with 370 normal X-rays and 383 TB-positive cases. Healthy patients (370) are 61.9\% male, with mean age of 36.1 years and standard deviation of 12.3 years. Meanwhile, TB patients (383) are 69.2\% male, and have a mean age of 40 years, with standard deviation of 15.8 years. The o.o.d. dataset X-rays were acquired in Montgomery County, USA, and Shenzen, China, locations that did not contribute to our training database. Therefore, the probability of both datasets sharing cases is minimal. Comparing evaluation results in the i.i.d. and o.o.d. test datasets shall allow us to analyze if the problem of tuberculosis detection using mixed databases also fosters shortcut learning. Moreover, we will be able to evaluate if ignoring the background with the ISNet improves generalization, boosting performance on the o.o.d. test database.

To produce segmentation masks for our training and validation datasets we employed the same method that we used for COVID-19 detection: we applied the U-Net created in a previous study\cite{bassi2021covid19}. Its outputs were binarized according to a threshold of 0.4. Again, this model was not used inside the alternative segmentation-classification pipeline, whose segmenter we trained from scratch, employing the same tuberculosis database used for classification.  Most images in the o.o.d. test dataset (663) have manually created segmentation targets \cite{ShenMasks}\textsuperscript{,}\cite{ChineseDataset1}, which we can use as a second gold standard to test our U-Nets' segmentation performance. Finally, we observe that the two classes in our training dataset have similar demographics. 

\subsection{X-ray Datasets' Limitations}

Due to the small number of pediatric patients in the COVID-19 and TB datasets, we removed all pediatric patients from our data. Moreover, our COVID-19 datasets contain samples from the first semester of 2020. Classification performance may be diverse for present-day X-rays, as the disease's main symptoms change, especially considering vaccination and the prevalence of more recent variants of the virus. Radiologists detected COVID-19 signs in all datasets' COVID-19+ X-rays. This characteristic facilitates DNN training and avoids false positives, but the database may under-represent COVID-19 cases with very mild or hard to detect radiological manifestations. Moreover, the COVID-19 and pneumonia images reflect a hospital environment, under-representing asymptomatic patients and people who did not seek medical attention. 

In the TB dataset, most of the X-rays were captured in a clinical setting. Thus, severe cases are more represented. Moreover, since our training dataset focuses on drug-resistant TB, its percentage of drug-resistant TB cases is higher than the one found among new TB patients\cite{TBData}. Furthermore, our TB databases do not contain sick patients without tuberculosis. We were able to find only one database that contained the classes healthy, TB-positive and sick but TB-negative. Their authors claim that it better simulates a clinical scenario, reducing the number of false positives in a real-world situation\cite{TBX11K}. However, not having a second source of sick but not TB cases, we are not able to create an o.o.d. test database with the three classes. We did not get sick but not TB samples from a general chest X-ray database (e.g., CheXPert\cite{irvin2019chexpert} or ChestX-ray14\cite{chex14}), because they do not have the TB class. However, they may contain tuberculosis cases and they do present TB manifestations among their classes (e.g., infiltration, pneumonia, atelectasis and pleural effusion\cite{TBSigns}). Out-of-distribution evaluation is essential to assess shortcut learning. Although possibly reducing the usefulness of our model in a clinical setting, with two classes we will be able to achieve our major objective: to evaluate if the task promotes background bias, and to measure the ISNet effectiveness in hindering shortcut learning and improving generalization. Moreover, classifying normal and TB-positive makes our work in line with the vast majority of tuberculosis detection papers and datasets\cite{TBNat}\textsuperscript{,}\cite{TBReview}\textsuperscript{,}\cite{TBRev2}.

In summary, the X-ray classification tasks in this paper must be seen as demonstrations of the ISNet's potential of reducing shortcut learning caused by background bias, not as indications of clinical performance. Without clinical tests to ensure adequate real-world results, we cannot claim diagnosis performance or point our methodology as a substitute for RT-PCR in COVID-19 detection. 

\subsection{Facial Attribute Estimation}

The CelebA database\cite{celebA} is composed of natural images, in three channels, displaying people in a wide variety of poses, and containing background clutter. Using these in the wild images, the ISNet will classify facial features, while focusing only on the face region, which can assume many shapes, sizes, and locations in the figures. Thus, identifying this region is challenging. Facial attribute estimation is a multi-label classification problem, and it is fundamentally different from COVID-19 or TB detection in chest X-rays. We make the problem even more challenging by adding synthetic background bias and not employing a machine learning pipeline that localizes, aligns, and crops around the face before performing classification (a common approach to increase accuracy\cite{celebA}).

For the task of facial attribute estimation, we employed images from the Large-scale CelebFaces Attributes Dataset, or CelebA\cite{celebA}. The database has 10000 identities, each one presenting 20 images. Binary labels were created by a professional labeling company, according to 40 facial attributes\cite{celebA}. The CelebAMask-HQ dataset\cite{CelebAMaskHQ} presents a subset of 30000 high-quality images from CelebA, cropped and aligned. Furthermore, these images have manually created segmentation masks, which indicate the face regions\cite{CelebAMaskHQ}. For this study, we selected the CelebA images that produced CelebAMask-HQ. Afterwards, we created their segmentation masks by applying translation, rotation and resizing to the CelebAMask-HQ masks; with this operation we reverted the CelebAMask-HQ crop and align procedure (described in\cite{CelebAHQ}, appendix C). We employ the dataset split suggested by the CelebAMask-HQ dataset authors\cite{CelebAMaskHQ}, which assigns 24183 samples for training, 2993 for hold-out validation and 2824 for testing. The proposed splits are subsets of the official CelebA training, validation and test datasets, which were randomly defined. 

The CelebA dataset authors\cite{celebA} show that their DNN naturally focuses more on the persons' faces when more facial attributes are classified. For this reason, we believe that the ISNet implicit segmentation task will be more difficult when it classifies a small number of attributes. Thus, we chose to work with 3 classes, to better assess the ISNet attention mechanism potential, and to better visualize the architectures' benefits. If we choose attributes that also have features outside of the face (e.g., gender), the ISNet architecture will change the natural classification strategy for one that ignores features outside of the face. This effect can be desirable or not, depending on the researcher's objective. Here we opted to classify three attributes that are exclusively present in the face: rosy cheeks, high cheekbones and smiling. The first two are considered identifying attributes, i.e., they can be used for user identification. Therefore, avoiding bias in their classification is a security concern. In the training images, the three attributes, rosy cheeks, high cheekbones and smiling, appear in 2694, 11124, and 11363 images, respectively. In the validation dataset, they appear in 352, 1354, and 1390 figures, respectively. Finally, in testing, the three attributes are exhibited in 333, 1369, and 1339 images, respectively.

As in COVID-19 detection, we also created an artificially biased dataset for facial attribute estimation. The second application constitutes a multi-label problem. Thus, we marked our images with a square in the lower-right corner, a circle in the lower-left, and a triangle in the upper-left, to indicate rosy cheeks, high cheekbones and smiling, respectively. The presence of the geometrical shape is correlated with the presence of the attribute. As before, the bias was added to all training and hold-out validation data. Refer to main article Figure 1 for an example of an image with three positive labels. For the deceiving bias evaluation, we invert the correlation of attributes and bias: now, the geometrical shape will appear when the attribute is not present. For example, the upper-left corner triangle will be present when the person is not smiling. This configuration simulates the worst-case scenario for deceiving bias in a multi-label classification problem. When a classifier is considering the geometrical shapes in its decisions, this scenario will reduce its accuracy the most.

\subsection{Dog Breed Classification}

Dog breed classification is the other natural RGB image classification task in this study. It has a very recognizable region of interest (foreground), the dogs. We utilized a subset of the Stanford Dogs Dataset\cite{StanfordDogs}. The original dataset is composed of 20580 images, displaying 120 different dog breeds. Its images and annotations were originally extracted from the ImageNet database\cite{imagenet}, excluding images smaller than 200x200\cite{StanfordDogs} and manually evaluating the samples. The dataset represents a fine-grained classification problem and is challenging due to a few aspects: there is little inter-class variation; there is large intra-class variation; dogs are presented in different poses, occlusion/self-occlusion, ages, and color; and the pictures present large background variation, including man-made environments\cite{StanfordDogs}.

The ISNet training time increases linearly with the number of classes in the application (Appendix \ref{speed}). Therefore, considering all 120 Stanford Dogs categories would require excessively long training time. Moreover, it would make the reproduction of our results more expensive and time consuming, especially when powerful computational resources are not available. Therefore, we work with a subset of Stanford Dogs, considering all images for three randomly selected classes: Pekingese, Tibetan Mastiff, and Pug.

The subset has 100 training images per class, which we randomly split, separating 20 images per class for hold-out validation. The test dataset has the images from the original Stanford Dogs test split (which was randomly created by the dataset authors\cite{StanfordDogs}): 49 Pekingese samples, 52 Tibetan Mastiff and 100 Pug. Our objective is to evaluate the ISNet capacity of focusing on images' foregrounds (the dogs). With only 240 training samples, this dataset will display the model's ability of ignoring the background when trained with small databases. Meanwhile, the other tasks showcase its behavior when trained in a different dataset scale, with thousands to tens of thousands of images.

The Stanford Dogs dataset has bounding boxes, but not semantic segmentation masks, which are required to train the ISNet and many of the benchmark DNNs. Therefore, we have created the ground-truth dog segmentation targets with DeepMAC\cite{deepMAC}, a pretrained general-purpose semantic segmenter. The model produces segmentation outputs according to images and their bounding boxes, and it is specialized in novel class segmentation. I.e., segmenting objects whose classes were not seen during training. The trained model is openly available for download\cite{deepMAC}. Its outputs were binarized according to a threshold of 0.5, producing the final dog segmentation masks. Upon visual inspection, the masks were highly accurate.

Stanford Dogs is not a dataset known for containing background bias, and our objective is to evaluate the neural networks' capacity of focusing on the foreground (the dogs) and avoiding the shortcut learning caused by background bias. Therefore, we employ a synthetically biased version of the dataset. Here, we include a triangle in the top-left image corner for the Pug class, a circle in the bottom-left corner for Tibetan Mastiff, and a square in the bottom-right corner for Pekingese. Examples are provided in main article Figure 1. The synthetic background bias was inserted in all training and hold-out validation images. For the deceiving bias evaluation, the circle became associated to the Pekingese, the triangle to the Tibetan Mastiff, and the square to the Pug. The shapes' locations did not change (e.g., circles were still in the bottom-left image corner during deceiving bias evaluation).

\section{Training and Implementation Details}
\label{ImpDetails}

\subsection{Data Processing and Augmentation}
\label{ProcessingAgumentation}

We loaded the X-rays as grayscale images to avoid the influence of color variations in the dataset. We then performed histogram equalization to further reduce dataset bias. Next, we re-scaled the pixel values between 0 and 1, reshaped the images (which can assume various sizes in the databases) to 224x224 and repeated them in 3 channels. The shape of (3,224,224) is the most common input size for the DenseNet121, reducing computational costs while still being fine enough to allow accurate segmentation and lung disease detection\cite{bassi2021covid19}\textsuperscript{,}\cite{chexnet}\textsuperscript{,}\cite{bassiCovid}. Single-channel images can be used but, without a profound change in the classifier architecture, they would provide marginal benefits in the training time. Test images were made square (by the addition of black borders) before resizing to avoid any bias related to aspect ratio. The technique was not used for training because the models without segmentation-based attention mechanisms could learn to identify the added borders.

Since the DenseNet121 classifier is a very deep model and our X-ray datasets are not massive, we used data augmentation in the training datasets to avoid overfitting. The chosen procedure was: random translation (up to 28 pixels up/down, left/right), random rotations (between -40 and 40 degrees), and flipping (50\% chance). Besides the overfitting prevention, the process also makes the DNN more resistant to natural occurrences of the operations. We used online augmentation and substituted the original samples by the augmented ones. 

For facial attribute estimation, we loaded the photographs in RGB, re-scaled values between 0 and 1, and reshaped the images to 224x224, which is a common input size for the DenseNet121 and for natural image classification. We employed the same online image augmentation procedure used with the X-rays.

The dog breed classification images were loaded and processed like the facial attribute estimation samples. However, we did not employ data augmentation. The data augmentation procedure can partially or totally remove the synthetic background bias from the images. By being entirely present in all training images, the background bias capacity of influencing the classifier is maximized. Accordingly, this scenario can act as an extreme test of the ISNet capacity of avoiding background attention and the consequent shortcut learning.

\subsection{Training Procedure}
\label{training}

We employed a similar classification training procedure for all DNNs, to allow a fairer comparison. All networks were randomly initialized with standard PyTorch initialization procedures. Please refer to Appendix \ref{TransferLearning} for the reasons why we did not use transfer learning. The chosen optimizer was stochastic gradient descent, using momentum of 0.99, and a mini-batch of 10 to 16 images. Only the ISNet Grad*Input in COVID-19 detection used 0.9 momentum, as it could to some extent improve the model's problematic loss convergence. We employed gradient clipping to limit gradient norm to 1, making the training procedure more stable. The learning rate was set to 10$^{-3}$. In COVID-19 detection, TB detection and dog breed classification, we utilized cross-entropy as the classification loss, and for facial attribute estimation we used binary cross-entropy, as it is a multi-label problem. The network later evaluated on the test dataset is the one achieving the best hold-out validation performance during training. We trained all DNNs for 48 epochs in COVID-19 detection, and 96 in TB detection. We used 144 epochs for the ISNet in facial attribute estimation, but in this task, we trained the alternative methodologies for 96 epochs, because they were already showing overfitting at that point. All DNNs were trained for 200 epochs for dog breed classification. The classification thresholds for facial attribute estimation were chosen to maximize validation maF1 in the trained DNNs. We could have used the same procedure to find a decision threshold in TB classification, as it is a binary classification problem. However, in preliminary tests we discovered that this methodology could make results in the o.o.d. test dataset worse and more unpredictable. Therefore, we decided to just select the output neuron with the highest activation as the winning one (i.e., employing the standard threshold of 0.5 after softmax activation). To produce the classification labels, we considered normal as class 0, pneumonia as class 1, and COVID-19 as class 2. In TB detection, class 0 is normal, and class 1 is tuberculosis. In facial attribute estimation, class 0 is rosy cheeks, class 1 high cheekbones, and class 2 smiling. In Stanford Dogs, the class order is Pekingese, Tibetan Mastiff and Pug. 

For the alternative segmentation-classification pipeline, we began by training the U-Net for segmentation. We employed the same dataset, data augmentation and preprocessing steps used in the ISNet training procedure. We used stochastic gradient descent, with mini-batches of 10 to 16 samples and momentum of 0.99. The learning rate was set to 10$^{-4}$, except for the dogs segmentation task, where we used 10$^{-3}$. We employed the cross-entropy loss function. We trained using hold-out validation, until overfitting could be observed. Accordingly, we employed 600 epochs in the lung segmentation task, in face segmentation, 400, and in Stanford Dogs, 1000. After training, we select the DNN that maximized validation performance.

Appendix \ref{hyperParameterTuning} explains in detail the hyper-parameter tuning procedure we used to define loss hyper-parameters. Here, we briefly describe the final parameter settings. For the ISNet loss, we set P=0.7 and d=0.9 in COVID-19 and TB detection, P=0.5 and d=0.996 in facial attribute estimation, and P=0.7 and d=1 in dog breed classification. The heatmap loss E, $w_{1}$, and $w_{2}$ hyper-parameters were always set to 1, 1, and 3, respectively. Except for CheXPert classification (Appendix \ref{CheXpertClassification}), where we also set $w_{2}=1$. During the hyper-parameter tuning procedure, we noticed that using weight decay makes it much more complicated to find the ISNet's P parameter. This is because it seems to strongly favor a zero solution (i.e., the network minimizing its parameters to generate null heatmaps, optimizing the heatmap background loss and minimizing the parameters' L2 norm, but ignoring the classification task). Thus, we did not use weight decay. The ISNet Grad*Input parameters are: P=0.7 and d=0.996 in Stanford Dogs, P=0.7 and d=0.9 in COVID-19 detection, and P=0.5 and d=0.996 in facial attribute estimation. The final choices for RRR's right reasons loss weight\cite{RRR} ($\lambda_{1}$) were: $10^{6}$ in Stanford Dogs and COVID-19 detection, $10^{7}$ in facial attribute estimation, and $10^{8}$ in tuberculosis detection. The multi-task U-Net employed a hyper-parameter ($P$) to balance the classification ($L_{C}$) and segmentation ($L_{S}$) losses. Therefore, we minimize a loss term, $L$, defined as $L=P.L_{S}+(1-P).L_{C}$. We ended up with a P of 0.75 in both facial attribute estimation and COVID-19 detection, 0.9 in TB detection, and P=0.999 in Stanford Dogs. To linearly combine the three GAIN losses, we utilized the same weights employed in the original GAIN paper\cite{GAIN}, i.e., 1 for the classification loss, 1 for attention mining, and 10 for the external supervision, as other configurations did not significantly improve results and generalization. The same hyper-parameters were used in the experiments with and without synthetic background bias.

Clipping the gradient norm to 1, we were able to train with mixed precision, a configuration that reduces stability, but increases training speed. Mixed precision training was used with all DNNs except for RRR, GAIN and the ISNet Grad*Input (as the models directly work with gradients, re-scaling it with mixed precision makes their implementation more complicated). ISNet training instability can easily be identified, due to high training losses and lack of proper convergence; in this case, decreasing learning rate and increasing the d hyper-parameter in the heatmap loss can help convergence. Moreover, we explicitly chose to use deterministic operations during training in PyTorch. This was to ensure better correspondence between the LRP block and the classifier. It can be accomplished with a single PyTorch command, use\_deterministic\_algorithms(True).

\subsection{Hyper-parameter tuning}
\label{hyperParameterTuning}

Some architecture-specific loss hyper-parameters are important for the mitigation of shortcut learning. In the ISNet, the most important parameters are P and d. We tuned them using the experiments with synthetic background bias. We trained the networks in the artificially biased training datasets, and we set hyper-parameters to maximize maF1 in a validation dataset without the artificial bias, while also minimizing the performance difference between this validation set and its artificially biased version (containing the same bias as the training data). Following this strategy, we used a grid search to set the P and d hyper-parameters. It considered P settings of 0.2, 0.4, 0.6 and 0.8, and d values of 0.9, 0.996, and 1. After the grid search, we refined the P parameter, using a finer search around the best configuration found in the grid search. This finer stage kept d unchanged and added or subtracted 0.1 from P. When two hyper-parameter settings resulted in very similar validation maF1 (strong confidence interval overlap), we preferred the one with higher P and lower d (prioritizing the higher P). Thus, our search selects the parameters that provide the highest background bias resistance but preserve performance and training stability.

\begin{figure}[!h]
\includegraphics[width=0.6\textwidth]{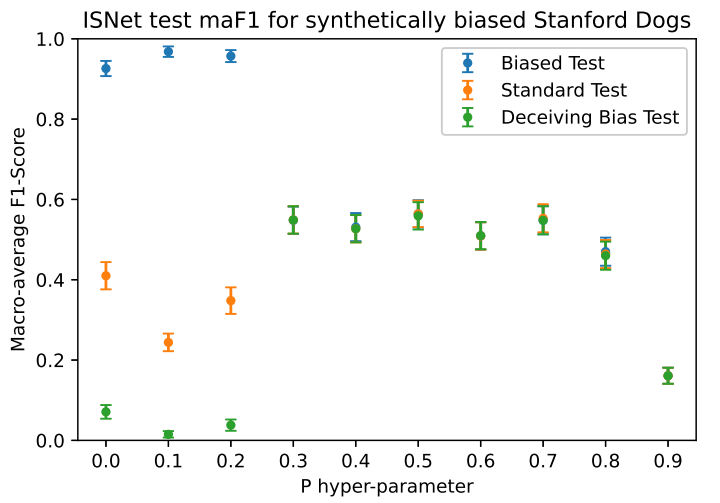}
\centering
\caption{\textbf{ISNet test maF1 (macro-average F1-Score) on the synthetically biased Stanford Dogs dataset, for 10 ISNets trained with diverse values of the P loss hyper-parameter (from 0 to 0.9, in intervals of 0.1).} Source data are provided as a Source Data file. Error bars indicate standard deviation, estimated with a Bayesian model, using a test dataset of n=201 images. Appendix \ref{statisticalMethods} explains the statistical analysis}
\label{Pchange}
\end{figure}

After the two stages, further finer searches did not significantly improve performance. For example, in dog breed classification, the two-stage search led to P=0.7 and d=1, which resulted in average F1-Scores (with standard deviation) of 0.548 +/-0.035 in the biased test, 0.553 +/-0.035 in the standard test, and 0.548 +/-0.035 in the deceiving bias test (main article Table 1). Meanwhile, an ISNet with P=0.65 and d=1 obtained maF1 of 0.542 +/-0.034, 0.542 +/-0.034, and 0.537 +/-0.034 in the biased, standard, and deceiving tests, respectively, and the network with P=0.75 and d=1 got 0.539 +/-0.034 maF1 in the three testing scenarios. Accordingly, the results with P=0.65, P=0.7, and P=0.75 were very similar, and no model presented a significant advantage over the others. Figure \ref{Pchange} exemplifies how test performances change with P, considering training on the synthetically biased Stanford Dogs dataset, d=1, and the training scheme described in Appendix \ref{training}. P changes in 0.1 increments, from 0 to 0.9. Before P=0.3, there is considerable variability in the standard test maF1 for adjacent data points (representing P=0, P=0.1, and P=0.2), a large gap between the performances on the diverse testing scenarios (standard, biased, and deceiving bias), and low standard and deceiving bias maF1. These results indicate that, while shortcut learning is still not effectively hindered, the synthetic background bias influences the DNN, and its capacity to analyze the foreground features is unpredictable and subpar. At about P=0.3, the gap between the three testing scenarios closed, and the deceiving and standard tests' maF1 significantly increased, indicating a minimization of shortcut learning. From P=0.3 to P=0.7, performances are stable, with confidence intervals' overlap and small variation with P. They begin to drop at around P=0.8. All networks were trained for the same time, 200 epochs. Thus, the final performance fall reflects that very high P makes training slower.

We did not tune the ISNet loss E, $w_{1}$, and $w_{2}$ hyper-parameters, because preliminary tests indicated low sensibility to them. As an example of this low sensibility, we could not get any significant accuracy improvement by setting E to 0.1 or 10, instead of 1. We choose the constants $C_{1}$ and $C_{2}$ for the ISNet foreground loss by analyzing a few heatmaps from a commonly trained classifier (not an ISNet), $\mathbf{H_{bk}^{standard}}$. $C_{1}$ is defined around the minimum observed value of $\mathrm{Sum}(\mathrm{abs}(\mathbf{H_{bk}^{standard}}))$, the total absolute heatmap relevance ($\mathrm{Sum}(\cdot)$ sums all elements in a tensor), while $C_{2}$ is set around the maximum observed value for the function. Accordingly, [$C_{1}$,$C_{2}$] should reflect a natural range of total absolute heatmap relevance. In this case, the ISNet background loss minimizes the background relevance, while the foreground loss enforces the LRP total absolute foreground relevance ($g(\mathbf{H_{bk}})$, Equation \ref{gLoss}) to stay within a natural range ([$C_{1}$,$C_{2}$]). In all experiments with the DenseNet121 backbone we set $C_{1}=1$ and $C_{2}=25$, except for Appendix \ref{CheXpertClassification}, where we reduced $C_{1}$ to 0.1. For the VGG-19 backbone (dog breed classification task), we employed $C_{1}=0.05$ and $C_{2}=0.5$. The ISNet does not seem very sensible to the choice of these hyper-parameters, as tuning them beyond the specified values did not lead to significant accuracy improvements.

In the experiments without synthetic background bias, we employed the same hyper-parameters that we tuned using the artificially biased training data. The idea behind this decision is that hyper-parameters tuned to hinder background attention in a dataset with synthetic background bias should also hinder background focus in the non-synthetically biased version of the same dataset because its background bias is less extreme. Moreover, due to the similarity between the two tasks, we used the hyper-parameters tuned for COVID-19 detection in the tuberculosis detection application. If one does not want to resort to synthetic background bias to define hyper-parameters, P and d can also be tuned to maximize performance on an out-of-distribution validation dataset. This procedure may result in less aggressive hyper-parameter settings (e.g., lower P in the ISNet) since hyper-parameters would not be tuned in a database with extreme synthetic background bias. However, the alternative procedure requires an o.o.d. validation dataset. We tested other hyper-parameter settings using a grid search in the applications without synthetic bias, applying the previously explained two-stages search protocol. Still, they could not significantly improve generalization, confirming the adequacy of our hyper-parameter tuning strategy. As an illustration, we would expect the tuning strategy to overestimate P in datasets without background bias. Appendix \ref{supApplications} shows applications where quantitative results indicate weak or no background bias (facial attribute estimation without synthetic bias and CheXPert\cite{irvin2019chexpert} classification). The ISNets trained in such applications had their hyper-parameters tuned on synthetically biased versions of the datasets using the procedure described in this section. The resulting parameters were P=0.7 and d=1 in CheXPert and P=0.5 and d=0.996 in facial attribute estimation. In both cases, the ISNet's average test performance had confidence interval overlaps with a standard DenseNet121 (Tables \ref{FacesPerformance} and \ref{chexpertAUCs}), equivalent to an ISNet with P=0. Thus, the performance gain obtained by reducing P in the dataset without strong background bias was not large.

The ISNet Grad*Input, RRR, multi-task U-Net, and GAIN also have loss hyper-parameters. We tuned them using the same strategy we employed to define the ISNet P and d hyper-parameters. I.e., relying on the experiments with synthetic background bias and a validation dataset without the artificial bias. The RRR $\lambda_{1}$ parameter controls the strength of RRR's right reasons loss\cite{RRR} term. The original paper presents experiments with large variations of the parameter, up to $10^{7}$. We searched for $\lambda_{1}$ considering powers of 10, from $1$ to $10^{20}$, when the classification task was ignored, and the model fell to the accuracy levels of random guessing. The multi-task U-Net minimizes a loss term, $L$, defined as  $L=P.L_{S}+(1-P).L_{C}$, where $L_{S}$ is the segmentation loss, and $L_{C}$ the classification loss. We first searched for P in intervals of 0.1, from 0.1 to 0.9. Afterwards, we used a fine search around the previous best P value, varying it by 0.05. If the best result was 0.95, we further tried 0.99, 0.999 and 0.9999 (when the model ignored the classification task). GAIN utilizes three losses, the classification loss, attention mining loss, and external supervision loss. We kept the classification loss weight as 1, and we employed a grid search for the other two weights, considering values of 1, 10, 100 and 1000.

\subsection{Considerations on Transfer Learning}
\label{TransferLearning}

In this study, all DNNs were initialized with PyTorch's standard random weight initialization procedures. In this section, we discuss the possible effects of pretraining on shortcut learning and attention to background bias. Accordingly, we justify why we chose not to utilize transfer learning in this study. In summary, transfer learning consists in training a neural network on one dataset (pretraining), and then training it again (fine-tuning) on another database, which is usually smaller than the first one. Its main objective is to use generalizable knowledge learned in the first dataset to improve performance and generalization in the task represented by the second database.

Pretrained neural networks have already learned to analyze image features that appeared in their pretraining datasets. Therefore, we expect a pretrained model to begin the fine-tuning procedure with a predilection to focus on features that are similar to what it learned in the pretraining dataset. If the pretraining data is more similar to the fine-tuning dataset’s background bias than to the database’s foreground features, we expect the neural network to start fine-tuning with a predilection to focus on the bias, increasing shortcut learning. Instead, if the pretraining dataset is more related to the fine-tuning database’s foreground features, transfer learning could reduce shortcut learning. If the pretraining data features are unrelated to both the foreground and the background in fine-tuning, transfer learning should have little effect over shortcut learning. In case pretraining features are related to both the foreground and the background, pretraining may help the DNN analyze both the foreground features and the background bias, making it difficult to predict what will have the strongest influence over the classifier. In summary, the consequence of pretraining on foreground focus and shortcut learning is highly dependent on the pretraining and fine-tuning datasets’ characteristics. Accordingly, if we employ experiments with pretrained models to analyze shortcut learning, the experiments’ results and conclusions will lose generality and become more data-dependent. Therefore, pretraining will make it more difficult to draw general conclusions about different architectures’ resistance to shortcut learning.

We do not want to draw conclusions that depend on the pretraining data being more similar to the foreground features than to background bias, because we have no guarantee that this assumption will hold in many realistic scenarios. For example, we cannot guarantee that ImageNet (a natural image dataset) pretraining will prompt attention to the lungs (foreground) in X-ray classification. Past studies have shown remarkable attention to background bias (e.g., text and markers) and strong shortcut learning in neural networks pretrained on ImageNet and fine-tuned for COVID-19 classification with chest X-rays\cite{ShortcutCovid}. A previous work observed a concerning quantity of LRP relevance outside of the lungs, even though their classifier was pretrained on ImageNet, then trained again on a large single-source X-ray dataset, and finally fine-tuned for COVID-19 classification in a mixed source database\cite{FirstPaperCovid}. For example, in the study, LRP revealed that words exclusively present in the COVID-19 class (e.g., Italian words) strongly attracted the classifier's attention\cite{FirstPaperCovid}. In summary, an ImageNet-pretrained model could show little shortcut learning in an experiment where foreground features are very similar to ImageNet features. This result could lead to the conclusion that the network generalizes well and is resistant to background bias. However, if this strong generalization is not inherent to the architecture but results from pretraining, it may not hold in the COVID-19 detection task. Thus, the conclusion from the first experiment would have limited generality.

Even in the specific cases where pretraining causes foreground focus, we have no guarantee that this predilection will persist throughout the entire fine-tuning procedure. Decision rules based on background bias may represent a much lower classification loss than rules based on foreground features. Thus, without specific strategies to avoid background attention, a pretrained classifier may eventually arrive at the lower loss minima corresponding to biased decision rules. This possibility further limits the generality of shortcut learning analyses in pretrained models: we could conclude that a pretrained network hinders background bias attention, but different fine-tuning hyper-parameters (e.g., learning rate) could allow it to find loss minima representing biased decision rules.

In summary, the pretraining influence on shortcut learning depends on the characteristics of the pretraining and fine-tuning datasets. Moreover, we cannot guarantee that any reduction in shortcut learning caused by pretraining will not be lost during the fine-tuning procedure. In this study, we introduced a deep neural architecture and designed experiments to assess its inherent resistance to background bias relative to the state-of-the-art. To maximize the generality of the experiments’ conclusions, we initialized all models with standard random parameter initializations, ensuring that the networks begin with no data-dependent predilection to focus on the images' background or foreground. Accordingly, the experiments’ results produced consistent patterns from which we could draw more robust and general conclusions about the ISNet. The study of how pretraining influences shortcut learning in multiple deep neural network architectures is an important research topic for future work, which requires careful consideration of the datasets used and the generality of the conclusions drawn.

We finish this section with a quantitative example of how the effect of pretraining on shortcut learning is highly data-dependent. We compare the consequences of pretraining on the Vision Transformer (ViT-B/16) in two tasks, facial attribute estimation and COVID-19 detection, both with synthetic background bias. We consider the standard PyTorch’s ImageNet-1K pretrained model. We do not employ dog breed classification because the Stanford Dogs images were originally extracted from ImageNet\cite{StanfordDogs}. Thus, considering the rare scenario where the pretraining and fine-tuning datasets come from the same data distribution would lead to even less general conclusions.

In the synthetically biased facial attribute estimation task, the randomly initialized Vision Transformer obtained macro-average F1-Scores (with 95\% confidence intervals, refer to Appendix \ref{statisticalMethods} for details about the statistical analysis) of 0.675 +/-0.023, 0.645 +/-0.03, and 0.531 +/-0.023  in the biased, standard and deceiving bias test datasets, respectively (main article Table 1). The same model pretrained on ImageNet (and fine-tuned following the specifications in Appendix \ref{training}) achieved maF1 of 0.904 +/-0.019, 0.726 +/-0.04, and 0.456 +/-0.018 in the three testing scenarios, respectively. The improvement in the standard test dataset (which has no synthetic bias) result indicates that pretraining allowed the network to better analyze the foreground features (faces). This finding is expected, given that faces are also common in ImageNet. However, the worse result in the deceiving bias evaluation shows that the model’s decisions are still strongly influenced by the synthetic background bias. Indeed, the higher F1-Score in the biased test dataset indicates that pretraining even helped the DNN better analyze the background bias.

In synthetically biased COVID-19 detection, the randomly initialized Vision Transformer achieved macro-average F1-Score (with standard deviation) of 0.685 +/-0.009, 0.496 +/-0.009, and 0.327 +/-0.009 in the biased, standard, and deceiving bias test datasets, respectively (main article Table 1). With ImageNet pretraining (and fine-tuning as specified in Appendix \ref{training}), the scores changed to 0.963 +/-0.004, 0.421 +/-0.009, and 0.031 +/-0.003. As in facial attribute estimation, pretraining helped the Vision Transformer analyze the synthetic background bias, improving the biased test performance and reducing the deceiving bias F1-Score. However, pretraining did not improve the standard test result in COVID-19 detection. Thus, it could not help the network analyze the foreground (lung) features. Conversely, in facial attribute estimation, pretraining improved the analysis of foreground features (faces), increasing the standard test F1-Score. The diverse consequences of pretraining on the two tasks are justified by the fact that ImageNet (a natural image dataset) features are much more similar to the faces in CelebA than to the lung features in the COVID-19 X-ray dataset. Accordingly, these differences exemplify the fact that using pretrained models would make the results of our experiments less general and more data-dependent.

\subsection{Challenges and Implementation Notes}
\label{ImpNotes}

Fundamentally, the ISNet architecture restrains the classifier during training, conditioning it to analyze only the relevant part of the image. This is the case even when there are undesired background features that could allow the model to reduce the classification loss rapidly and easily. Therefore, any change in the ISNet architecture needs to be cautiously implemented, because it can create a new and unintended way for the classier to minimize the heatmap loss. Namely, as a flexible model, the classifier will look for the easiest way to reduce $L_{LRP}$, which may be by cheating and finding a strategy to hide the undesired relevance in the heatmaps.

The ISNet performance is significantly affected by the definition of the heatmap loss. New LRP-based loss functions, tailored for specific applications, are a promising path for future research. However, some functions may prompt the aforementioned cheating solutions. For example, we may create a simple heatmap loss formulation, penalizing a ratio between the undesired relevance and the relevance inside the region of interest. If there is no mechanism to avoid this solution, the classifier may minimize $L_{LRP}$ by artificially boosting the relevance inside the region of interest. We tested this alternative heatmap loss, which was still based on the absolute value of the heatmaps (to minimize positive and negative undesired relevance). To optimize the function, the DNN created a fine and regular chessboard pattern inside the zones of interest, alternating between strong positive and strong negative relevance. The artificial pattern allows the region of interest to equally affect each class in the classifier, making it meaningless for the classification task. However, the high absolute values in the pattern strongly reduce the ratio in the alternative heatmap loss, allowing its minimization while the classifier focuses on the image background with a comparatively small quantity of absolute relevance. 
 
Regarding implementation, we need to be careful with in-place operations inside the classifier, since they may change values that will later be processed by the LRP block. Using PyTorch's forward hooks, we can easily store all relevant variables during the forward propagation through the classifier, allowing their later access by the LRP block. This is an effortless way to create the skip connections between the classifier and the LRP block, even if the classifier is already defined and instantiated.

The reliability of the standard segmentation-classification pipeline depends on the proper function of its segmenter. Naturally, if the segmenter removes parts of the image foreground, the classification scores will be compromised. Moreover, the improper removal of the background also affects the classifier performance. E.g., our experiments demonstrate that, if we skip the segmentation step from the pipeline after training, classification performance drops dramatically (main article section ``COVID-19 detection''). Analogously to the traditional segmentation-classification approach, the ISNet relies on the proper operation of its implicit segmentation. I.e., to accurately classify an image, the ISNet must be able to identify its foreground.

The ISNet backbone architectures in this study are the VGG-19\cite{vggOriginal} (only in dog breed classification) and DenseNet121\cite{DenseNet} (remaining applications). We modified the original DenseNet121 classifier by substituting its last layer with a layer containing the same number of neurons as the number of classes in the application, preceded by dropout of 50\%. The only change to the VGG-19 was in its last layer, using 3 neurons to classify the 3 possible dog breeds. The same backbones were used for RRR, GAIN, HAM, ISNet Grad*Input, segmentation-classification pipeline, and the standard classifier (Appendix \ref{BaselineImplementations}).

The ISNet code is mainly based on PyTorch (version 1.11.0), PyTorch Lightning (1.6.3), and Python (3.9). All benchmark models were also based on the same packages. Additional supporting and required Python packages were CUDA (11.3.1), cuDNN (8.2.0), NumPy (1.21.5), TorchVision (0.12.0), Matplotlib (3.5.1), opencv-python (4.5.2), SciPy (1.7.3), scikit-image (0.19.2), scikit-learn (0.23.2), pandas (1.4.2), dill (0.3.4), Jupyter Notebook (4.7.1), and PyTorch Grad-CAM (1.4.6). Bayesian analyses were performed with PyMC3 (3.1). The DeepMAC\cite{deepMAC} network used to create segmentation ground-truth masks for the Stanford Dogs dataset employed TensorFlow (2.10.0).


\section{Supplementary Applications: better Delimiting the ISNet Use-case}
\label{supApplications}

This Section presents experiments that are designed to better delimit the ISNet's use case scenario. The model was built to avoid the influence of background bias on classifier's decisions, hindering shortcut learning and improving o.o.d. generalization. Here we show two applications that are not characterized by background bias. The first is facial attribute estimation without synthetic background bias. We use it to simulate an in-domain task, with i.i.d. training and testing datasets. The second is lung disease classification with o.o.d. testing, but learning from a X-ray dataset collected from a single hospital. As the database is not mixed, it should have less background bias than the COVID-19 or tuberculosis datasets.

\subsection{Facial Attribute Estimation}

Past works did not necessarily search for, or found, an accuracy improvement when using face segmentation before facial attribute estimation. However, they pointed out that background features may influence classifiers\cite{FaceBias}\textsuperscript{,}\cite{FacesSmallSets}. Therefore, minimizing background attention can reduce bias and improve security, which is crucial when the classified attributes will be employed for user identification\cite{FaceBias}. In this work, the results on the synthetically biased facial attribute estimation dataset showed that the ISNet can precisely contain its attention inside the foreground when evaluating natural images, considering a dataset where finding the images' foreground (faces) would constitute a challenging semantic segmentation problem. Here, we train the neural networks in the same facial attribute estimation dataset (Appendix \ref{dataset}), but without the synthetic bias. Again, we classify only 3 facial attributes (rosy cheeks, high cheekbones and smiling), increasing the difficulty of learning an adequate attention profile\cite{celebA}. Training procedure, data processing, hyper-parameters, and neural network architectures are exactly those used in the synthetically biased facial attribute estimation application. This experiment will lead to an analysis of the impact of segmentation-based attention mechanisms in an in-domain application, with standard i.i.d. testing. Thus, it will better delimit the ISNet's objective and use-case scenario. 

Table \ref{FacesPerformance} shows the test performance of all DNNs in the facial attribute estimation task. The table cells report the metrics' mean and error, considering 95\% confidence. Appendix \ref{statisticalMethods} describes the statistical methods used. The U-Net in the alternative segmentation-classification pipeline achieved a test IoU of 0.925, and the face segmentation masks seemed adequate upon the visual inspection. In facial attribute estimation, removing the U-Net before testing the alternative segmentation methodology causes a devastating effect, as it did in COVID-19 detection: it reduced maF1 from 0.806 +/-0.027 to 0.543 +/-0.191. 

\begin{table}[!h]
\centering
\small
\caption{Performance metrics for the deep neural networks in facial attribute estimation}
\label{FacesPerformance}
\begin{tblr}{
  width = \linewidth,
  colspec = {Q[290]Q[150]Q[179]Q[150]Q[167]},
  hlines,
  vlines,
}
Model and Metric                            & Rosy Cheeks    & High Cheekbones & Smiling        & {Mean\\(macro-average)} \\
ISNet Precision                             & 0.59 +/-0.048  & 0.843 +/-0.019  & 0.937 +/-0.013 & 0.79 +/-0.027           \\
U-Net+DenseNet121 Precision                 & 0.615 +/-0.048 & 0.863 +/-0.018  & 0.905 +/-0.016 & 0.794 +/-0.027          \\
DenseNet121 Precision                       & 0.624 +/-0.05  & 0.876 +/-0.018  & 0.926 +/-0.014 & 0.809 +/-0.027          \\
Multi-task U-Net Precision                  & 0.652 +/-0.049 & 0.909 +/-0.016  & 0.937 +/-0.013 & 0.833 +/-0.026          \\
AG-Sononet Precision                        & 0.545 +/-0.044 & 0.896 +/-0.017  & 0.911 +/-0.015 & 0.784 +/-0.025          \\
Extended GAIN Precision                     & 0.644 +/-0.052 & 0.851 +/-0.019  & 0.902 +/-0.016 & 0.799 +/-0.029          \\
RRR Precision                               & 0.59 +/-0.05   & 0.839 +/-0.019  & 0.836 +/-0.019 & 0.755 +/-0.029          \\
{Vision Transformer (ViT-B/16) \\Precision} & 0.476 +/-0.046 & 0.664 +/-0.023  & 0.693 +/-0.023 & 0.611 +/-0.031          \\
ISNet Recall                                & 0.71 +/-0.04   & 0.828 +/-0.02   & 0.876 +/-0.018 & 0.805 +/-0.027          \\
U-Net+DenseNet121 Recall                    & 0.707 +/-0.045 & 0.843 +/-0.019  & 0.912 +/-0.015 & 0.821 +/-0.026          \\
DenseNet121 Recall                          & 0.671 +/-0.048 & 0.822 +/-0.021  & 0.897 +/-0.016 & 0.797 +/-0.028          \\
Multi-task U-Net Recall                     & 0.695 +/-0.048 & 0.792 +/-0.023  & 0.914 +/-0.015 & 0.8 +/-0.029            \\
AG-Sononet Recall                           & 0.79 +/-0.036  & 0.816 +/-0.021  & 0.926 +/-0.014 & 0.844 +/-0.024          \\
Extended GAIN Recall                        & 0.617 +/-0.053 & 0.818 +/-0.021  & 0.901 +/-0.016 & 0.779 +/-0.03           \\
RRR Recall                                  & 0.65 +/-0.049  & 0.836 +/-0.02   & 0.913 +/-0.014 & 0.8 +/-0.028            \\
{Vision Transformer (ViT-B/16) \\Recall}    & 0.63 +/-0.045  & 0.801 +/-0.019  & 0.786 +/-0.021 & 0.739 +/-0.028          \\
ISNet F1-Score                              & 0.644 +/-0.047 & 0.835 +/-0.02   & 0.905 +/-0.016 & 0.795 +/-0.028          \\
U-Net+DenseNet121 F1-Score                  & 0.658 +/-0.047 & 0.853 +/-0.019  & 0.908 +/-0.016 & 0.806 +/-0.027          \\
DenseNet121 F1-Score                        & 0.647 +/-0.049 & 0.848 +/-0.02   & 0.911 +/-0.015 & 0.802 +/-0.028          \\
Multi-task U-Net F1-Score                   & 0.673 +/-0.049 & 0.846 +/-0.02   & 0.925 +/-0.014 & 0.815 +/-0.028          \\
AG-Sononet F1-Score                         & 0.645 +/-0.043 & 0.854 +/-0.019  & 0.918 +/-0.015 & 0.806 +/-0.026          \\
Extended GAIN F1-Score                      & 0.63 +/-0.053  & 0.834 +/-0.02   & 0.901 +/-0.016 & 0.788 +/-0.03           \\
RRR F1-Score                                & 0.619 +/-0.05  & 0.837 +/-0.02   & 0.873 +/-0.017 & 0.776 +/-0.029          \\
{Vision Transformer (ViT-B/16) \\F1-Score}  & 0.542 +/-0.047 & 0.726 +/-0.022  & 0.737 +/-0.022 & 0.668 +/-0.03           \\
ISNet AUC                                   & 0.936 +/-0.011 & 0.922 +/-0.01   & 0.97 +/-0.005  & 0.943 +/-0.009          \\
U-Net+DenseNet121 AUC                       & 0.942 +/-0.01  & 0.938 +/-0.008  & 0.979 +/-0.004 & 0.952 +/-0.008          \\
DenseNet121~AUC                             & 0.945 +/-0.01  & 0.939 +/-0.008  & 0.98 +/-0.004  & 0.955 +/-0.007          \\
Multi-task U-Net~AUC                        & 0.907 +/-0.018 & 0.912 +/-0.011  & 0.962 +/-0.007 & 0.927 +/-0.012          \\
AG-Sononet~AUC                              & 0.945 +/-0.01  & 0.939 +/-0.008  & 0.978 +/-0.004 & 0.954 +/-0.007          \\
Extended GAIN~AUC                           & 0.945 +/-0.011 & 0.939 +/-0.008  & 0.982 +/-0.004 & 0.955 +/-0.008          \\
RRR AUC                                     & 0.899~+/-0.018 & 0.924~+/-0.009  & 0.951~+/-0.007 & 0.925~+/-0.011          \\
{Vision Transformer (ViT-B/16) \\AUC}       & 0.898~+/-0.015 & 0.798~+/-0.016  & 0.821~+/-0.015 & 0.839~+/-0.015          
\end{tblr}
\end{table}

We observe that, in this task, the CNNs have similar performances, presenting strong overlap in the 95\% confidence intervals for macro-averaged F1-Score in Table \ref{FacesPerformance}. Forcing a model to not consider the background may reduce performance on the i.i.d. test database, by making the classifier ignore background bias that could artificially help classification. Indeed, models affected by shortcut learning perform well on the standard i.i.d. benchmarks\cite{ShortcutLearning}\textsuperscript{,}\cite{ShortcutCovid}. Table \ref{FacesPerformance} displays no strong positive or negative impact of segmentation-based attention mechanisms on performance scores (e.g., ISNet and segmentation-classification pipeline). Thus, the results indicate that the CelebA\cite{celebA} dataset background features do not have a strong correlation with the images' classes, making it easier for standard classifiers to naturally focus on the faces.

\begin{figure}[!h]
\includegraphics[width=1\textwidth]{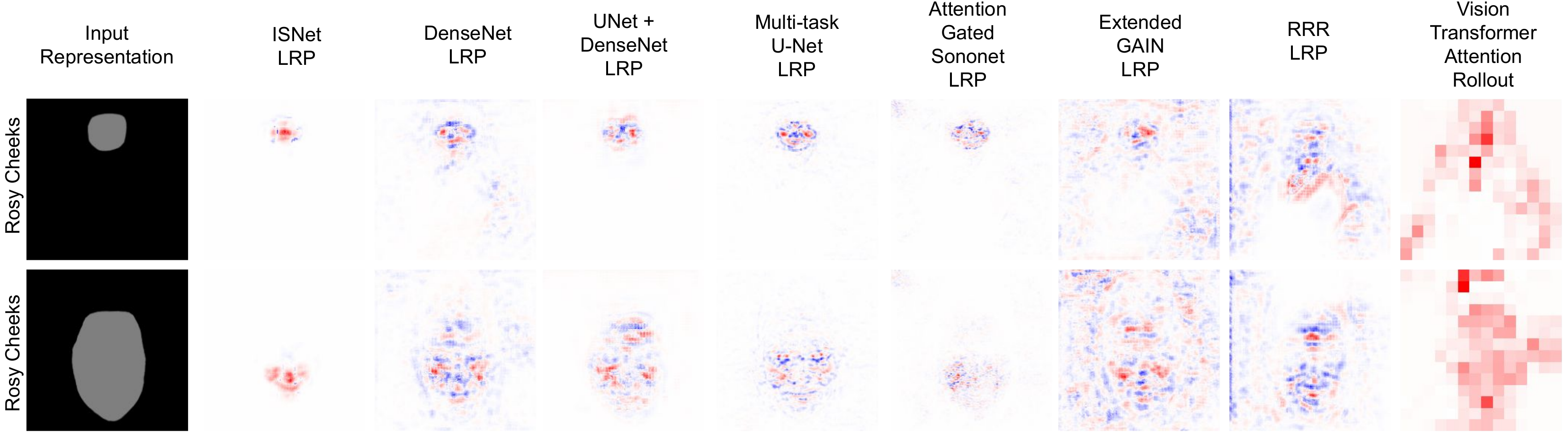}
\centering
\caption{\textbf{CelebA test images' heatmaps (attention rollout for Vision Transformer and Layer-wise Relevance Propagation/LRP for remaining networks).}  For LRP, red colors indicate areas that the DNN associated to the true class rosy cheeks, blue colors are areas that decreased the network confidence for the class. For attention rollout, red indicates the DNN attention. White represents areas with little influence over the classifiers. The two photographs in are labeled as containing the 3 considered attributes (rosy cheeks, high cheekbones and smiling). For privacy, the pictures were substituted by a representation of the faces' locations (gray), but classifiers received the real photographs}
\label{facesmaps}
\end{figure}

Accordingly, Figure \ref{facesmaps} shows that the DenseNet121 background attention is much less intense in this task than in the synthetic bias experiments, COVID-19 detection, or tuberculosis detection (main article Figures 1 and 2). We see that the standard DenseNet121 pays some attention to body features outside of the face, and to the remaining background. However, most of the model's focus is on the face. Attention gated CNNs were designed to be more resistant to background clutter\cite{AGNet}. Additionally, past studies suggest that, considering images with cluttered backgrounds, multi-task models learning classification and segmentation can achieve better performances for the two tasks\cite{MultiTask1}\textsuperscript{,}\cite{MultiTask2}. When comparing the two models with the DenseNet121, we observe that their attention is more contained inside the face regions. Thus, multi-task learning and attention gates can be useful to deal with background clutter that is not correlated with the samples' classes. Figure \ref{facesmaps} also exemplifies the wide diversity of segmentation mask sizes in the CelebA dataset.

 The ISNet was designed to minimize the influence of background features on classifiers, hindering shortcut learning and improving o.o.d. generalization. Accordingly, the model did not improve accuracy for an in-domain problem with little background bias. However, Figure \ref{facesmaps} shows that the ISNet precisely focused on the photographs' region of interest (the faces), considering natural images with cluttered backgrounds and a wide variety of segmentation masks' shapes, locations, and sizes. By accurately focusing on the faces, the ISNet represents a security improvement for facial attribute estimation. Indeed, when synthetic bias was added to the data, the ISNet did not allow it to influence the classifier's decisions. Thus, it became the best performing DNN, as benchmark neural networks lost performance (main article Table 1).

\subsection{CheXPert classification and comparison to Hierarchical Attention Mining}
\label{CheXpertClassification}

The two X-ray classification problems previously analyzed (COVID-19 and TB detection) clearly demonstrate background bias causing shortcut learning. Therefore, they exemplify situations where the ISNet use is very advantageous. Both problems consider mixed source training datasets, where X-rays from different classes were produced by diverse hospitals. In this section we examine if a large single-source X-ray database is also prone to background bias. Thus, we assess if the task fits in the ISNet main use case: to hinder background bias attention and the consequent shortcut learning.

We consider a subset of CheXPert\cite{irvin2019chexpert} as the training dataset. The utilization of a subset reduces the experiments' computational cost, without detracting from the validity of our findings. As discussed in Appendix \ref{dataset}, the database is exceptionally large, and all its X-rays come from the same source, the Stanford University Hospital (California, United States of America). We selected all CheXPert frontal X-rays that contained the lung conditions atelectasis, consolidation, edema, pneumonia, and pneumothorax, along with the X-rays with the no finding label. In total, 112179 images were gathered; 29795 radiographs were positive for atelectasis, 13015 for consolidation, 49717 for edema, 17000 for no finding, 4683 for pneumonia, and 17700 for pneumothorax. The subset patients' mean age was 59.7 years, with standard deviation of 18.6 years. They were 55.8\% male. Hold-out validation utilized 25\% of the images, randomly selected.

For o.o.d. testing, we chose the MIMIC-CXR-JPG database (v2.0.0)\cite{MIMIC}\textsuperscript{,}\cite{MIMIC1}\textsuperscript{,}\cite{PhysioNet}. It is another large open dataset, which contains a set of labels with the same structure as CheXPert, considered in this study. Its 377110 X-rays were gathered from the Beth Israel Deaconess Medical Center (Massachusetts, United States of America), between 2011 and 2016. Labels were also created with natural language processing, constructed according to radiologists' reports. We treated absent and uncertainty labels as negative, as in CheXPert. To create the external evaluation database, we selected a subset of MIMIC's original test database, consisting of all frontal X-rays that were labeled positive for atelectasis, consolidation, edema, no finding, pneumonia, or pneumothorax. According to this strategy, 2265 samples were gathered; 763 were positive for atelectasis, 210 for consolidation, 722 for edema, 593 for no finding, 340 for pneumonia, and 108 for pneumothorax. The MIMIC database does not inform patient age or sex.

We trained the ISNet, DenseNet121, segmentation-classification pipeline (U-Net followed by DenseNet121), multi-task U-Net, RRR, Vision Transformer and AG-Sononet using the CheXPert subset. The numerical labels used in training were defined according to the alphabetical order of the 6 possible classes. The DNNs' definitions and implementations were the same as for the COVID-19 and TB detection tasks (refer to Appendix \ref{BaselineImplementations} and main article section ``Background relevance minimization and ISNet''). However, CheXPert and MIMIC classification are multi-label problems. Thus, binary cross-entropy was used as the classification loss function. The X-ray processing, augmentation procedure, training and hyper-parameter tuning strategies followed the guidelines in Appendix \ref{ImpDetails}. Specific changes in relation to the strategy used for COVID-19 detection were training for 96 epochs and using mini-batches from 16 to 32 samples. Again, we set $10^{-3}$ as the learning rate. Like in the other X-ray classification tasks, we utilized the U-Net trained in a previous study\cite{bassi2021covid19} to produce the ground-truth lung masks for CheXPert, binarizing its output according to a threshold of 0.4. The U-Net in the segmentation-classification pipeline was trained for 400 epochs, using such targets and learning rate of $10^{-4}$. The ISNet loss hyper-parameters were: d=1, P=0.7, and w$_{1}$=w$_{2}$=1. The RRR right reasons\cite{RRR} loss weight ($\lambda_{1}$) was 10. The multi-task U-Net loss balancing hyper-parameter was P=0.75.

We did not train GAIN\cite{GAIN} for this task, as the model training time was exceedingly long, due to the large dataset and the multi-label classification task, which greatly reduced the model's training speed (refer to Appendix \ref{speed}). Instead, we used the new application to implement a Hierarchical Attention Mining\cite{HAM} (HAM) DNN. Like GAIN, HAM utilizes Grad-CAM optimization to guide classifier attention. Moreover, both architectures' main objective is to produce classifiers' Grad-CAM explanations that are useful for weakly-supervised localization. However, HAM's loss functions are tailored for the task of lung lesion localization, especially in datasets containing lesion annotations. In the previous experiments, we observed that DNNs that do not learn from segmentation targets were ineffective against background bias. For HAM to consider such targets, lesion annotations must be available. Such annotations are rare for X-ray datasets and are not available for large TB and COVID-19 databases, such as the ones we used. For this reason, HAM was not implemented for the previous X-ray classification tasks. However, the creators of HAM openly provided 6099 bounding-boxes (created by a radiologist) for 2345 CheXPert images\cite{HAM}. Therefore, we compare HAM to the ISNet in the task of CheXPert classification. To create more precise segmentation targets for the lesions (used only in HAM training), we employed the intersection of the provided bounding-boxes and the lung masks we produced for CheXPert. For the CheXPert subset, there were a total of 1769 lesion segmentation targets, 501 for atelectasis, 255 for consolidation, 543 for edema, 195 for pneumonia, and 275 for pneumothorax.

HAM combines multiple strategies to control the classifier attention. First, the model has a foreground attention block (FAB), a self-attention mechanism based on channel and spatial attention\cite{HAM}\textsuperscript{,}\cite{attentionSurvey}. As the AG-Sononet, such mechanism does not rely on segmentation targets, thus learning to focus on image regions that improve classification loss. HAM is a hierarchical classifier, it has one output classifying if there is any abnormality in the X-ray, along with outputs for each possible abnormal condition. A Grad-CAM heatmap of the binary output is called positive attention map, while the Grad-CAM heatmaps of the other outputs are named abnormality maps. The attention bound loss\cite{HAM} enforces the abnormality maps to be contained inside the positive attention map. The attention union loss\cite{HAM} compels the union of the abnormality maps to match the positive map. Finally, adaptive mean square error (AMSE) loss\cite{HAM} penalizes differences between the available lesion segmentation targets and the abnormality maps. Our HAM implementation followed the official public implementation\cite{HAM}, which uses a ResNet50\cite{ResNet} backbone. Unlike the original paper, we employed the input size of 224x224, matching the other DNNs. Training parameters also followed the strategy used for the other models, allowing a fairer comparison. HAM loss hyper-parameters followed the values in the original paper\cite{HAM}.

\subsubsection {Results and Discussion}

We express the o.o.d. test results (MIMIC dataset) in terms of AUC (Area Under the ROC Curve), as it is the standard for reporting MIMIC and CheXPert performances\cite{irvin2019chexpert}\textsuperscript{,}\cite{HAM}. Table \ref{chexpertAUCs} displays per-class and macro-average AUC scores, considering 95\% confidence intervals. Statistical methods were the same as those used for facial attribute estimation (Appendix \ref{statisticalMethods}), as both tasks are multi-label classification problems. The ISNet achieved 0.735 +/-0.009 macro-average AUC for the i.i.d. validation dataset (CheXPert), the segmentation-classification pipeline 0.779 +/-0.008, the DenseNet121 0.802 +/-0.008, the multi-task U-Net 0.812 +/-0.007, the AG-Sononet 0.804 +/-0.008, HAM 0.792 +/-0.008, RRR 0.759 +/-0.008, and the Vision Transformer 0.733 +/-0.009. The validation performances may be seen as an optimistic approximation of the i.i.d. test performance, as the loss in the validation dataset was utilized to select the best performing DNNs.

\begin{table}[!h]
\centering
\small
\caption{Test AUC for the deep neural networks trained with CheXPert and evaluated on MIMIC}
\label{chexpertAUCs}
\begin{tblr}{
  width = \linewidth,
  colspec = {Q[127]Q[104]Q[125]Q[85]Q[104]Q[108]Q[133]Q[150]},
  hlines,
  vlines,
}
Model                               & Atelectasis        & Consolidation      & Edema              & No Finding         & Pneumonia          & Pneumothorax       & {Mean\\(macro-average)} \\
ISNet                               & {0.633 \\+/-0.024} & {0.656 \\+/-0.038} & {0.773 \\+/-0.022} & {0.764 \\+/-0.02}  & {0.606 \\+/-0.034} & {0.69 \\+/-0.054}  & {0.687 \\+/-0.032}      \\
{U-Net+\\DenseNet121}               & {0.64\\+/-0.023}   & {0.688\\+/-0.039}  & {0.795\\+/-0.019}  & {0.774\\+/-0.022}  & {0.651\\+/-0.033}  & {0.777\\+/-0.041}  & {0.721 \\+/-0.03}       \\
DenseNet121                         & {0.64 \\+/-0.023}  & {0.663 \\+/-0.04}  & {0.804 \\+/-0.018} & {0.782 \\+/-0.022} & {0.645 \\+/-0.033} & {0.783 \\+/-0.043} & {0.72 \\+/-0.03}        \\
{Multi-task \\U-Net~}               & {0.686 \\+/-0.022} & {0.713 \\+/-0.037} & {0.807 \\+/-0.018} & {0.788 \\+/-0.022} & {0.666 \\+/-0.031} & {0.798 \\+/-0.046} & {0.743 \\+/-0.029}      \\
AG-Sononet                          & {0.645 \\+/-0.023} & {0.624 \\+/-0.040} & {0.805 \\+/-0.018} & {0.792 \\+/-0.022} & {0.646 \\+/-0.032} & {0.771 \\+/-0.05}  & {0.714 \\+/-0.031}      \\
HAM                                 & {0.629 \\+/-0.023} & {0.675 \\+/-0.039} & {0.795 \\+/-0.018} & {0.791 \\+/-0.022} & {0.637 \\+/-0.033} & {0.755 \\+/-0.049} & {0.714 \\+/-0.031}      \\
RRR                                 & {0.618\\+/-0.024}  & {0.659\\+/-0.039}  & {0.793\\+/-0.019}  & {0.769\\+/-0.022}  & {0.616\\+/-0.034}  & {0.692\\+/-0.049}  & {0.691\\+/-0.031}       \\
{Vision \\Transformer~\\(ViT-B/16)} & {0.569\\+/-0.025}  & {0.630\\+/-0.038}  & {0.766\\+/-0.02}   & {0.751\\+/-0.023}  & {0.601\\+/-0.033}  & {0.619\\+/-0.049}  & {0.656\\+/-0.031}       
\end{tblr}
\end{table}

In Table \ref{chexpertAUCs}, all DNNs show similar o.o.d. test AUCs, and almost all macro-average AUCs have some overlap in their 95\% confidence intervals. Our previous experiments demonstrated that the ISNet and the segmentation-classification pipeline are the two DNNs that better ignore background bias, and the two neural networks have the smallest gaps between the CheXPert validation AUC (i.i.d.) and the MIMIC test AUC (o.o.d.). However, the two models' average AUCs could not surpass a standard classifier (DenseNet121). Thus, ignoring the CheXPert background was not exceedingly beneficial for o.o.d. accuracy, unlike what we observed in COVID-19 or TB detection.

CheXPert classification produced relatively small gaps between the i.i.d. validation AUCs (on CheXPert) and the o.o.d. test AUCs (on MIMIC). For the common DenseNet121 classifier, we observed a macro-average AUC drop from 0.802 +/-0.008 to 0.72 +/-0.03. Meanwhile, when comparing i.i.d. and o.o.d. evaluation performance in tuberculosis detection, the same DenseNet121 AUC dropped from 0.999 +/-0.001 in the i.i.d. test to 0.576 +/-0.04 in the o.o.d. evaluation. Past studies demonstrate similarly large gaps for COVID-19 detection\cite{ShortcutCovid}. The drastic differences in the performance gaps indicate that CheXPert training is much less prone to shortcut learning than the two previous X-ray classification problems, supporting the conclusion that the dataset contains less background bias. Moreover, the positive multi-task DNN results (Table \ref{chexpertAUCs}) indicate that CheXPert background features are more representative of clutter than bias, i.e., they have little correlation with the classes. Past studies pointed out that multi-task classification with segmentation may improve results when background clutter is present\cite{MultiTask2}, while our previous experiments indicated that the multi-task model is not resistant to background bias.

Having these considerations in mind, with CheXPert we cannot assess whether HAM is robust to background bias. Thus, we resorted to adding artificial background bias to the CheXPert subset images. We utilized the same geometrical shapes used in the facial attribute estimation task (triangle, circle, and square, as exemplified in main article Figure 1), and placed them in the image corners. As the subset contains 6 possible classes, we vertically divided the 3 geometrical shapes in half, and each new geometrical shape was correlated with one class. After training on the artificially biased database, we tested the neural network on the MIMIC test subset. When the geometrical shapes were present in the test data, HAM's macro-averaged AUC was 0.951 +/-0.009. After the artificial bias removal, the score dropped to 0.627 +/-0.032. Therefore, the classifier decisions were clearly affected by the background bias. Like we observed for GAIN (main article Figure 1), the HAM LRP heatmaps clearly show the attention paid to the geometrical shapes, while its Grad-CAM explanations indicate little focus on the bias. Overall, HAM's Grad-CAM explanations match well with the segmentation targets. Again, the strong influence of the geometrical shapes on the DNN's AUC proves that the LRP explanations' fidelity to the classifier's behavior surpasses Grad-CAM's. HAM and GAIN are two different architectures, whose attention mechanisms are both based on Grad-CAM optimization. Accordingly, both failed in avoiding background bias attention, and produced spurious Grad-CAM explanations. The similar behavior indicates that their failure is caused by fundamental weaknesses in the Grad-CAM methodology, corroborating with our analysis in Appendix \ref{GAINComparison}.

For comparison, we also trained the ISNet on the artificially biased dataset. Again, it minimized the influence of background bias: macro-average AUC was 0.7 +/-0.031 when the test data had the geometrical shapes. When they were removed, the score did not change. Moreover, the performance is remarkably similar to the value in Table \ref{chexpertAUCs}. As in previous experiments, the ISNet's explanations (LRP and Grad-CAM) show no attention to the artificial bias. Finally, as a baseline, we trained a DenseNet121 on the artificially biased CheXPert subset. It had 0.97 +/-0.006 macro-average AUC with the geometrical shapes, and 0.622 +/-0.032 without them. Once more, in the presence of background bias, the ISNet strongly surpassed the other DNNs. HAM's macro-average AUC scores are remarkably similar to the standard DenseNet121's. Moreover, the LRP heatmaps for both DNNs show strong focus on bias. In the other experiments, GAIN's LRP heatmaps revealed attention to bias and the region of interest (main article Figure 1), resulting in a smaller performance gap between evaluation with and without the geometrical shapes. Thus, we conclude that GAIN's attention mining loss (absent in HAM) made the DNN pay attention to input features in the foreground and the background. Meanwhile, the two losses that compare Grad-CAM heatmaps to segmentation targets (GAIN's external supervision and HAM's AMSE) could not effectively avoid background attention.

\begin{figure}[!h]
\includegraphics[width=1\textwidth]{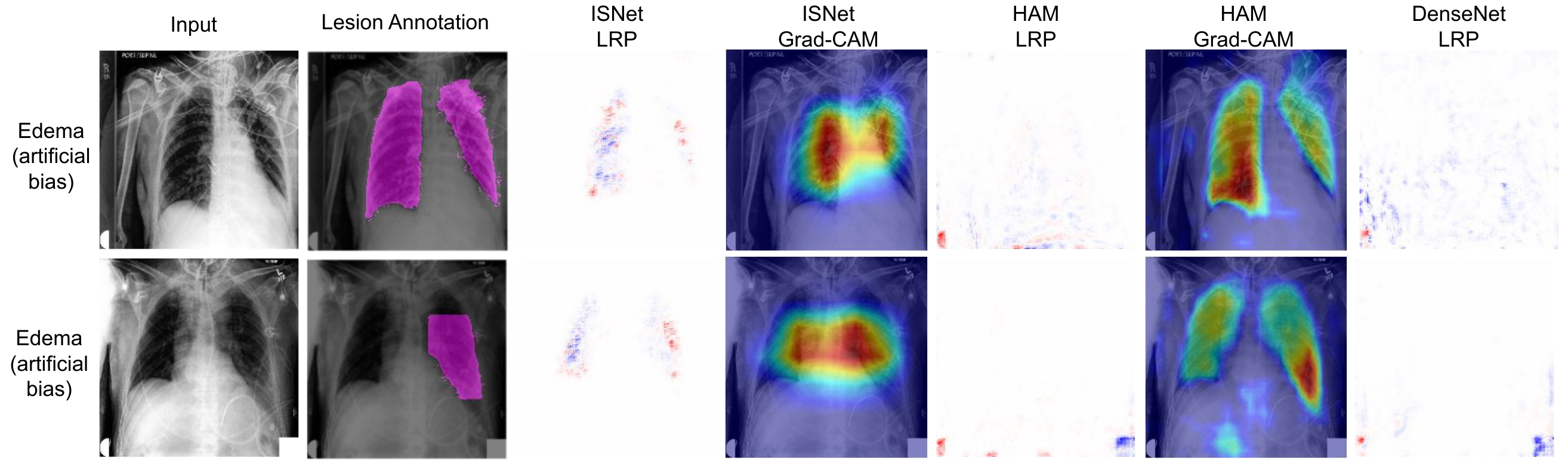}
\centering
\caption{\textbf{Artificially biased validation X-rays, their lesion (edema) annotations (in magenta), and heatmaps for multiple DNNs.} Images are positive for the class edema, associated with the geometrical shape in the X-ray bottom left corner. In LRP (Layer-wise Relevance Propagation) heatmaps, red indicates areas that the DNN associated with edema. Blue areas reduced the edema classification score or were more representative of other conditions. For Grad-CAM (Gradient-weighted Class Activation Mapping), the redder the region, the more it influenced the classifier's decision}
\label{chexBias}
\end{figure}

Figure \ref{chexBias} provides examples of the DNNs' heatmaps for artificially biased images. It shows CheXPert validation X-rays. When it exists, attention to the geometrical shapes should be observable in training, test and validation samples. By choosing validation X-rays, we can display its lesion annotations (created from radiologists' bounding boxes\cite{HAM}). The two X-rays are positive for edema, and the magenta annotations refer to this condition. The semicircle is the geometrical shape associated with edema. The bottom X-ray in Figure \ref{chexBias} is also positive for atelectasis and consolidation (indicated by two adjacent rectangles, which form a square). Notice that the HAM Grad-CAM heatmaps have higher resolution than the ISNet Grad-CAM, because HAM's last convolutional feature map is larger. Attention to the semi-circle is visible in all LRP heatmaps, except for the ISNet's. Moreover, HAM's LRP explanations are similar to the common DenseNet121's. Again, we observe that HAM's Grad-CAM heatmaps hid the attention to the geometrical shapes. They are similar to the lesion annotations, but LRP and HAM's test AUC scores reveal that the model mostly focused on bias. In summary, HAM results with synthetic bias are consistent with spurious mapping (Appendix \ref{GAINComparison}). The other benchmark DNNs were already tested with artificially biased datasets, considering both single-label and multi-label classification (main article Figure 1).

We must note that HAM's authors do not claim robustness against background bias, nor tested this functionality\cite{HAM}. Furthermore, for a fair comparison to the other DNNs in Table \ref{chexpertAUCs}, we trained HAM with subset of the CheXPert dataset, and employed a smaller input size and higher learning rate in relation to those used in the original HAM paper\cite{HAM}, which may have reduced the model's performance. We do not state that Grad-CAM-based methodologies cannot improve attention in any application. Instead, the results in this study show that Grad-CAM-based models are less reliable than LRP optimization (ISNet) the presence of background bias. Please refer to Appendix \ref{GAINComparison} for an in-depth analysis.

The results from the experiments with the unaltered CheXPert images (without the geometrical shapes) point out that strong background bias and the consequent shortcut learning are not intrinsic to X-ray classification. However, many X-ray classification tasks require dataset mixing, as large databases may not contain all classes of interest. The experiments in this study indicate mixing as the main cause of background bias. In mixed X-ray datasets, images from different classes come from dissimilar sources (hospitals), which can present unique background characteristics. Due to mixing, these characteristics become correlated to the classification labels, becoming background bias. In line with these conclusions, we find studies analyzing applications with mixed X-ray datasets and showing large o.o.d. generalization gaps, shortcut learning and strong background bias\cite{ShortcutCovid}. Meanwhile, other works have trained DNNs on large single-source chest X-ray datasets and displayed smaller o.o.d. generalization gaps\cite{LabelShift}. Moreover, they pointed out other causes for such gaps, like label shift between different datasets\cite{LabelShift}.

The multiple experiments in this study clearly delimit the ISNet best use case. First, datasets without background bias are not the ideal application for the model, as shown by the results in this section. Furthermore, the ISNet is not meant to boost i.i.d. accuracy. As clearly seen in the tests with artificially biased datasets, focus on background bias improves i.i.d. performances. Thus, by avoiding such attention, the ISNet can reduce accuracy on i.i.d. evaluation databases, to enhance real-world performance. The ideal use for the ISNet, according to our experiments, is to improve generalization when the training dataset presents background bias. The occurrence of background bias is common in practical applications, as was demonstrated by the tasks of COVID-19 detection and TB detection. The ISNet strongly improved o.o.d. generalization for the tasks. The architecture's efficiency in dealing with background bias was further exemplified by the datasets containing artificial background bias. The presence of background bias can be revealed by large gaps between i.i.d. and o.o.d. performances\cite{ShortcutCovid}\textsuperscript{,}\cite{ShortcutLearning}, along with strong background attention in a standard classifier's explanation heatmaps (main article Figures 1 and 2). However, we note that shortcut learning is not only caused by background bias\cite{ShortcutLearning}, e.g., foreground features may also be biased. Moreover, besides shortcut learning, generalization gaps can have other causes, such as label shift between databases\cite{LabelShift}.

\section{The Denoising Property of LRP-$\varepsilon$}
\label{lrpVsGradInput}

An early DNN explanation technique is saliency maps\cite{saliency}, which consist of the absolute value of the DNN logit's gradient with respect to the input image ($\mathrm{abs}(\nabla_{\mathbf{X}}(y_{c}))$). Basically, large values in the absolute input gradient indicate features whose perturbation would cause large logit changes. Thus, saliency maps consider input pixels with large gradients (in terms of magnitude) as important for the DNN decision. However, saliency maps are considered noisy, and the absolute value operation avoids the differentiation of positive and negative class evidence\cite{LRPvsGrad}. Gradient*Input explanations\cite{GradInput} create heatmaps with an element-wise multiplication between the input gradient and the input itself, i.e., $\mathbf{X} \odot \nabla_{\mathbf{X}}(y_{c})$. The explanations should improve sharpness over saliency maps and differentiate positive and negative evidence\cite{LRPvsGrad} (i.e., input features that would increase or decrease the class c logit, $y_{c}$, if they were higher). It has been proposed\cite{LRPBook} that LRP-0\cite{LRP} heatmaps are equivalent to Gradient*Input explanations\cite{GradInput} when the only non-linearity in the DNN is the ReLU function. In Appendix \ref{equivalence}, we formally demonstrate the equivalence.

Here, we point out a notation inconsistency. Some works mention an equivalence between LRP-$\varepsilon$ and Gradient*Input, but they demonstrate such equivalence by setting $\varepsilon=0$. Thus, the equality is being proved for LRP-0, not LRP-$\varepsilon$. LRP-0 can be seen as the extreme case of LRP-$\varepsilon$ (with $\varepsilon$ set as its smallest possible value, 0). However, calling LRP-0 as LRP-$\varepsilon$ is confusing, especially considering the importance of the $\varepsilon$ constant, it is not a term that can be ignored.  In relation to LRP-0, LRP-$\varepsilon$ explanations are less noisy and more coherent\cite{LRPBook}. From the deep Taylor decomposition standpoint, they represent a reduction of the first-order Taylor approximation residuum\cite{LRPBook} (main article section ``ISNet theoretical fundamentals''). Such differences are made clear by the superior results of optimizing LRP-$\varepsilon$ (ISNet) in comparison Gradient*Input optimization (ISNet Grad*Input), as shown in main article Table 1. 

In Appendix \ref{LRPDenoise}, we provide an alternative view of LRP-$\varepsilon$: we define it as a modified gradient backpropagation procedure, followed by an element-wise multiplication of the back-propagated quantity and the neural network input. This view clarifies the effect of the $\varepsilon$ hyper-parameter over the signal being back-propagated, making the LRP-$\varepsilon$'s denoising quality explicit: the higher the $\varepsilon$, the more attenuated the impact of small neuron activations over the back-propagated signal. If $\varepsilon=0$, the modified procedure matches standard gradient backpropagation, and the final heatmaps match LRP-0 or, equivalently, Gradient*Input. In summary, we show that the hyper-parameter $\varepsilon$ controls a special denoising process, justified by the framework of deep Taylor decomposition, and important for the stable convergence of the ISNet optimization.

\subsection{LRP-0 and Gradient*Input Equivalence}
\label{equivalence}
The equivalence between LRP-0 and Gradient*Input is valid for neural networks with ReLU activations. Moreover, we must consider that no division by zero is performed over the entire LRP-0 procedure; i.e., none of the neural network layers outputs a zero element $z_{k}$ (main article Equation 2). In practice, the LRP-0 propagation over all DNN layers is numerically unstable. Assuming stability, we demonstrate the two explanations' equivalence for a fully-connected neural network. Such demonstration does not lack generality, because convolutional layers can be expressed as fully-connected layers with sparse parameters, while batch normalization and pooling can be also reformulated as equivalent convolutions or dense layers.

We start by rearranging the LRP-0 rule (main article Equation 2). We now include superscripts to indicate the layer number. The value $w_{jk}^{L}$ refers to the weights connecting layer L input $a_{j}^{L}$ to its output $z_{k}^{L}$ (before the ReLU activation), $R_{k}^{L+1}$ is the LRP relevance of layer L output $z_{k}^{L}$ (which is also the relevance at the layer L+1 input, $a_{k}^{L+1}$), and $R_{j}^{L}$ is the relevance of layer L's input, $a_{j}^{L}$. The rearranged LRP-$\varepsilon$ rule is:

\begin{gather}
\label{lrp0}
G_{j}^{L}=\sum_{k}\frac{w_{jk}^{L}R_{k}^{L+1}}{z_{k}^{L}}\\
\label{q}
\mbox{where: } G_{j}^{L}=R_{j}^{L}/a_{j}^{L}
\end{gather}

Considering the ReLU activation, we express the forward-pass of a dense layer in Equation \ref{fwd}. Notice that the ReLU output for layer L is also the input of layer L+1 ($a_{k}^{L+1}$).

\begin{gather}
\label{fwd}
a_{k}^{L+1}=\mathrm{ReLU}(z_{k}^{L})=
\begin{cases}
z_{k}^{L}, \mbox{ if } z_{k}^{L} > 0 \\
0,  \mbox{ otherwise }
\end{cases}\\
\mbox{where: } z_{k}^{L}=\sum_{j} w_{jk}^{L}a_{j}^{L}
\end{gather}

For layer L+1, we utilize the subscript k to refer to its inputs, as they are element-wise functions (ReLU) of layer L outputs, which also use subscripts k. Considering Equations \ref{q} and \ref{fwd}, we can reformulate main article Equation 2:

\begin{equation}
\label{lrp0b}
G_{j}^{L}=\sum_{k}\frac{w_{jk}^{L}R_{k}^{L+1}}{z_{k}^{L}}=
\sum_{k}\frac{w_{jk}^{L}G_{k}^{L+1}a_{k}^{L+1}}{z_{k}^{L}}=
\sum_{k}w_{jk}^{L}G_{k}^{L+1}\frac{\mathrm{ReLU}(z_{k}^{L})}{z_{k}^{L}}
\end{equation}

We simplify Equation \ref{lrp0b} by using the unit step function, $H(\cdot)$:

\begin{gather}
\label{QRule}
G_{j}^{L}=\sum_{k}w_{jk}^{L}H(z_{k}^{L})G_{k}^{L+1}\\
\mbox{where: } H(z_{k}^{L})=
\label{unitStep}
\begin{cases}
1, \mbox{ if } z_{k}^{L} > 0 \\
0, \mbox{ if } z_{k}^{L} < 0 
\end{cases}
\end{gather}

Equation \ref{QRule} expresses a recursive propagation rule (from the neural network logit to the input layer) of the quantity $G_{j}^{L}$, which is associated to the layer 
L input $a_{j}^{L}$. The new propagated signal is not conservative ($\sum_{j} G_{j}^{L}$ is not the same for all layers L). However, the semi-conservative\cite{LRP} LRP relevance, $R_{j}^{L}$, can always be reconstructed with the element-wise multiplication of the $G_{j}^{L}$ quantity and the layer L input value itself, $a_{j}^{L}$ (Equation \ref{q}). Therefore, if we propagate $G_{j}^{L}$ until the neural network input layer (producing the tensor $\mathbf{G^{0}}$, with elements $G_{j}^{0}$), we can obtain the LRP-0 heatmap ($\mathbf{R^{0}}$, with elements $R_{j}^{0}$) by element-wise multiplying $\mathbf{G^{0}}$ and the DNN input image ($\mathbf{X}$, with elements $X_{j}=a_{j}^{0}$). Thus, the LRP-0 propagation is equivalent of the propagation of $G_{j}^{L}$ followed by an element-wise multiplication with the DNN input ($\mathbf{G^{0}}\odot \mathbf{X}$).

\begin{equation}
\label{finalMap}
R_{j}^{0}=G_{j}^{0}X_{j}
\end{equation}

Now, to prove the equivalence between LRP-0 and Gradient*Input heatmaps, we must simply show that $G_{j}^{0}$ is the gradient of the DNN logit with respect to the input element $X_{j}$, i.e., $G_{j}^{0}=\partial y_{c}/\partial X_{j}$. Consider $y_{c}$ the neural network logit for class c, and assume that we are propagating the gradient (and the LRP-0 relevance) for class c. The logit gradient backpropagation (backward-pass from output to input) through the fully-connected layer L can be derived according to Equation \ref{chainRule}, which follows the calculus' chain rule.

\begin{equation}
\label{chainRule}
    \frac{\partial y_{c}}{\partial a_{j}^{L}}= \sum_{k} \frac{\partial z_{k}^{L}}{\partial a_{j}^{L}} \frac{\partial y_{c}}{\partial z_{k}^{L}}= 
    \sum_{k}  w_{jk}^{L} \frac{\partial y_{c}}{\partial z_{k}^{L}}=
    \sum_{k}  w_{jk}^{L} \frac{\partial a_{k}^{L+1}}{\partial z_{k}^{L}}\frac{\partial y_{c}}{\partial a_{k}^{L+1}}=
    \sum_{k}  w_{jk}^{L} \frac{\partial \mathrm{ReLU}(z_{k}^{L})}{\partial z_{k}^{L}}\frac{\partial y_{c}}{\partial a_{k}^{L+1}}
\end{equation}

The derivative of the ReLU function is the unit step function, $H(\cdot)$. Thus, we have a simple recurrence rule for the gradient backpropagation in DNNs composed of fully-connected layers with ReLU activation:

\begin{equation}
\label{gradRecurrent}
    \frac{\partial y_{c}}{\partial a_{j}^{L}}= \sum_{k} w_{jk}^{L} H(z_{k}^{L}) \frac{\partial y_{c}}{\partial a_{k}^{L+1}}
\end{equation}

We see that the recurrent backpropagation rule for the gradient $\partial y_{c}/\partial a_{j}^{L}$ (Equation \ref{gradRecurrent}) is the same as the rule for the quantity $G_{j}^{L}$ (Equation \ref{QRule}). Now we compare the initial values of $G_{j}^{L}$ and $\partial y_{c}/\partial a_{j}^{L}$, i.e., the G value and gradient for the outputs of the DNN last layer. The DNN logits are indicated as $y_{k}=z_{k}^{Lmax}$, where Lmax indicates the DNN last layer. Starting with the gradient, we have:

\begin{gather}
\label{initialGRad}
\frac{\partial y_{c}}{\partial y_{k}}=
\begin{cases}
1, \mbox{ if } k=c \\
0,  \mbox{ otherwise }
\end{cases}
\end{gather}

LRP-0 considers that the logit relevance is the logit value for the explained class logit ($R_{c}^{logits}=y_{c}$), and 0 otherwise ($R_{c' \neq c}^{logits}=0$)\cite{LRP}. Thus, considering Equation \ref{q}, we can define the G quantity for the logits, $G_{k}^{logits}$:

\begin{gather}
\label{initialQ}
G_{k}^{logits}=
\begin{cases}
1, \mbox{ if } k=c \\
0,  \mbox{ otherwise }
\end{cases}
\end{gather}

As we are explaining logits, we disconsider the last layer non-linear activation, viewing it as a linear layer. Thus, for back-propagating gradients ($\partial y_{c}/\partial y_{k}$) or $G_{k}^{logits}$ through it, we ignore the unit step functions in Equations \ref{gradRecurrent} and \ref{QRule}, producing:

\begin{equation}
\label{iGrad}
    \frac{\partial y_{c}}{\partial a_{j}^{Lmax}}=  w_{jc}^{Lmax}
\end{equation}

\begin{equation}
\label{iQ}
G_{j}^{Lmax}=w_{jc}^{Lmax}
\end{equation}

Equations \ref{iGrad} and \ref{iQ} prove that the starting values for $\partial y_{c}/\partial a_{j}^{L}$ and $G_{j}^{L}$ are the same. Because the two quantities are initially identical and follow the same propagation rules (Equations \ref{gradRecurrent} and \ref{QRule}), they are equal throughout the entire propagation procedure:

\begin{equation}
\label{equals}
G_{j}^{L}=\frac{\partial y_{c}}{\partial a_{j}^{L}}
\end{equation}

Taking into account Equations \ref{equals} and \ref{q}, we conclude that Gradient*Input ($X_{j}\partial y_{c}/\partial X_{j}$) explanations are equivalent to LRP-0 heatmaps ($R_{j}^{0}$), considering the requirements that ReLU is the only non-linearity in the DNN, and that LRP-0 is numerically stable:

\begin{equation}
X_{j}\frac{\partial y_{c}}{\partial X_{j}}=X_{j}G_{j}^{0}=a_{j}^{0}G_{j}^{0}=R_{j}^{0}
\end{equation}

\subsection{LRP-$\varepsilon$ Denoising Property and Advantages over LRP-0 and Gradient*Input}
\label{LRPDenoise}

We must emphasize the differences between LRP-$\varepsilon$ and LRP-0. Besides ensuring numerical stability (by avoiding divisions by zero), the $\varepsilon$ stabilizer (main article Equation 3) reduces the explanation noise, improves heatmap coherence and contextualization, and reduces the Taylor approximation error\cite{LRP}\textsuperscript{,}\cite{LRPBook} (main article section ``ISNet theoretical fundamentals''). When $\varepsilon$ is not zero, Gradient*Input is not equivalent to LRP-$\varepsilon$. In this case, Equation \ref{LRP-e_Q} substitutes Equation \ref{lrp0b}. In summary, the LRP-$\varepsilon$ heatmap can be obtained by backpropagating the quantity G, according to the recursive application of Equation \ref{LRP-e_Q}, considering the starting values in Equation \ref{initialQ}, and finalizing with the element-wise multiplication between $\mathbf{G^{0}}$ and the input image (Equation \ref{finalMap}). 

\begin{gather}
\label{LRP-e_Q}
G_{j}^{L}=\sum_{k}\frac{w_{jk}^{L}R_{k}^{L+1}}{z_{k}^{L}+\mathrm{sign}(z_{k}^{L})\varepsilon}=
\sum_{k}\frac{w_{jk}^{L}G_{k}^{L+1}a_{k}^{L+1}}{z_{k}^{L}+\mathrm{sign}(z_{k}^{L})\varepsilon}=
\sum_{k}w_{jk}^{L}G_{k}^{L+1}\frac{\mathrm{ReLU}(z_{k}^{L})}{z_{k}^{L}+\mathrm{sign}(z_{k}^{L})\varepsilon}=
\sum_{k}w_{jk}^{L}A(z_{k}^{L})G_{k}^{L+1} \\
\label{AS}
\mbox{where: } A(z_{k}^{L})=\frac{\mathrm{ReLU}(z_{k}^{L})}{z_{k}^{L}+\mathrm{sign}(z_{k}^{L})\varepsilon}=
\begin{cases}
\frac{z_{k}^{L}}{z_{k}^{L}+\varepsilon}, \mbox{ if } z_{k}^{L} \geq 0 \\
0,  \mbox{ otherwise }
\end{cases}
\end{gather}

\begin{figure}[!h]
\includegraphics[width=0.6\textwidth]{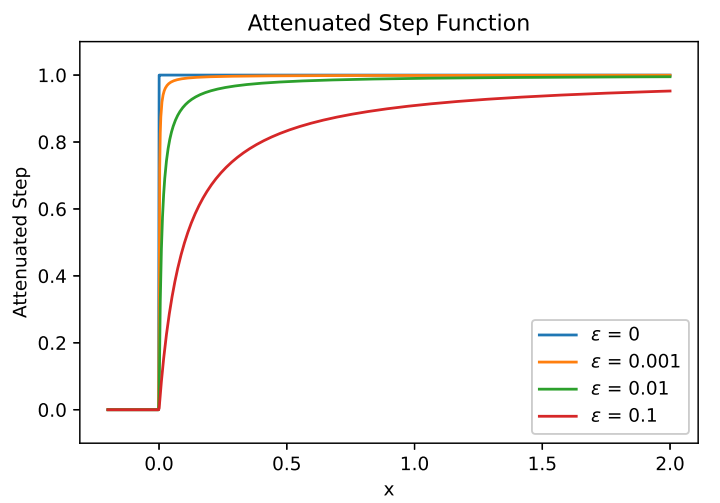}
\centering
\caption{\textbf{Illustration of the attenuated unit step function implicitly used by LRP-$\varepsilon$ to denoise the signal backpropagation.} Different lines correspond to diverse values of $\varepsilon$, indicated in the image's bottom right corner. The attenuated step is defined as $\mathrm{ReLU}(x)/(x+\mathrm{sign}(x)\varepsilon)$, where ReLU is the rectified linear unit and sign($\cdot$) the sign function}
\label{atStep}
\end{figure}

The function $A(z_{k}^{L})$ (Equation \ref{AS}) is not the unit step (Equation \ref{unitStep}) if $\varepsilon \neq 0$. Therefore, the propagation rule for the G quantity in LRP-$\varepsilon$ (Equation \ref{LRP-e_Q}) is different from the rule for the gradient backpropagation (Equation \ref{gradRecurrent}). Accordingly, for LRP-$\varepsilon$, G is not the gradient. Thus, the resulting LRP-$\varepsilon$ heatmap is not equivalent to a Gradient*Input explanation. Equation \ref{AS} can be seen as an attenuated unit step function ($\varepsilon$ is a positive constant, empirically set to 0.01 in our experiments). It is a non-linear continuous function, which is zero for negative $z_{k}^{L}$ (like the unit step), but that monotonically increases for positive $z_{k}^{L}$, saturating at 1 (which is the unit step output for all positive $z_{k}^{L}$). The lower the parameter $\varepsilon$, the faster $A(z_{k}^{L})$ approximates 1, matching the discontinuous unit step function when $\varepsilon=0$. The function approximates the unit step for large values of $z_{k}^{L}$ ($z_{k}^{L}>>\varepsilon$), but its outputs are smaller for small $z_{k}^{L}$. Accordingly, the propagation rule in Equation \ref{LRP-e_Q} approximates the gradient backpropagation (Equation \ref{gradRecurrent}) when $G_{k}^{L+1}$ is associated to a large or negative neuron output ($z_{k}^{L}>>\varepsilon$ or $z_{k}^{L}<0$). However, the attenuated step reduces the impact of small neuron activations ($z_{k}^{L}$) on the propagated quantity ($G_{j}^{L}$). I.e., for small but positive $z_{k}^{L}$, the resulting small $A(z_{k}^{L})$ makes the $w_{jk}^{L}A(z_{k}^{L})G_{k}^{L+1}$ term (Equation \ref{LRP-e_Q}) significantly smaller than $w_{jk}^{L}H(z_{k}^{L})G_{k}^{L+1}$, the equivalent term in gradient backpropagation (Equation \ref{QRule} or \ref{gradRecurrent}). The attenuated unit step function is illustrated in Figure \ref{atStep} for multiple values of $\varepsilon$. Higher $\varepsilon$ more strongly attenuates the importance of small neuron activations on $G_{j}^{L}$, and $G_{j}^{L}$ matches the gradient when $\varepsilon=0$, representing no attenuation. Therefore, we demonstrated that $\varepsilon$ controls a denoising process, which is justified by the deep Taylor decomposition (DTD) framework; from the DTD point of view, $\varepsilon$ can reduce the Taylor first-order approximation error, and lead to more coherent and contextualized explanations\cite{LRPBook}.

The ISNet Grad*Input is an ablation experiment, where the ISNet LRP heatmaps were substituted by Gradient*Input explanations. Its comparison to the ISNet empirically demonstrated that the effect caused by $\varepsilon$ is important for the stability and convergence of the ISNet loss functions, resulting in higher bias resistance, easier and more stable training, and better accuracy, especially for deeper architectures (refer to the main article Table 1, main article section ``ISNet theoretical fundamentals'', and Appendix \ref{baselineComparisons}).

\section{Speed and Efficiency Analysis}
\label{speed}

Models were trained with an NVIDIA Tesla V100 graphics processing unit (GPU), from a HPC (High Performance Computing) structure, or with an NVIDIA RTX 3080. The performances reported in this subsection consider the RTX 3080 and mixed precision. In facial attribute estimation, an epoch considered 24183 training images and 2993 validation samples, loaded in mini-batches of 10 figures. An epoch took about 2300 s for the ISNet Grad*Input, 1300 s for the ISNet, 1500 s for GAIN, 900 s for RRR, 320 s for the alternative segmentation model, 240 s for the standalone DenseNet121, 250 s for the multi-task U-Net, 160 seconds for the Vision Transformer, and 60 s for the AG-Sononet. It is worth noting that the alternative pipeline also required training a U-Net, which needed 400 epochs of about 180 s. Considering this additional training procedure, the ISNet's training time was about 82\% longer than the alternative segmentation-classification pipeline's. Attention gates are not a computationally expensive methodology during both training and run-time. Meanwhile, the multi-task model requires sensible additional computation to create the segmentation outputs. It was faster than the U-Net followed by the DenseNet121, but we must consider that its classification path is based on a different architecture (VGG-16). The ISNet needs to generate one heatmap per class, while GAIN produces one Grad-CAM explanation for each ground-truth class in the image, uses it to mask the original sample, and classifies it again. Thus, GAIN training tends to be slower than the ISNet optimization in multi-label problems, but faster in single-label tasks. Both models have similar run-time speed, as does RRR. In the remainder of this subsection, we will mostly compare the ISNet performance to the alternative segmentation-classification pipeline's, since they are the tested methods that proved most usable for the objective of avoiding shortcut learning caused by background bias. Furthermore, both methods were based on the same classifier (we now consider the DenseNet121-based models), allowing a fairer comparison.

During training, the ISNet is slower than a traditional pipeline of segmentation and classification. The model's biggest limitation may be the need to generate one heatmap per class. Treating the different heatmap calculations as a propagation of different batch elements allows parallelism, but training time and memory consumption still increases with more classes. The worst-case scenario for this problem is if memory is not sufficient for the batch approach. In this case, the heatmaps will need to be created in series, causing training time to increase linearly with the number of classes. Therefore, more efficient implementations for the LRP block are a promising path for future developments.

At run-time, the ISNet shows clear benefits because the LRP block can be removed from the model, leaving only the classifier. Thus, the DenseNet121-based ISNet matches the speed of a DenseNet121 at run-time. With mixed precision and using mini-batches of 10 images, the standard pipeline of a U-Net followed by a DenseNet121 classifies an average of 207 samples per second, while the DenseNet121-based ISNet classifies an average of 353. Utilizing the same mini-batch size, but without mixed precision, the alternative pipeline classifies 143 samples per second, and the ISNet, 298. Therefore, the ISNet is about 70\% to 108\% faster in this configuration. However, there is an even greater advantage regarding the model's size: the ISNet has about 8M parameters (the same as the DenseNet121), while the combined U-Net and DenseNet121 have 39M. Thus, the model size is reduced by a factor of almost 5. Naturally, the smaller the classifier model in relation to the segmentation network, the stronger the performance benefit provided by the ISNet architecture.

\section{Statistical Methods for Results' Analysis}
\label{statisticalMethods}

Dog breed classification and the classification of chest X-rays as COVID-19, normal or pneumonia constitute multi-class, single-label problems. To find interval estimates for the classification performance metrics we employed Bayesian estimation. We utilized a Bayesian model\cite{bayesianEstimator} that takes a confusion matrix (as those displayed in main article Table 4) as input and estimates the posterior probability distribution of class and average precision, recall and F1-Score. The model was expanded in a subsequent study\cite{bassi2021covid19}, to also estimate class and average specificity. For the posteriors' estimation we used Markov chain Monte Carlo (MCMC), implemented with the Python library PyMC3\cite{pymc}. We employed the No-U-Turn Sampler\cite{NUTS}, using 4 chains and 110000 samples, being the first 10000 for tuning. The estimated distributions allow us to report the means, standard deviations, and the 95\% highest density intervals of the performance metrics. The estimation's Monte Carlo error is below 10$^{-4}$ for all scores. We adopted the same priors as the Bayesian model creators\cite{bayesianEstimator}. We utilized a macro-averaging and pairwise approach to calculate the overall area under the ROC curve (AUC), a technique developed for multi-class single-label problems\cite{MulticlassAUC}. We do not present interval estimates for this metric, because there is not an established methodology to calculate it (its authors\cite{MulticlassAUC} suggest bootstrapping, but it would not be feasible with the deep models and cross-dataset evaluation used in this study). We utilize the standard one-versus-rest methodology to calculate the per-class ROC-AUC. Thus, we can employ an already established non-parametric approach\cite{DeLongAUC} to obtain the class AUCs' 95\% confidence intervals.

Facial attribute estimation is a multi-class multi-label classification problem, as is CheXPert classification (Appendix \ref{CheXpertClassification}). Therefore, the aforementioned Bayesian model\cite{bayesianEstimator} is not adequate for it. To create the interval estimates for the performance metrics in this scenario, we employ the Wilson Score Interval. Working with a multi-label task, we calculate the AUC 95\% confidence intervals with the same non-parametric technique we previously utilized\cite{DeLongAUC}, as we consider the standard one-versus-rest AUC again.

In tuberculosis detection we deal with a binary classification task. We also utilize the Wilson Score Interval to create 95\% confidence intervals and use the aforementioned non-parametric technique to calculate the ROC-AUC interval estimate, considering the standard one-versus-rest AUC calculation once more\cite{DeLongAUC}. 

Statistics are calculated over the test datasets, which are explained in Appendix \ref{dataset}. In summary, the COVID-19 test database has n=3126 samples (406 normal, 1295 pneumonia, and 1515 COVID-19), the tuberculosis i.i.d. test dataset has n=1284 samples (642 tuberculosis and 642 normal), the tuberculosis o.o.d. dataset has n=753 (370 normal and 383 TB-positive), the facial attribute estimation test dataset has n=2824 samples (333 showing rosy cheeks, 1369 high cheekbones and 1339 smiling, the dataset is multi-label), the Stanford Dogs evaluation set has n=201 images (100 Pug, 52 Tibetan Mastiff, 49 Pekingese), and the MIMIC test dataset had n=2265 samples (763 atelectasis, 210 consolidation, 722 edema, 593 no finding, 340 pneumonia, and 108 pneumothorax, the dataset is multi-label).

\end{document}